\documentclass[a4paper,fleqn,usenatbib]{mnras}

\usepackage[T1]{fontenc}
\usepackage{ae,aecompl}

\usepackage{graphicx}	
\usepackage{amsmath}	
\usepackage{amssymb}	
\usepackage{booktabs}
\hypersetup{pageanchor=false}

\title[Central H \textsc{ii} region in NGC 4449 ]{A GMOS-N IFU study of the  central H \textsc{II} region in the blue compact dwarf galaxy NGC 4449: Kinematics, Nebular Metallicity and Star-Formation\thanks{Based on observations obtained at the Gemini Observatory (processed using the Gemini IRAF package), which is operated by the Association of Universities for Research in Astronomy, Inc., under a cooperative agreement with the NSF on behalf of the Gemini partnership: the National Science Foundation (United States), the National Research Council (Canada), CONICYT (Chile), Ministerio de Ciencia, Tecnolog\'{i}a e Innovaci\'{o}n Productiva (Argentina), and Minist\'{e}rio da Ci\^{e}ncia, Tecnologia e Inova\c{c}\~{a}o (Brazil).}}

\author[Kumari et al.]{
Nimisha Kumari$^{1}$\thanks{E-mail: nkumari@ast.cam.ac.uk(NK)},
Bethan L. James$^{1,2,3}$,
Mike J. Irwin$^{1}$\\
$^{1}$Institute of Astronomy, University of Cambridge, CB3 0HA UK \\
$^{2}$ Kavli Insititute of Cosmology, Cambridge UK\\
$^{3}$Space Telescope Science Institute, 3700 San Martin Dr, Baltimore, MD 21218}

\date{Accepted 2017 June 06 . Received 2017 May 24 ; in original form 2017 February 18}

\pubyear{2016}

\begin{document}
\label{firstpage}
\pagerange{\pageref{firstpage}--\pageref{lastpage}}
\maketitle

\begin{abstract}

\indent We use integral field spectroscopic (IFS) observations from the Gemini North Multi-Object Spectrograph (GMOS-N) to study the central H \textsc{ii} region in a nearby blue compact dwarf (BCD) galaxy NGC 4449. The IFS data enable us to explore the variation of physical and chemical conditions of the star-forming region and the surrounding gas on spatial scales as small as 5.5 pc. Our kinematical analysis shows possible signatures of shock ionisation and shell structures in the surroundings of the star-forming region. The metallicity maps  of the region, created using direct T$_e$ and indirect strong line methods (R$_{23}$, O3N2 and N2), do not show any chemical variation. From the integrated spectrum of the central H \textsc{ii} region, we find a metallicity of 12 + log(O/H) = 7.88 $\pm$ 0.14 ($\sim$ 0.15$^{+0.06}_{-0.04}$ Z$_{\odot}$) using the direct method. Comparing the central H \textsc{ii} region metallicity derived here with those of H \textsc{ii} regions throughout this galaxy from previous studies, we find evidence of increasing metallicity with distance from the central nucleus. Such chemical inhomogeneities can be due to several mechanisms, including gas-loss via supernova blowout, galactic winds, or metal-poor gas accretion. However, we find that the localised area of decreased metallicity aligns spatially with the peak of star-forming activity in the galaxy, suggesting that gas-accretion may be at play here.  Spatially-resolved IFS data for the entire galaxy is required to confirm the metallicity inhomogeneity found in this study, and  determine its possible cause. 
   
\end{abstract}




\section{Introduction}
\label{intro}

\indent Galaxy formation and evolution is regulated by a complex interplay between star-formation, chemical abundance and gas dynamics. Stars form from cool gas, which is either formed in-situ in the galaxy or is accreted from the circum/inter-galactic medium or from the merger progenitors. Metals form in stars and are ejected into the interstellar medium (ISM) by subsequent supernova explosions. The metal-enriched gas can get dispersed in the ISM due to the internal gas dynamics of the ISM or get transported out of the galaxy via supernova-driven outflows or galactic winds. This intricate ballet of gas flows in, out and within galaxies can lead to chemical inhomogeneities in the gas across the star-forming galaxies \citep{Cresci2010, Kewley2010, Dave2011, Sanchez2013, Sanchez2014, Sanchez2015, Ceverino2016}, and manifest itself in the form of ordered/disordered kinematics and photo/shock-ionisation of the ISM. Detailed spatially-resolved kinematical and chemical studies of star-forming regions can hence unravel some of the secrets of the key mechanisms involved in galaxy formation and evolution. However a definitive picture can only be obtained by carrying out detailed analyses in star-forming systems across cosmic history -- at low and high redshifts. The limited sensitivity and precision of present technology prevent us from studying the high-redshift star-forming systems in detail. Fortunately, the local Universe hosts a variety of analogues to the high-redshift star-forming galaxies, such as blue compact dwarf galaxies (BCDs), tadpole galaxies, extremely metal poor galaxies (XMPs), green pea galaxies etc., which can be studied in great detail because of their proximity  and hence enable us to probe and predict the properties of high-redshift galaxies. BCDs are low metallicity \citep[1/40 -- 1/3 Z$_{\odot}$;][]{KunthOstlin2000} star-forming  galaxies, as such they are thought to be excellent local analogues/proxies for high-redshift galaxies \citep{SearleSargent1972} and hence provide the best local laboratories to study the primordial Universe in detail.  

\indent BCDs have been the focus of many imaging \citep[e.g.][]{Papaderos1996a, Papaderos1996b, Cairos2001a, Cairos2001b, GildePaz2003, McQuinn2010,Roychowdhury2012,Janowiecki2014}, spectroscopic \citep[e.g.][]{IzotovThuan1999,IzotovThuan2004,IzotovThuan2016, Bergvall2002, Wu2006, Guseva2011} and integral field spectroscopic (IFS) studies \citep[e.g.][]{James2009, James2010, James2013, Kehrig2016, Lagos2016} for over two decades. IFS is the best available technique to study these galaxies which host H \textsc{ii} (star-forming) regions, because it not only allows us to access information encoded in the emission lines from the star-forming regions, but also enables us to \textit{map} the spatial distribution of the information (i.e. physical and chemical properties) from them.

\indent NGC 4449 is a nearby ($\sim$ 3.79 $\pm$ 0.57 Mpc), spectacularly luminous (M$_B$ = $-$18.2) BCD \citep{Ann2015}, which has been studied extensively using multiwavelength data. In this paper, we present spatially-resolved observations of the central H \textsc{ii} region in NGC 4449 taken with the integral field unit (IFU) on the Gemini Multi Object Spectrograph North (GMOS-N). The IFS data has allowed us to study the chemical and physical properties of this star-forming region at a spatial scale of 5.5 pc (i.e. per spaxel).  In our work, we mainly explore the following three questions with respect to this particular star-forming region: (a) Do chemical inhomogeneities exist at this scale? (b) What are the possible ionisation mechanisms at play in the gas surrounding the star-forming region? (c) What is the age of the stellar population currently ionising the gas here? In addressing these questions, our aim is to spatially-resolve the chemical and kinematic patterns in the central star-forming region of NGC 4449, and explore the interplay between star-formation, chemical abundance and gas kinematics/dynamics.
  
\begin{figure}
 \centering
\includegraphics[width = 0.5\textwidth]{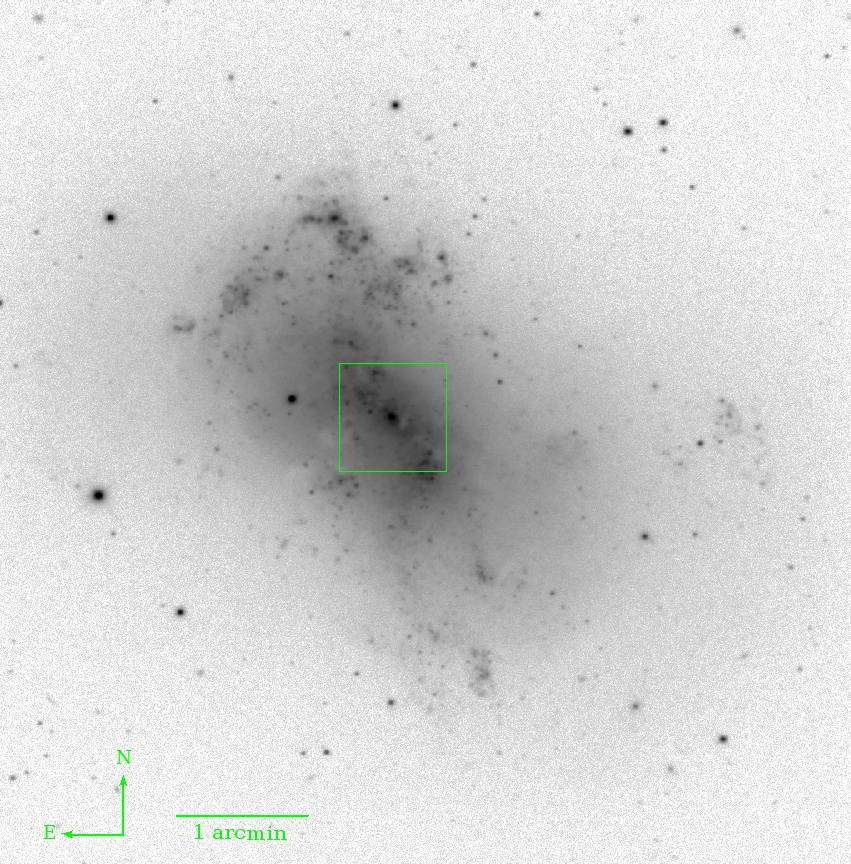}
\caption{SDSS image of NGC 4449 taken in r-band. The green square box denotes the region covered by the HST image shown in Figure \ref{HST}.}
\label{sdss}
\end{figure}

\begin{figure*}
 \centering
\includegraphics[width = \textwidth]{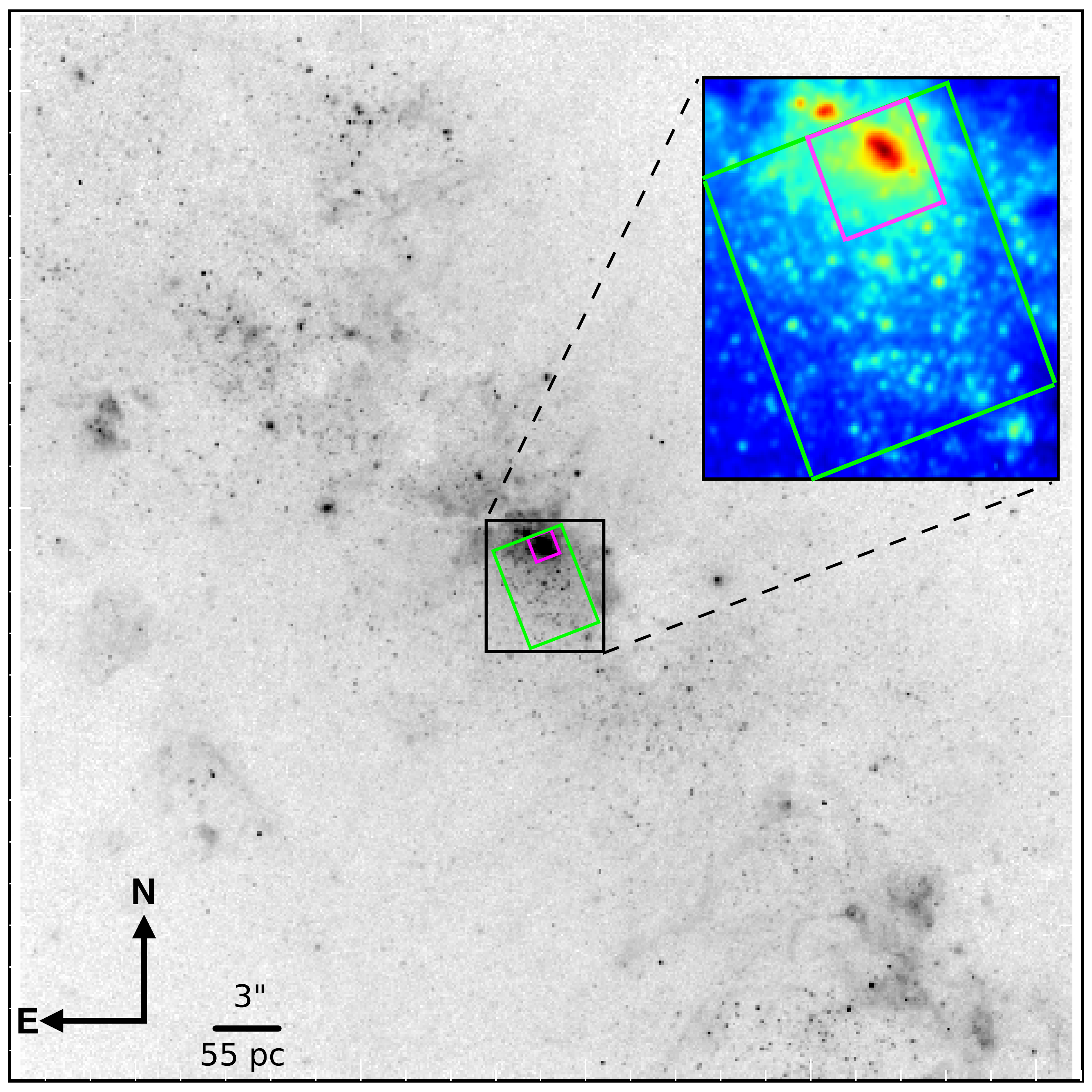}
\caption{HST image of NGC 4449 taken in the filter F660N centered at (RA, Dec) : (187.04657, 44.093941). The green box in the central region of the galaxy represents the GMOS aperture (3.5$\arcsec$ $\times$ 5$\arcsec$). The magenta box inside the green box shows the main emission region. The HST image has a spatial scale of 0.05$\arcsec$ pixel$^{-1}$ (0.92 pc pixel$^{-1}$) and are 51.2$\arcsec$ $\times$ 51.2$\arcsec$ in size. The upper-right inset is obtained by superimposing HST images in three filters F502N (blue), F550M (green) and F660N (red) and is in log scale.}
\label{HST}
\end{figure*}

\indent Figure \ref{sdss} shows the Sloan Digital Sky Survey (SDSS) image of NGC 4449 taken in r-band with the region of the HST cutout shown in Figure \ref{HST} hightlighted. General properties of NGC 4449 are given in Table \ref{tab properties}. It is also classified as a Magellanic-type irregular galaxy, with a high star formation rate (SFR) [current SFR $\sim$ 0.5--1.5 M$_{\odot}$ yr$^{-1}$, \citet{Hill1998, Hunter1998, Thronson1987}] and low metallicity [12 + log(O/H) $\sim$ 8.3, \citet{Lequeux1979, Martin1997}]. High-resolution imaging of this galaxy by the \textit{Hubble Space Telescope} (HST) (Figure \ref{HST}) reveals  populations of young and old stellar clusters which are indicative of an active star-formation history \citep{Annibali2011}. Long-slit \'echelle spectroscopy has revealed interesting kinematic features in NGC 4449 - expanding bubbles, supergiant shells and high velocity filaments, along with diffuse ionised gas \citep{Bomans2014,HunterGallagher1997}. The heavy-element (O \textsc{i}, Si \textsc{ii}, Fe \textsc{ii}) absorption lines of NGC 4449 suggest the occurrence of turbulent broadening within the neutral gas \citep{James2014}. Radio observations of this galaxy suggest that it is embedded in a large halo of H \textsc{i} gas which is rotating in the opposite direction to that of the main stellar component \citep{Bajaja1994}. The H \textsc{i} morphology and the velocity field indicates that the galaxy might have undergone an interaction with DDO 125 which is at an apparent separation of 41 kpc from the centre of NGC 4449) and is hypothesised to channel gas into the centre of NGC 4449 \citep{Hunter1998}. \citet{Martinez-Delgado2012} resolved a stream of red giant branch stars in the halo of NGC 4449 and thus speculated a scenario of ``stealth" merger which possibly triggered the starburst in this galaxy. The merger events can also possibly lead to variation in the metal content of the galaxy. Hence NGC 4449 is an excellent target to study the three aforementioned questions linking star-formation, metallicity and gas kinematics. 
   
\indent The paper is organised as follows: Section \ref{data} describes the observation and data reduction.  Section \ref{analysis} presents the preliminary procedures required for data analysis. Section \ref{kinematics} presents the spectral decomposition of the strong H$\alpha$ emission line, kinematics of different components, and their role in the ionisation of the surroundings of the H \textsc{ii} region are presented in Section \ref{ionisation}. In Section \ref{chemistry}, we present a detailed chemical abundance analysis by mapping metallicity using direct and indirect strong line methods; and also explore the chemical variation. Section \ref{stars} presents our analysis on age-dating the stellar population and star-formation. We present the interpretation of our analysis in Section \ref{discussion}. Section \ref{summary} summarises our major findings. Throughout this work, we have assumed the solar metallicity of 12 + log(O/H)$_{\odot}$ = 8.69 \citep{Asplund2009}. 

\begin{table}
\centering
\caption{General Properties of NGC 4449}
\label{tab properties}
\begin{tabular}{@{}lc@{}}
\toprule
Parameter                    & NGC 4449          \\ \midrule
Other designation            & UGC 07592         \\
Morphological Type           & IBm               \\
R.A. (J2000.0)               & 12:28:11.09      \\
DEC (J2000.0)                & 44:05:37.06       \\
Redshift (z)$^a$                 & 0.00069$\pm$0.00001 \\
Distance (Mpc)$^a$              & 3.79$\pm$0.57      \\
inclination ($^\circ$)$^a$        & 60\\
Helio. Radial Velocity(km s$^{-1}$)$^a$ & 207$\pm$4            \\
E(B-V)$^b$                       & 0.017$\pm$0.001    \\ \bottomrule
\multicolumn{2}{l}{$^a$ Taken from NED}\\
\multicolumn{2}{l}{$^b$ Foreground Galactic extinction \citep{Schlafly2011}}\\
\end{tabular}
\end{table}


\begin{figure*}
\centering
\includegraphics[width=0.95\textwidth]{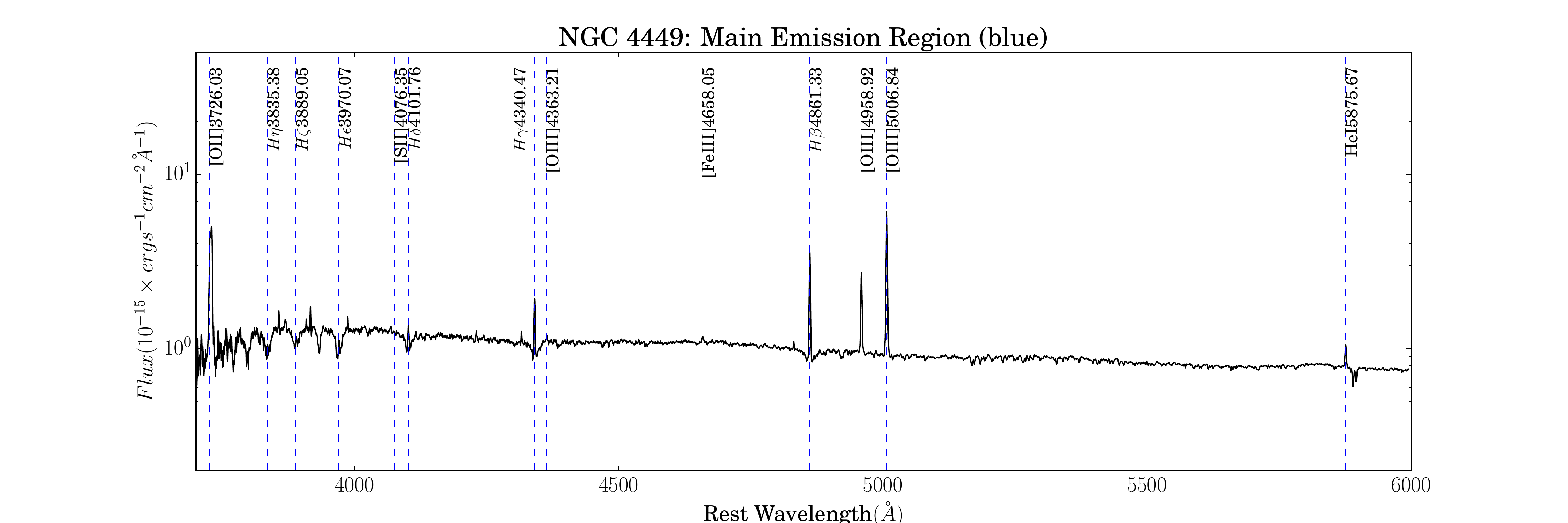}
\includegraphics[width=0.95\textwidth]{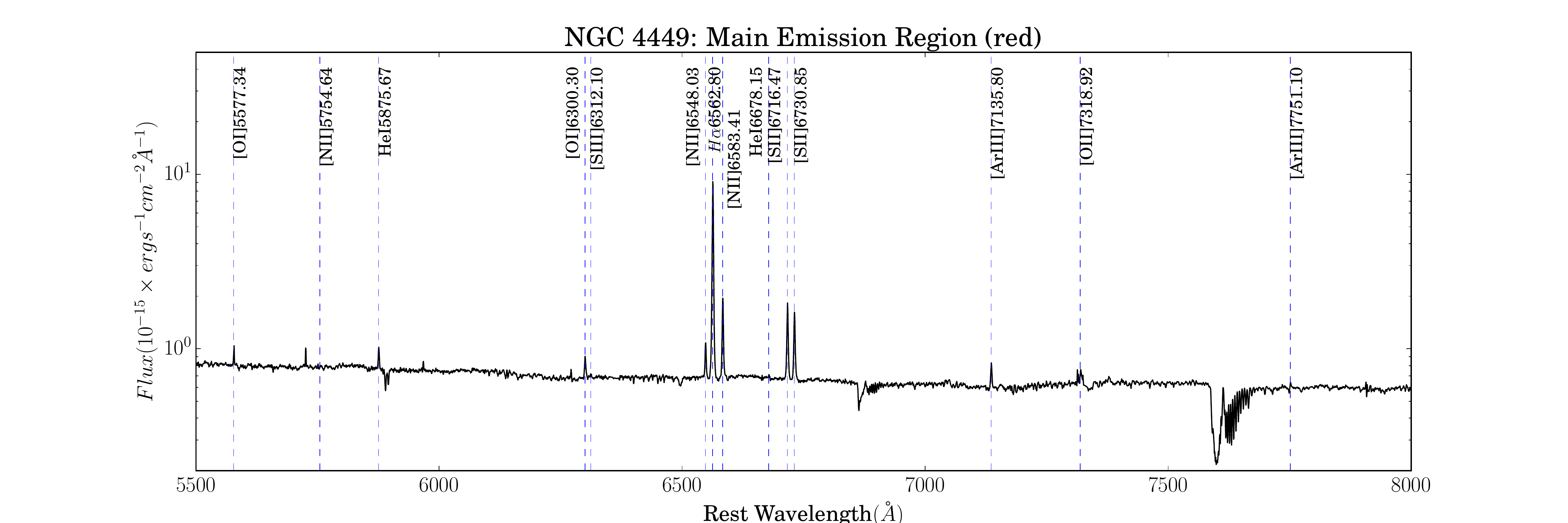}
\includegraphics[width=0.95\textwidth]{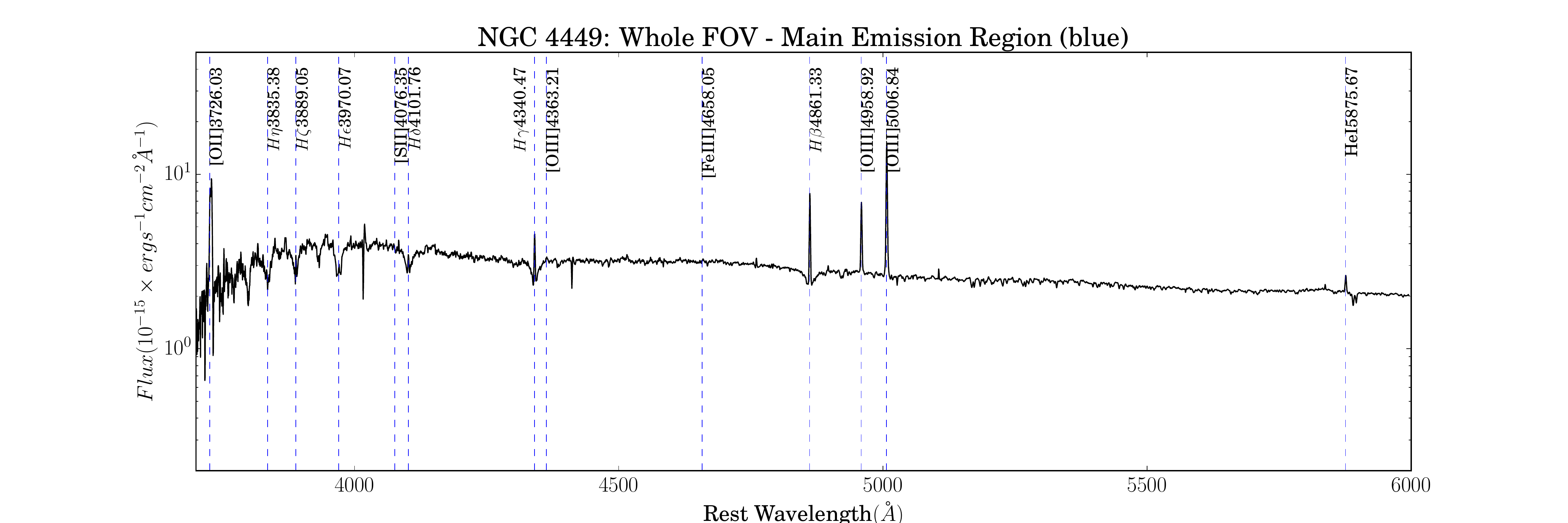}
\includegraphics[width=0.95\textwidth]{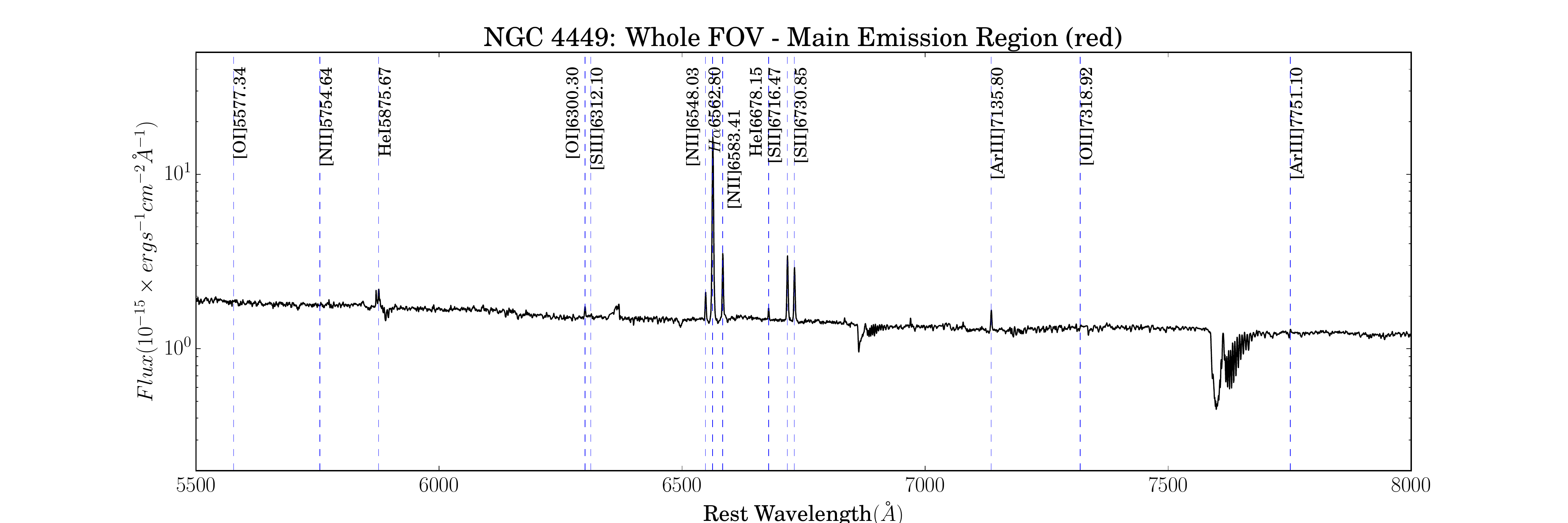}
\caption{GMOS-IFU integrated spectra of NGC 4449 integrated over the main (bright) emission region of 1.2$\arcsec$ $\times$ 1.2$\arcsec$ (upper two panels), and integrated over the remaining spaxels in the FOV (whole FOV -- main emission region, lower two panels) for each grating (red and blue) (see Figure \ref{HST}). The principal emission lines are over-plotted at their rest wavelengths.}
\label{spectra}
\end{figure*}

\begin{figure*}
\centering
\includegraphics[width = 0.48\textwidth]{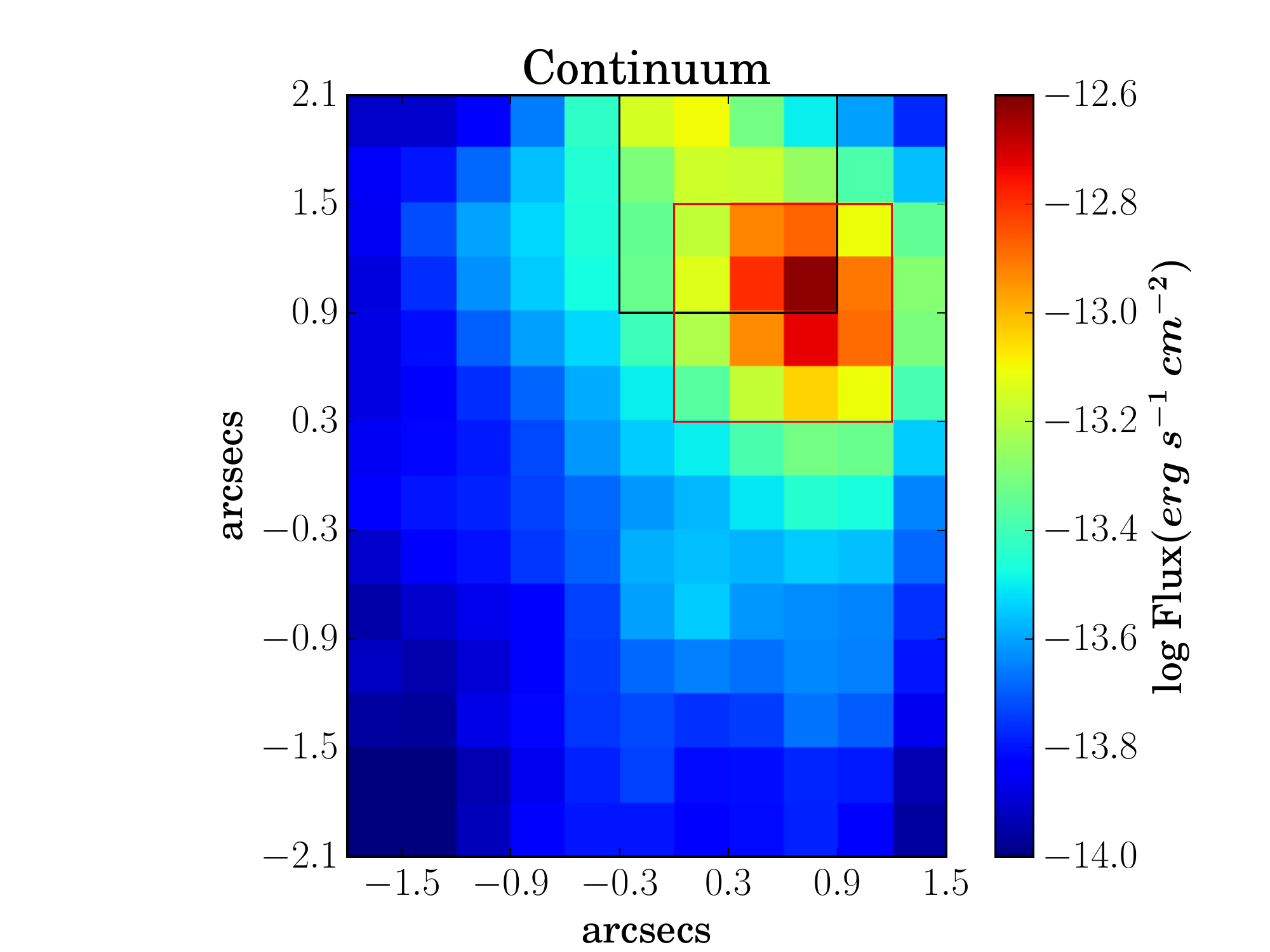}
\includegraphics[width = 0.48\textwidth]{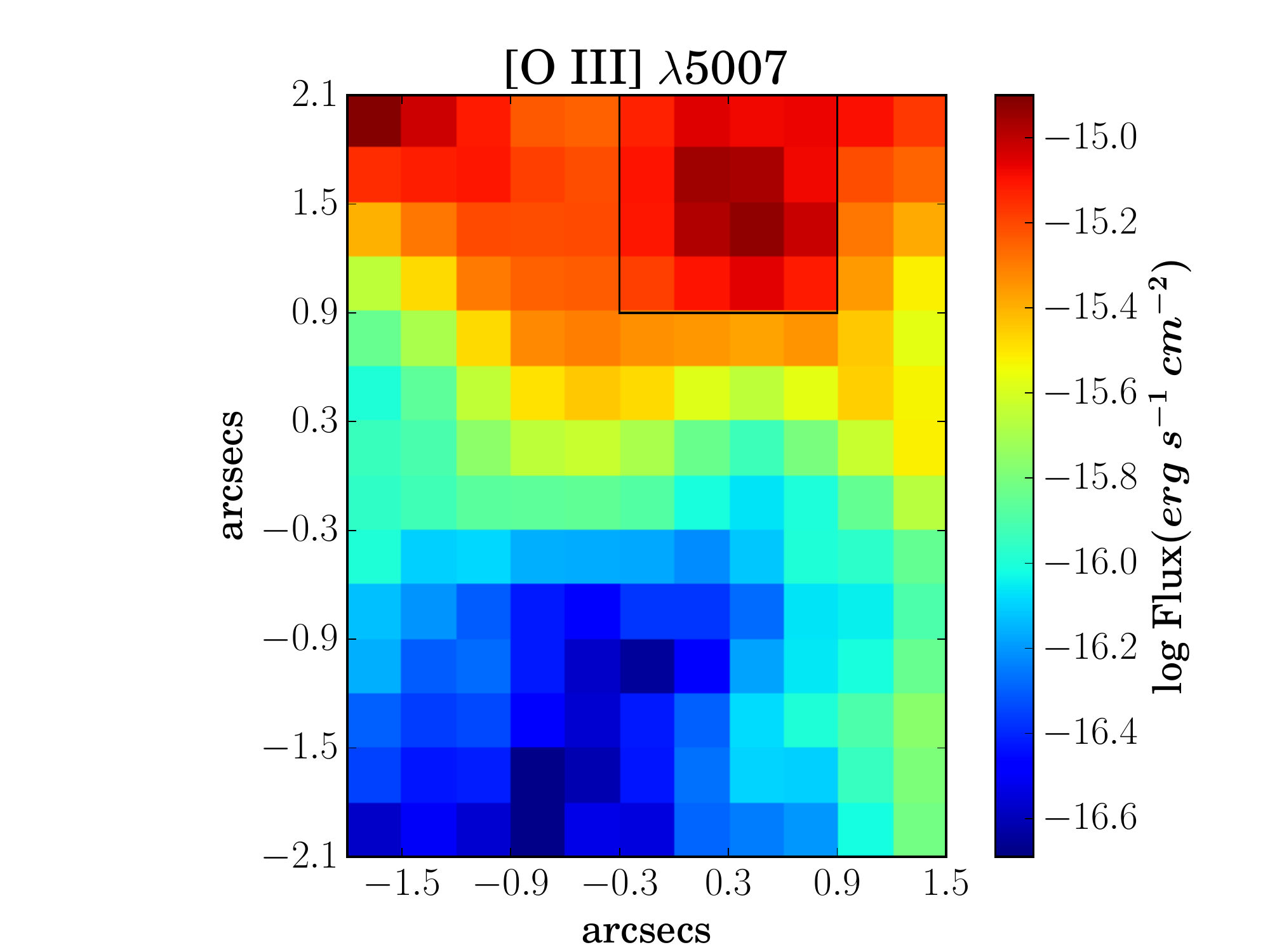}
\includegraphics[width = 0.48\textwidth]{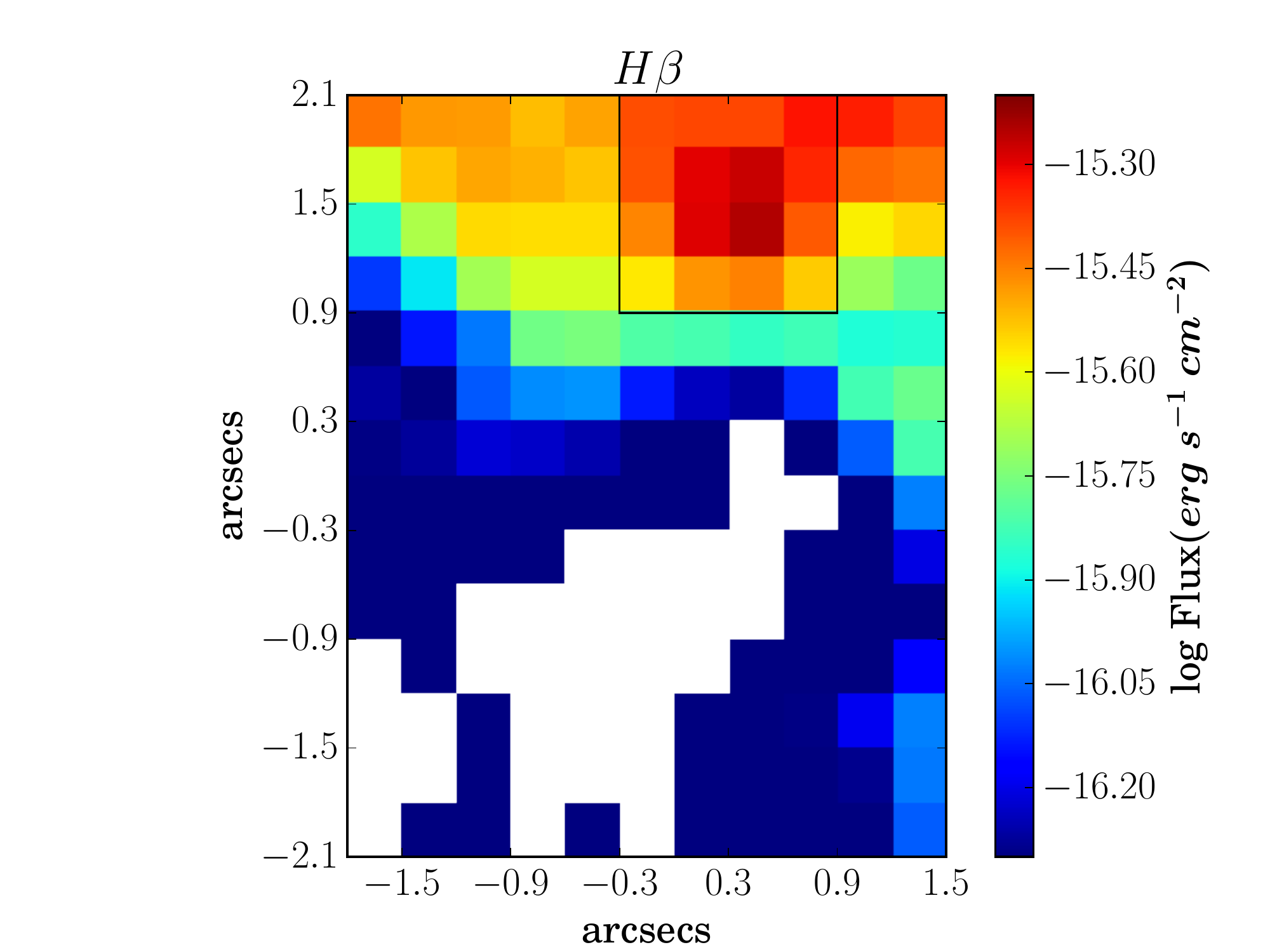}
\includegraphics[width = 0.48\textwidth]{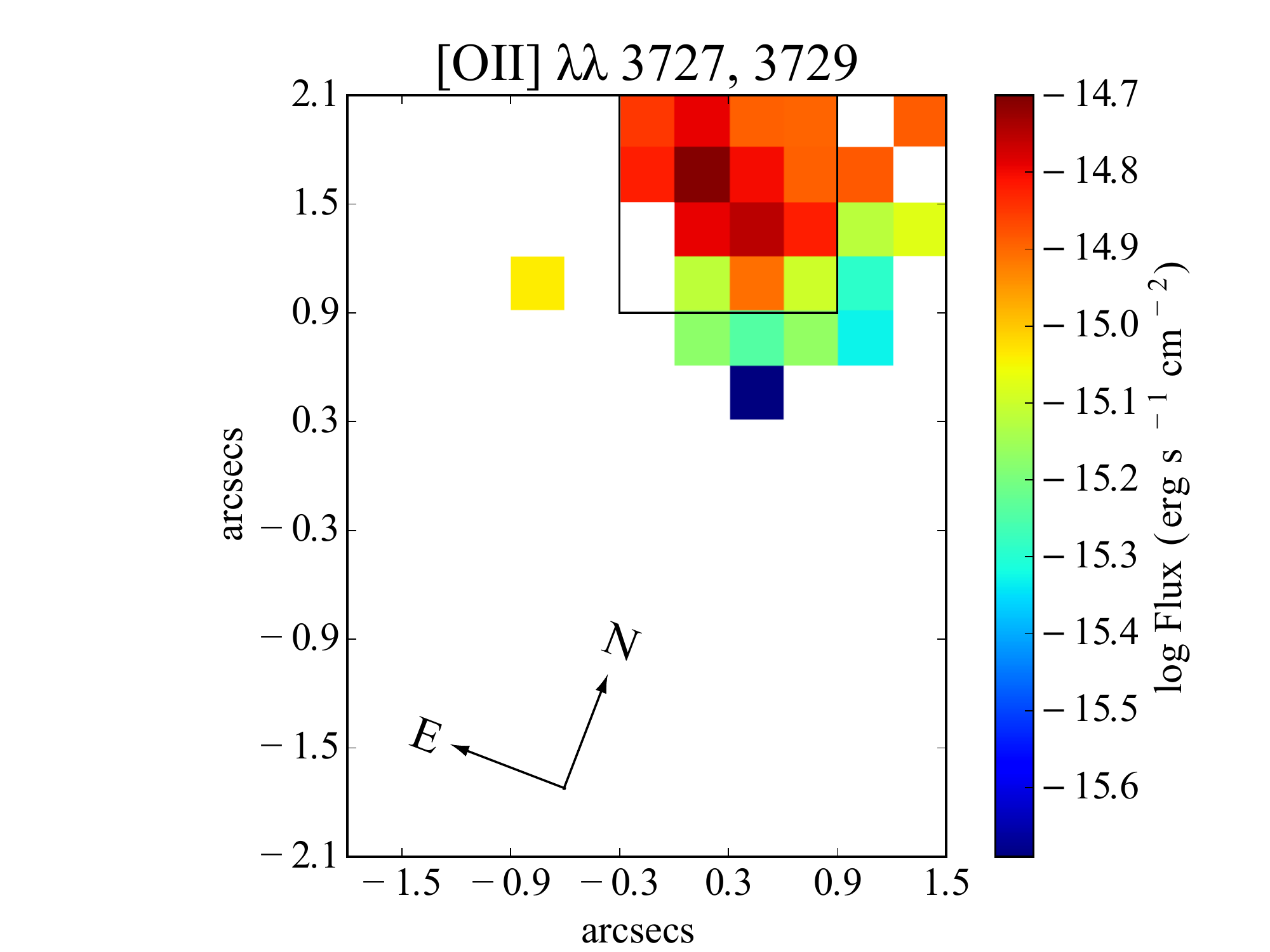}
\includegraphics[width = 0.48\textwidth]{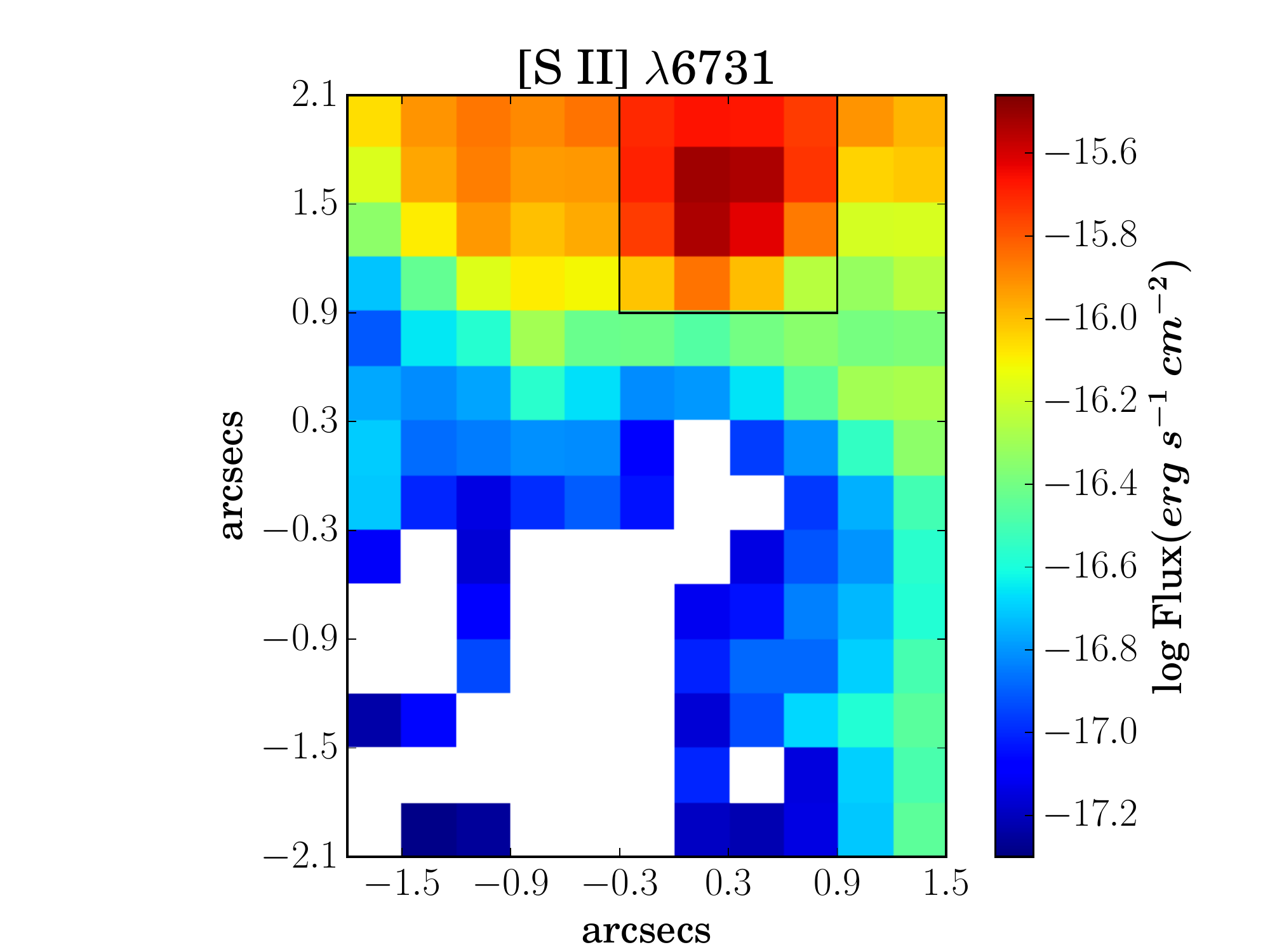}
\includegraphics[width = 0.48\textwidth]{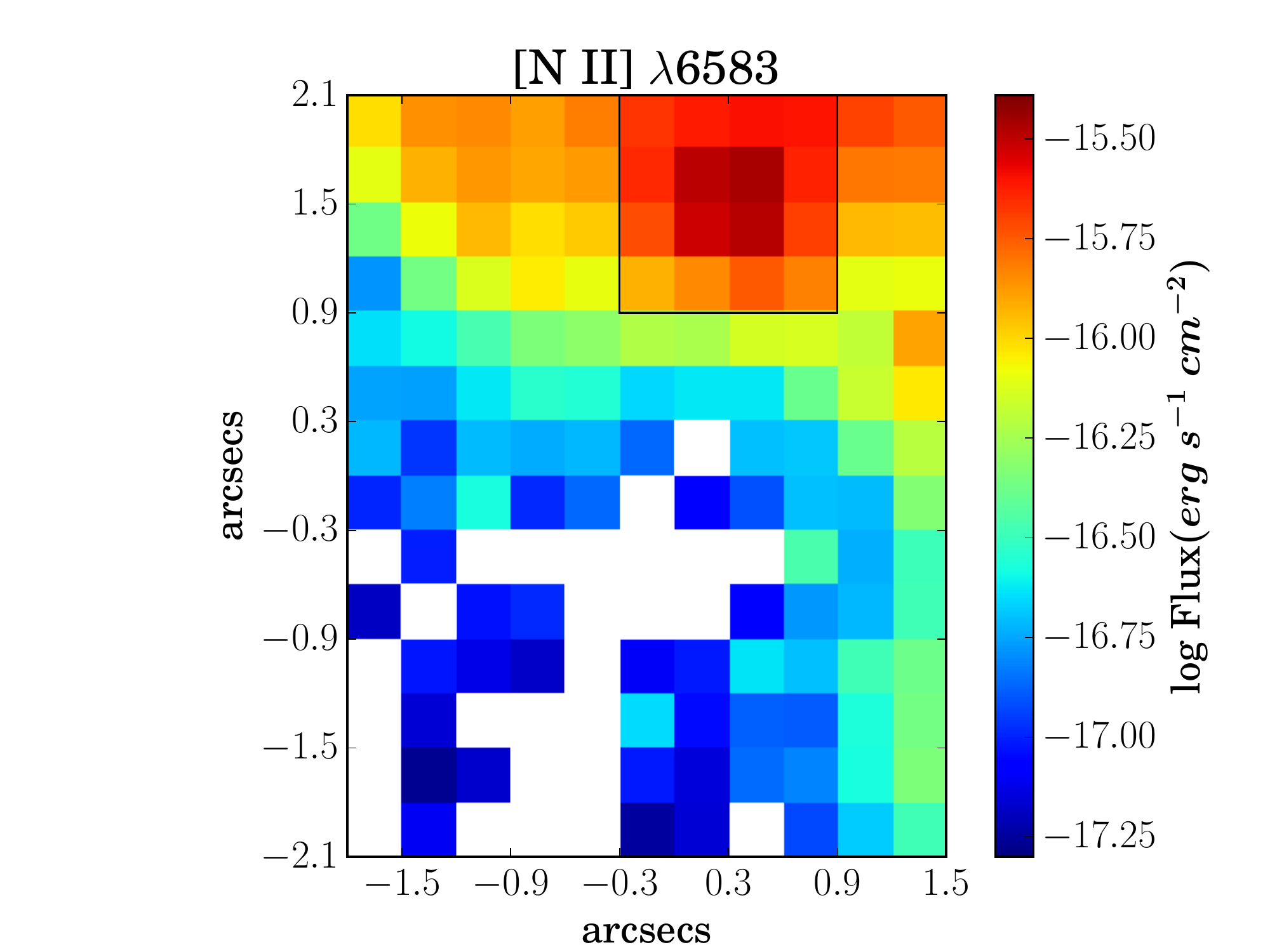}
\caption{Observed emission line flux maps of NGC 4449: Continuum (5216.03--5971.46 \AA), [O \textsc{iii}] $\lambda$5007, H$\beta$, [O \textsc{ii}] $\lambda\lambda$3726, 3729, [S \textsc{ii}]$ \lambda$6731, [N \textsc{ii}] $\lambda$6583.
 We observe an offset ($\sim$ 15 pc) in the continuum (red box) and  emission line (black box) peaks.  White spaxels correspond to the spaxels in which emission/flux have S/N $<$ 3.}
\label{observed flux}
\end{figure*}

\section{Observation \& Data Reduction}
\label{data}
\indent The central H \textsc{ii} region of NGC 4449 was observed by GMOS-N in one-slit queue-mode in 2012 as a part of a GMOS-IFU spectroscopy program (PI: B James) for seven star-forming galaxies. The one-slit mode of GMOS-N IFU provides a field-of-view (FOV) of 3.5\arcsec$\times$5\arcsec  sampled by 500 hexagonal lenslets of projected diameter  0.2\arcsec and 250 lenslets for background determination. Information from the data observing log are tabulated in Table \ref{log}.

\begin{table*}
\centering
\caption{GMOS-N IFU observing log for NGC 4449}
\label{log}
\begin{tabular}{@{}cccccc@{}}
\toprule
Grating      & \begin{tabular}[c]{@{}c@{}}Central wavelength\\ (\AA)\end{tabular} & \begin{tabular}[c]{@{}c@{}}Wavelength Range\\ (\AA)\end{tabular} & \begin{tabular}[c]{@{}c@{}}Exposure Time\\ (s)\end{tabular} & Average Airmass & Standard Star \\ \midrule
B600+\_G5307  & 4650                                                              &    3202 -- 6067                                          & 1200                                                        & 1.148           & Hz44          \\
B600+\_G5307 & 4700                                                               &    3252 -- 6119                                          & 2$\times$1200                                                  & 1.099, 1.108    & Wolf1346      \\
R600+\_G5304 & 6900                                                               &    5345 -- 8261                                           & 1100                                                        & 1.097           & Wolf1346      \\
R600+\_G5304 & 6950                                                               &    5397 -- 8314                                           & 1100                                                        & 1.103           & Wolf1346      \\ \bottomrule
\end{tabular}
\end{table*}
 
\indent  Observations were carried out in four different settings which involved two gratings each for blue (B600) and red (R600) observations. To  have a better wavelength coverage and limit the problems due to two chip gaps between the three detectors of GMOS-N IFU, two sets of observations were taken with spectral dithering of 50 \AA.  For each of the four settings, a set of observations for GCAL and twilight flats, CuAr lamp for wavelength calibration and standard star for flux calibration were taken.  The basic reduction steps including bias subtraction, flat-field correction, wavelength calibration and sky (background) subtraction were carried out using the standard GEMINI pipeline written in Image Reduction and Analysis Facility (IRAF)\footnote{IRAF is distributed by the National Optical Astronomy Observatory, which is operated by the Association of Universities for Research in Astronomy (AURA) under a cooperative agreement with the National Science Foundation.}. However the standard pipeline does not provide satisfactory results for some procedures and we therefore had to develop and implement our own codes. For example, the focal plane of GMOS-IFU uses three detectors, each readout from two amplifiers. At the junction of fourth, fifth and sixth amplifiers, reduced spectra were found to be offset by a significant number of counts. We statistically determined the offsets and corrected the spectra before flux calibration. The cosmic rays which could not be removed by the standard pipeline, were removed while combining multiple frames taken in the same setting through a sigma-clipping procedure. The spectra in each setting were corrected for differential atmospheric refraction and converted into data cubes with spatial sampling of 0.3\arcsec/spaxel, using the routine $gfcube$ in Gemini's IRAF reduction package. Spectral dithering and spatial offset between the cubes from the same grating were corrected while combining them together. The FOV covered by the red and blue gratings  showed a spatial offset of $\sim$0.3\arcsec on both axes. We however concentrated our analysis on the overlapping area of the FOVs covered by the two settings, and hence produced cubes and row-stacked spectra of the overlapping region for the two settings. The value of instrumental broadening (FWHM) was obtained by fitting a Gaussian profile to several emission lines of the extracted row stacked spectra of the arc lamp and was found to be $\sim$1.7 \AA\hspace{0.02in} for both blue and red setting. 

\begin{table}
\caption{Emission line measurements (relative to H$\beta$ = 100) for the summed spectrum of the main emission region. Line fluxes ($F_{\lambda}$) were extinction corrected to calculate $I_{\lambda}$ using the c(H$\beta$) shown at the bottom of the table.} 
\label{properties}
\resizebox{\columnwidth}{!}{%
\begin{tabular}{cccc}
\toprule
Line & $\lambda_{air}$ & $F_{\lambda}$ & $I_{\lambda}$ \\ \midrule
$[OII]$ & 3726.03 & $319.54 \pm 5.52 $ & $397.20 \pm 15.70 $ \\
$[OIII]$ & 4363.21 & $4.05 \pm 0.94 $ & $4.44 \pm 1.04 $ \\
$H\beta$ & 4861.33 & $100.00 \pm 0.66 $ & $100.00 \pm 2.26 $ \\
$[OIII]$ & 4958.92 & $71.03 \pm 0.76 $ & $69.95 \pm 2.24 $ \\
$[OIII]$ & 5006.84 & $214.22 \pm 1.58 $ & $209.20 \pm 6.46 $ \\
$[NII]$ & 6548.03 & $17.02 \pm 0.34 $ & $13.68 \pm 0.45 $ \\
$H\alpha$ & 6562.8 & $356.61 \pm 2.41 $ & $286.00 \pm 7.73 $ \\
$[NII]$ & 6583.41 & $55.93 \pm 0.51 $ & $44.77 \pm 1.24 $ \\
$[SII]$ & 6716.47 & $53.50 \pm 0.49 $ & $42.27 \pm 1.16 $ \\
$[SII]$ & 6730.85 & $45.90 \pm 0.45 $ & $36.20 \pm 1.00 $ \\ \midrule
c(H$\beta$)   &&$0.31\pm 0.01 $&\\
F(H$\beta$) &&$6.61 \pm 0.04$& $13.35 \pm 0.30$ \\

 \bottomrule
\end{tabular}%
}
Notes: F(H$\beta$) in units of $\times$ 10$^{-15}$ erg cm$^{-2}$ s$^{-1}$
\end{table}

\section{Observed and Intrinsic Fluxes}
\label{analysis}

 \begin{figure}
 \centering
\includegraphics[width = 0.48\textwidth]{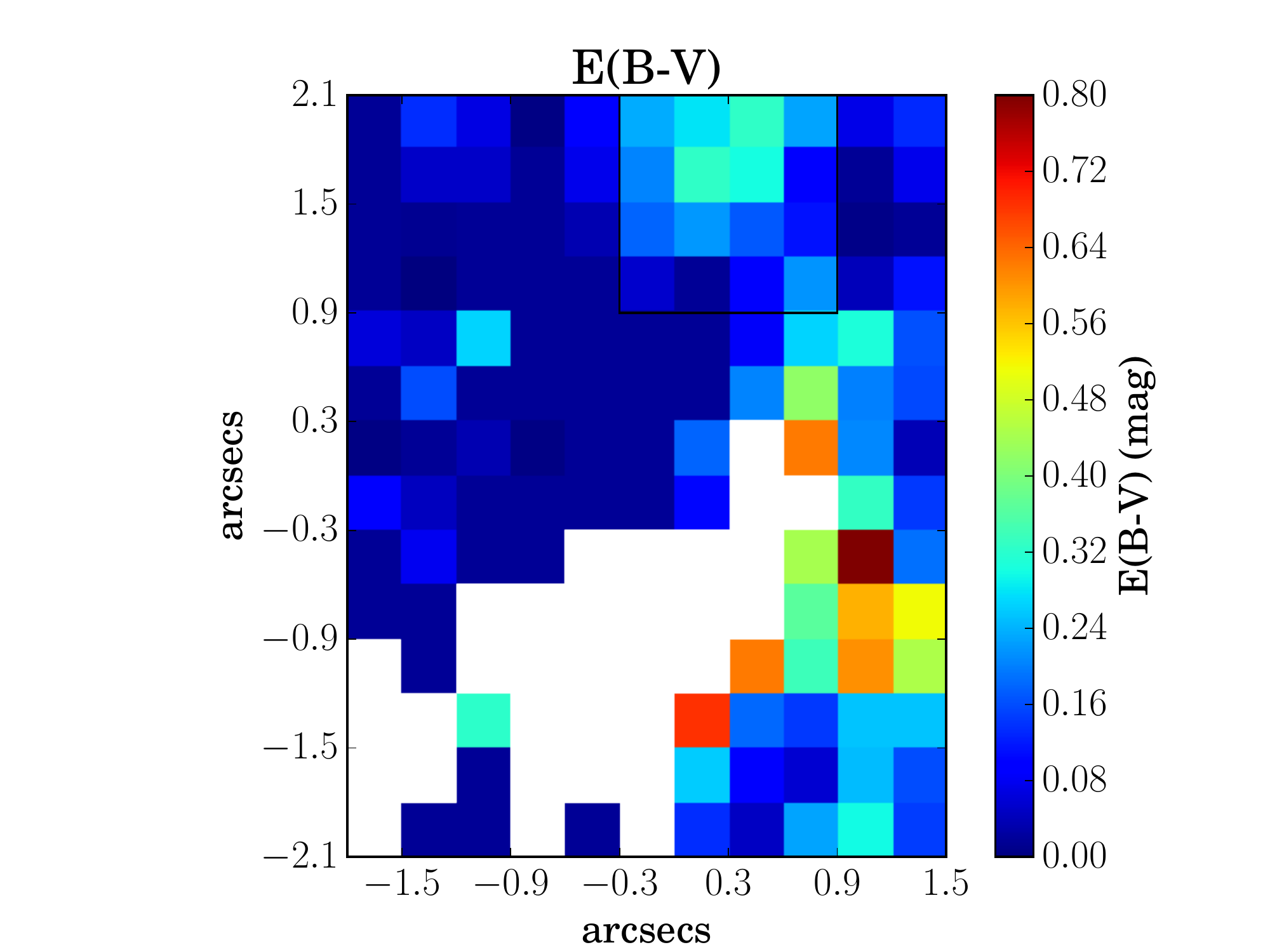}
\caption{E(B-V) map of the FOV, which varies from 0.2--0.4 mag in the main emission region (black square box).  E(B-V) goes as high as 0.8 mag to the south-west of the main emission region. White spaxels correspond to the spaxels in which emission line fluxes had S/N $<$ 3.}
\label{EBV}
\end{figure}

    \begin{figure*}
 \centering
\includegraphics[width = \textwidth, trim={0 2.5cm 0 1cm},clip]{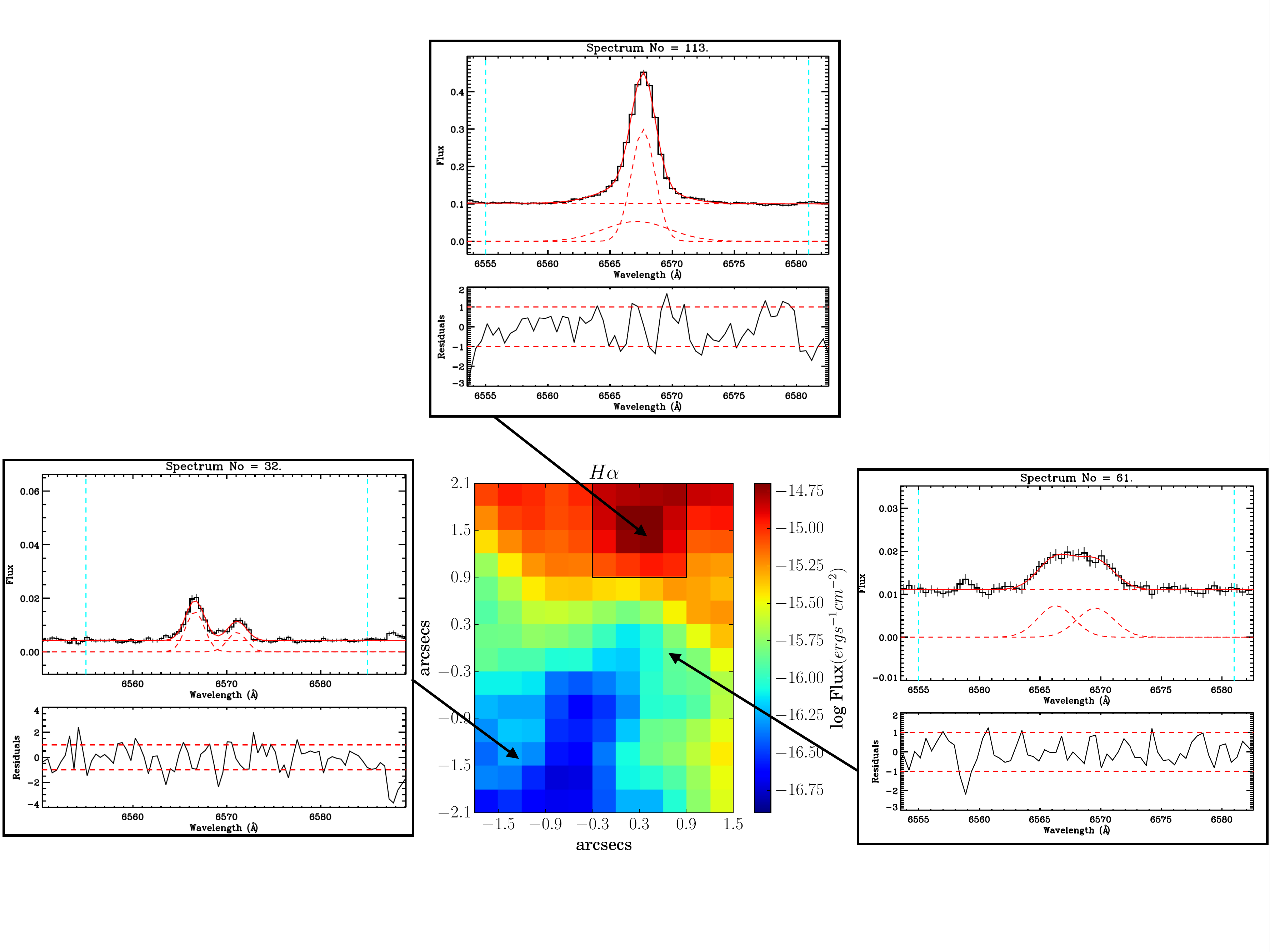}
\caption{Spectral decomposition of the H$\alpha$ emission line showing different kinematic structures at different locations of the FOV.  The H$\alpha$ emission line map is created by summing the flux in the spectral region of interest and takes into account the flux in both of the spectral components. The black square box on the H$\alpha$ emission line map shows the main emission region. In each of the panel showing the Gaussian fits, flux is in units of 10$^{-15}$ erg cm$^{-2}$ s$^{-1}$ \AA$^{-1}$, and residuals are normalised to be in $\sigma$-noise units.}
\label{velocity structure}
\end{figure*}

\subsection{Integrated Spectra \& Flux Maps}

\indent Figure \ref{HST} shows the HST image of NGC 4449 taken in the filter F660N. The green box in the central region of the galaxy represents the GMOS aperture (3.5$\arcsec$ $\times$ 5$\arcsec$) and the  magenta box inside the green box shows the main emission region covering an area of 1.2$\arcsec$ $\times$ 1.2 $\arcsec$ ($\sim$ 22 $\times$ 22 pc). In Figure \ref{spectra},  we show the GMOS-IFU integrated spectra of NGC 4449 integrated over the main (bright) emission region of 1.2$\arcsec$ $\times$ 1.2$\arcsec$ (upper two panels), and integrated over the remaining spaxels in the FOV (Whole FOV -- main emission region, lower two panels) to highlight the difference, e.g. some of the weaker lines e.g. [ O \textsc{i}] $\lambda$5577, He \textsc{i} $\lambda$6678, [O \textsc{ii}] $\lambda$ 7319) are present either in the main emission region or in the integrated spectra over the remaining spaxels of the FOV. The principal emission lines are over-plotted at their rest wavelengths in air.

\indent We measure the emission line fluxes for all the main recombination and collisionally excited lines within the spectra by integrating the fluxes within the line after subtracting the continuum and absorption features (in recombination lines) from the spectra.  The observed flux for the emission lines used in the present analysis for the main emission region is given in Table \ref{properties}.  Figure \ref{observed flux} shows the observed flux maps of NGC 4449 of continuum, [O \textsc{iii}] $\lambda$5007, H$\beta$ and [O \textsc{ii}] $\lambda\lambda$3726, 3729 (summed as they are blended), [S \textsc{ii}] $\lambda$6731 and [N \textsc{ii}] $\lambda$6583. The continuum map is created by integrating the blue cube in the emission-free wavelength range 5216.03--5971.46 \AA\hspace{0.01in} (V-band). A spatial offset ($\sim$ 15 pc) in the  continuum and emission line peaks is observed. 

\indent We checked for the following main sources of errors in our flux measurement:  photon noise, readout noise and any uncertainty in our data processing. Since the data acquisition was carried out in slow readout mode, the readout noise ($\sim$ 3 e$^{-}$ with respect to $\sim$ 500--1000 e$^{-}$ in emission lines including continuum) is negligible. The continuum levels of the same regions in the FOV (e.g. main emission region) obtained from the red and blue gratings, agree with each other which suggest that the error in defining the continuum level is negligible. Hence, for calculating the error on the observed flux, we assume that photon noise is the dominant source of error. Since photon noise is Poissonian, we calculate the relative error ($\sigma_{rel}$) as the inverse of square root of the uncalibrated flux ($f_{uncal}$ in e$^{-}$) and scale the corresponding calibrated flux to calculate the absolute error ($\sigma_{abs}$) on calibrated flux ($f_{cal}$ in ergs cm$^{-2}$ s$^{-1}$ \AA$^{-1}$), i.e. 
\begin{equation}
 \sigma_{abs} = \sigma_{rel} \times f_{cal} = \frac{\sqrt{f_{uncal}}}{f_{uncal}} \times f_{cal}
\end{equation}

The uncertainties on all quantities in this work are computed by propagating the error on the observed calibrated flux.

\subsection{Dust attenuation}
 \indent To estimate dust attenuation, we derive E(B--V) using the relationship between the nebular emission line colour excess and the Balmer decrement given by: 
\begin{multline}
 E(B-V) = \frac{E(H\beta - H\alpha)}{k(\lambda_{H\beta})-k(\lambda_{H\alpha})}\\
= \frac{2.5}{k(\lambda_{H\beta})-k(\lambda_{H\alpha})}log_{10}\bigg[\frac{(H\alpha/H\beta)_{obs}}{(H\alpha/H\beta)_{theo}}\bigg]
\end{multline}
 
\indent where $k(\lambda_{H\beta})$ and $k(\lambda_{H\alpha})$ are the values from the LMC (Large Magellanic Cloud) attenuation curve \citep{Fitzpatrick1999} evaluated at the wavelengths $H\beta$ and $H\alpha$ respectively\footnote{Note here that it is customary to calculate the total E(B-V) by averaging the E(B-V) from H$\alpha$/H$\beta$ and H$\gamma$/H$\beta$ in the ratio 3:1. However, we have only used H$\alpha$/H$\beta$ here as the H$\gamma$ emission line is excessively noisy.}, $(H\alpha/H\beta)_{obs}$ and $(H\alpha/H\beta)_{theo}$ are the observed and theoretical $H\alpha/H\beta$ line ratios respectively.  This curve is chosen because the metallicity of NGC 4449 is similar to that of LMC. (\citet{Lequeux1979} reports a metallicity of 12+log(O/H) = 8.3, which is close to that of LMC, 8.35$\pm$0.06 \citep{RussellDopita1992}). 

\indent Figure \ref{EBV} shows the map of E(B-V), which varies from 0.2 to 0.4 mag in the main emission region (black square box).  E(B-V) goes as high as 0.8 mag in the south-west of the main emission region.  Using the calculation for E(B-V) above, we calculate the extinction in magnitudes at wavelength $\lambda$, given by $A_{\lambda} = k(\lambda)E(B-V)$ and finally calculate the intrinsic flux maps  using the following equation: 
\begin{equation}
 F_{int}(\lambda) = F_{obs}(\lambda)\times 10^{0.4 A_\lambda}
\end{equation}
\indent Some spaxels away from the main emission region have negative values of E(B-V) due to stochastic error and shot noise \citep{Hong2013}. We force E(B-V) in these spaxels to the Milky Way ``foreground" value (Table \ref{tab properties}) before calculating intrinsic flux. The intrinsic emission line fluxes calculated from the integrated spectra of the main emission region are tabulated in Table \ref{properties}.

   \begin{figure*}
 \centering
\includegraphics[width = 0.48\textwidth]{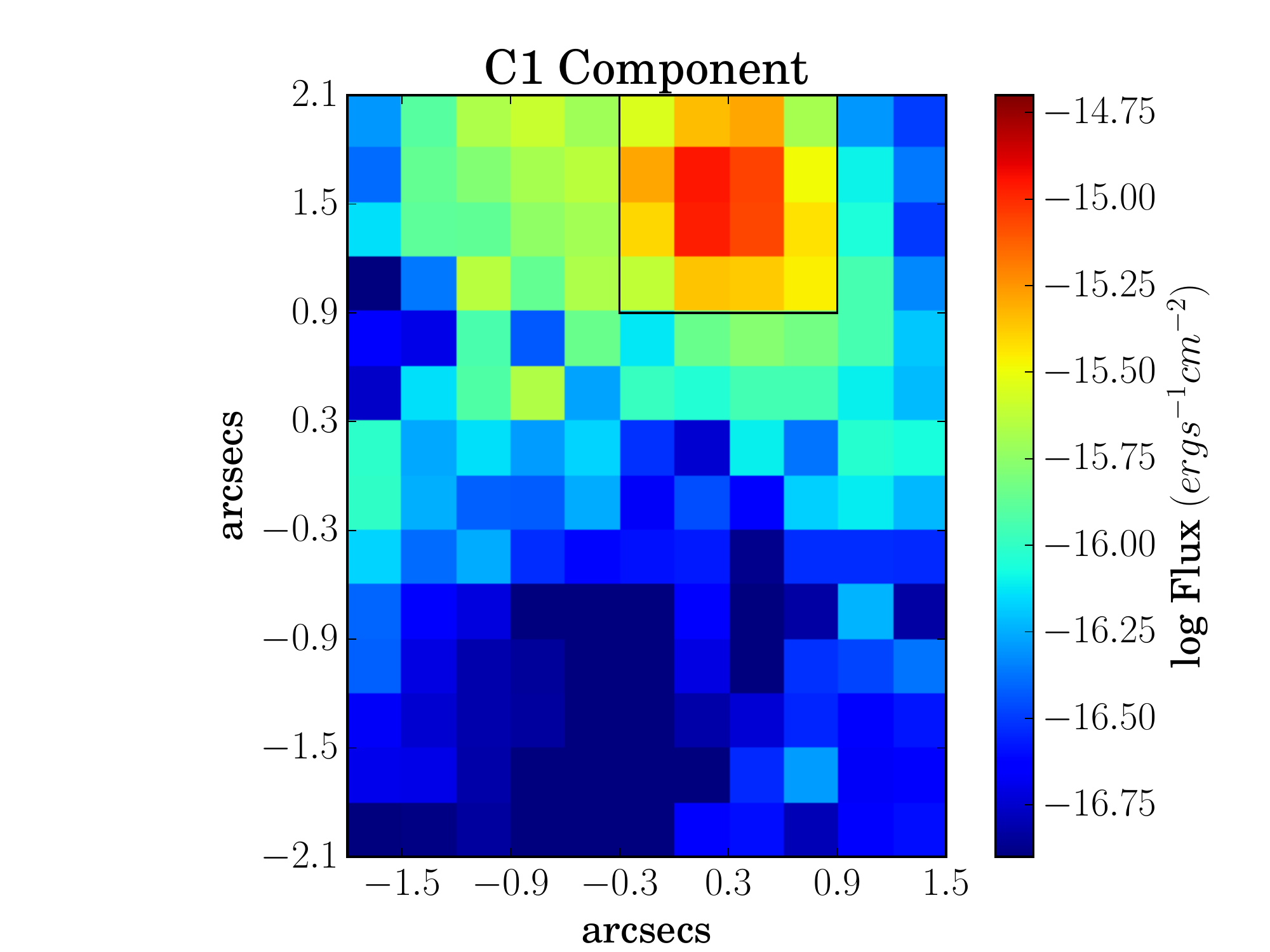}
\includegraphics[width = 0.48\textwidth]{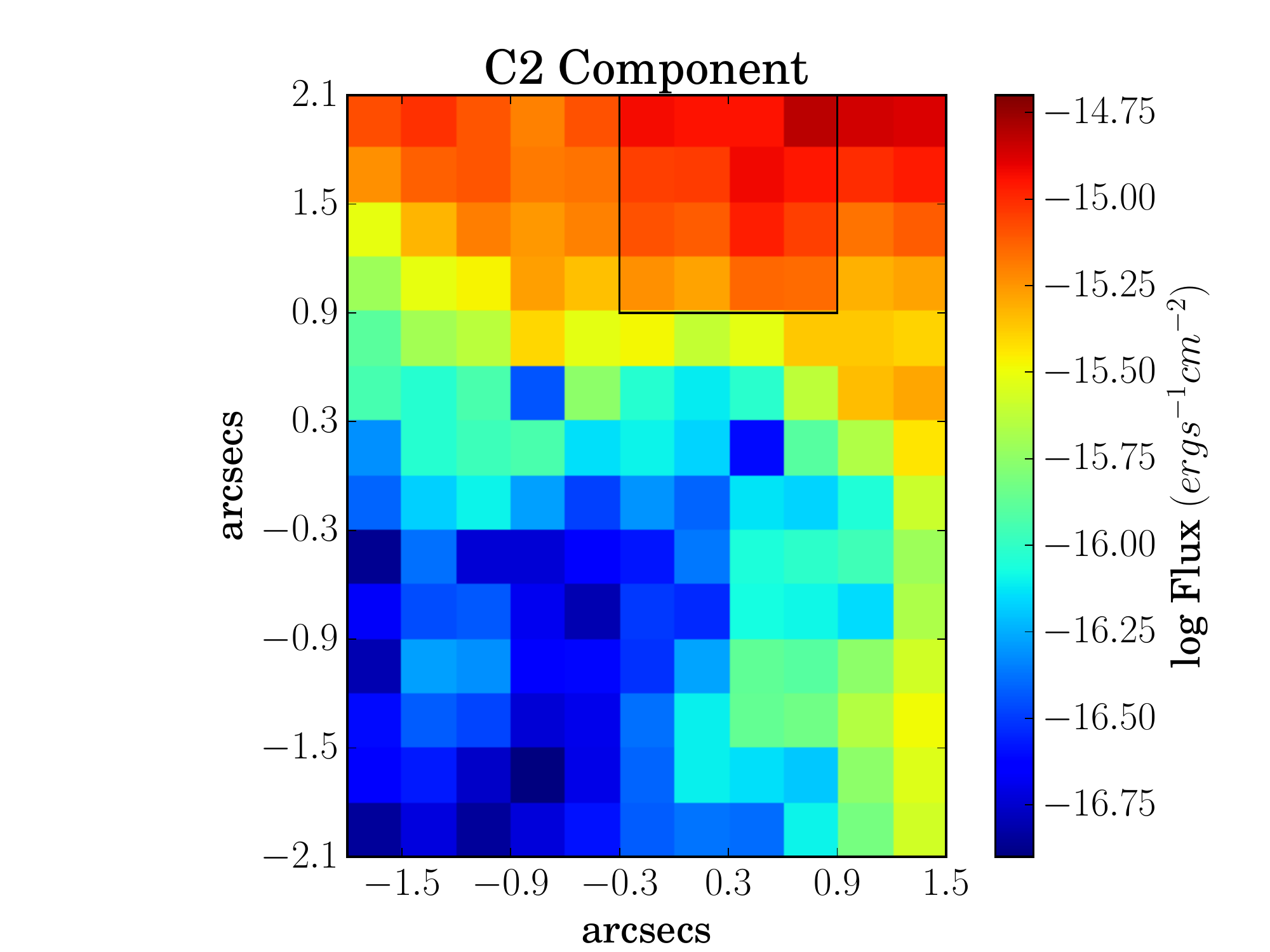}

\includegraphics[width = 0.48\textwidth]{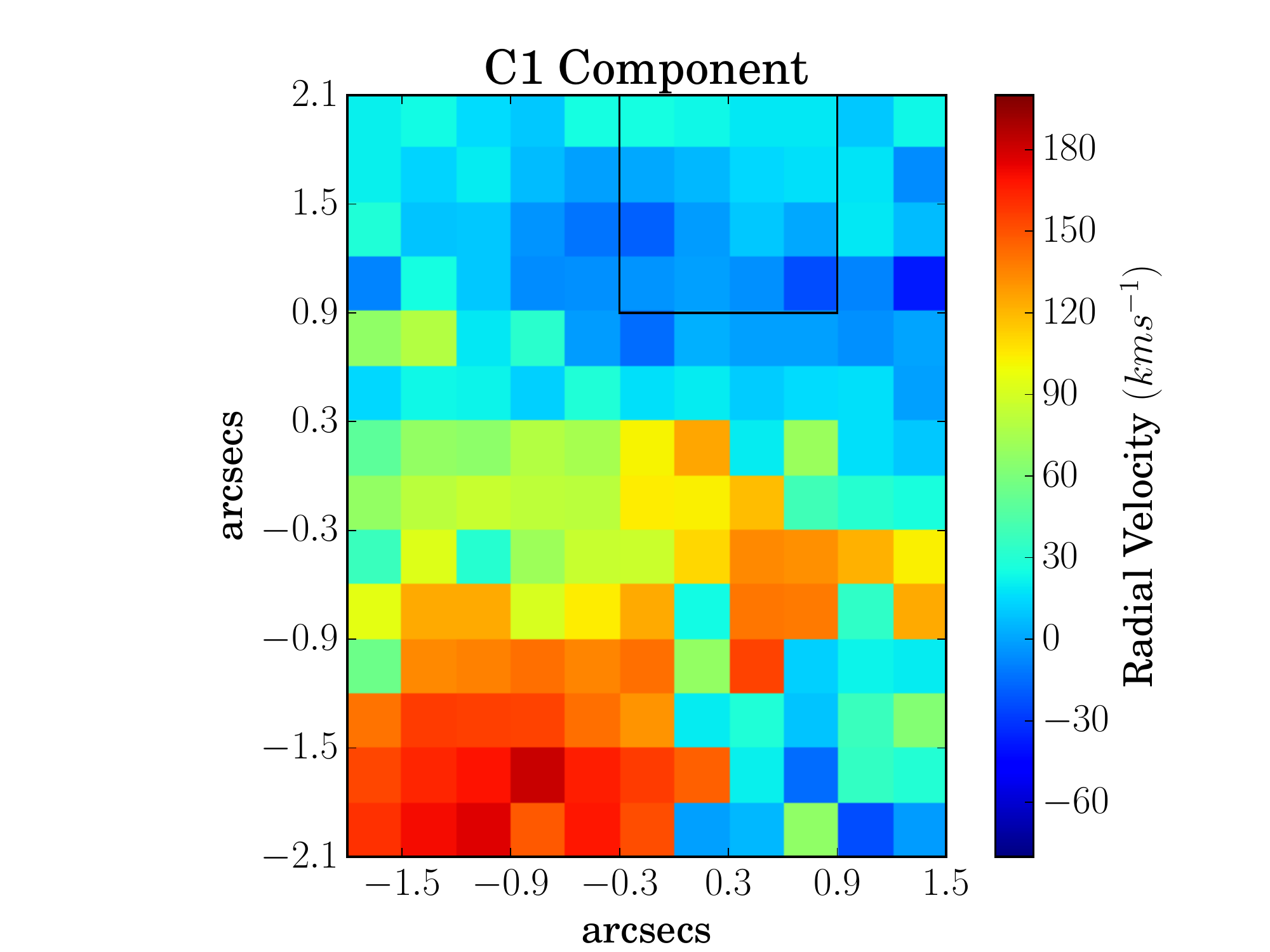}
\includegraphics[width = 0.48\textwidth]{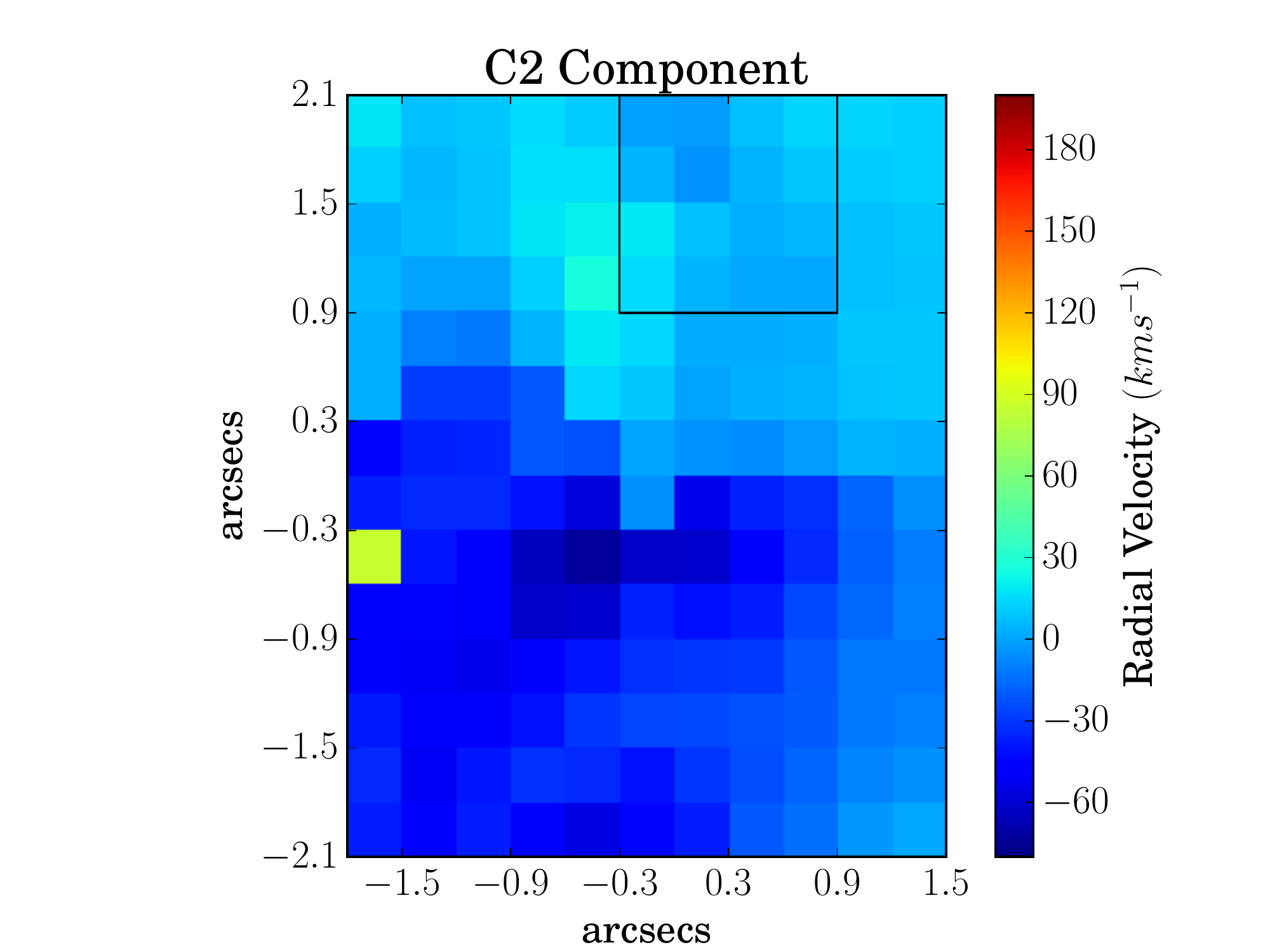}

\includegraphics[width = 0.48\textwidth]{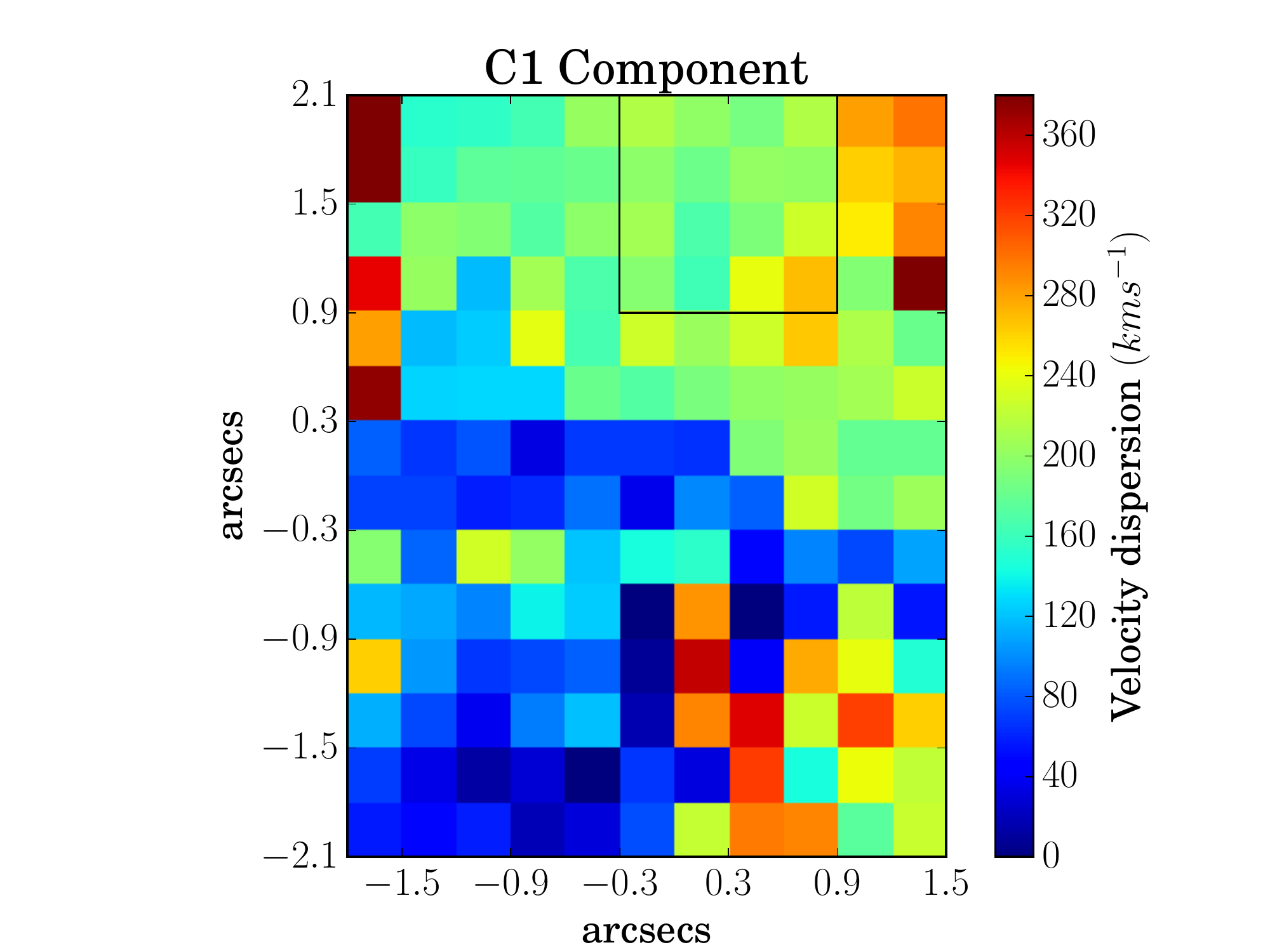}
\includegraphics[width = 0.48\textwidth]{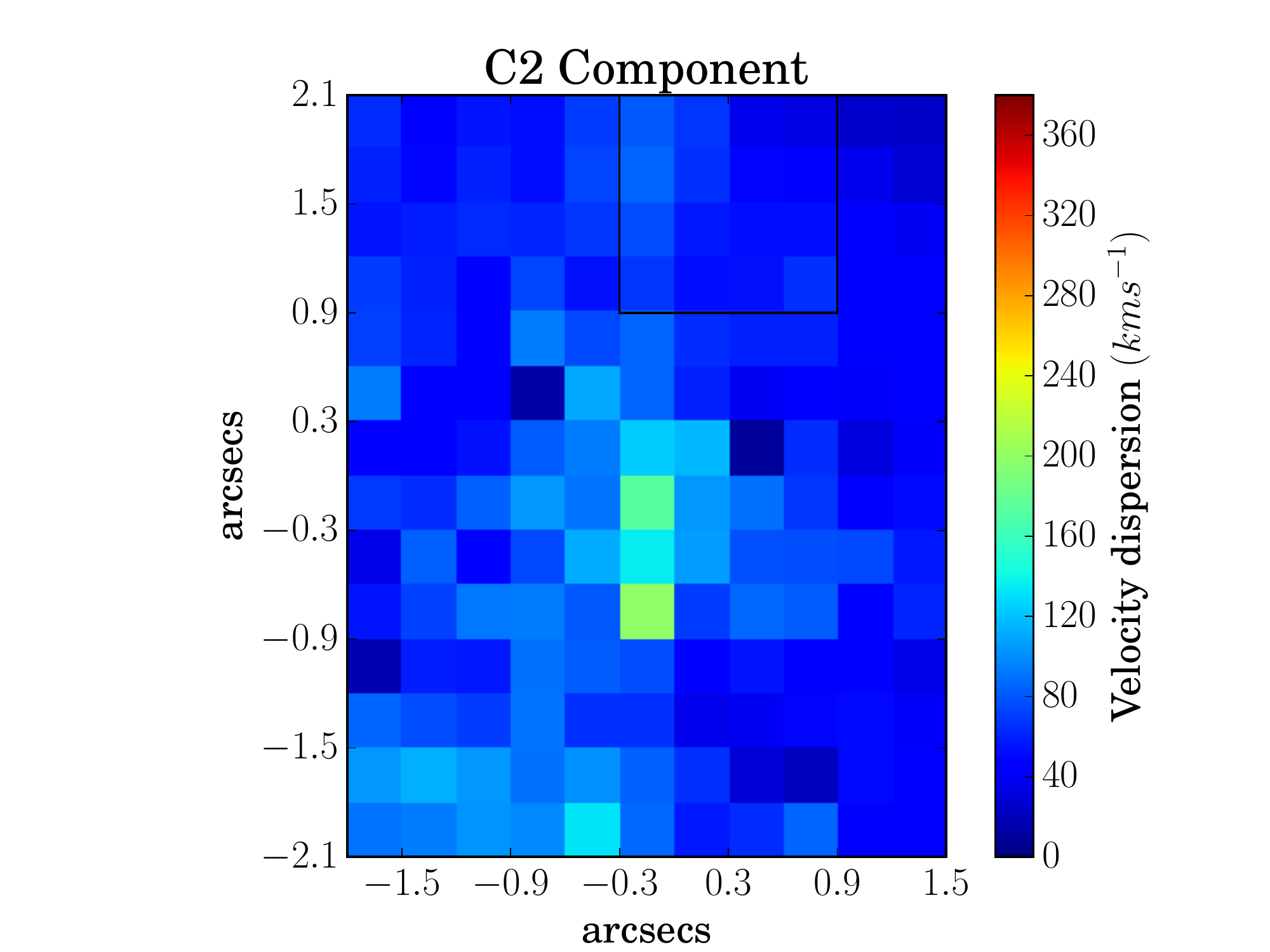}

\caption{H$\alpha$ maps of the two spectral components: C1(left) and C2(right). We assign the broad spectral component to C1 and the narrow spectral component to C2.  When spaxels have two spectral components with comparable velocity widths, we assign the component with higher radial velocity to C1 and the component with lower radial velocity to C2. Observed flux (upper panel); radial velocity (middle panel) relative to systemic velocity of v$_{sys}$ = 207 km s$^{-1}$; and velocity dispersion/FWHM (bottom panel) corrected for instrumental broadening. The black square box shows the main emission region.}
\label{spectral decomposition}
\end{figure*}

 \begin{figure*}
\centering
 \includegraphics[width = 0.48\textwidth]{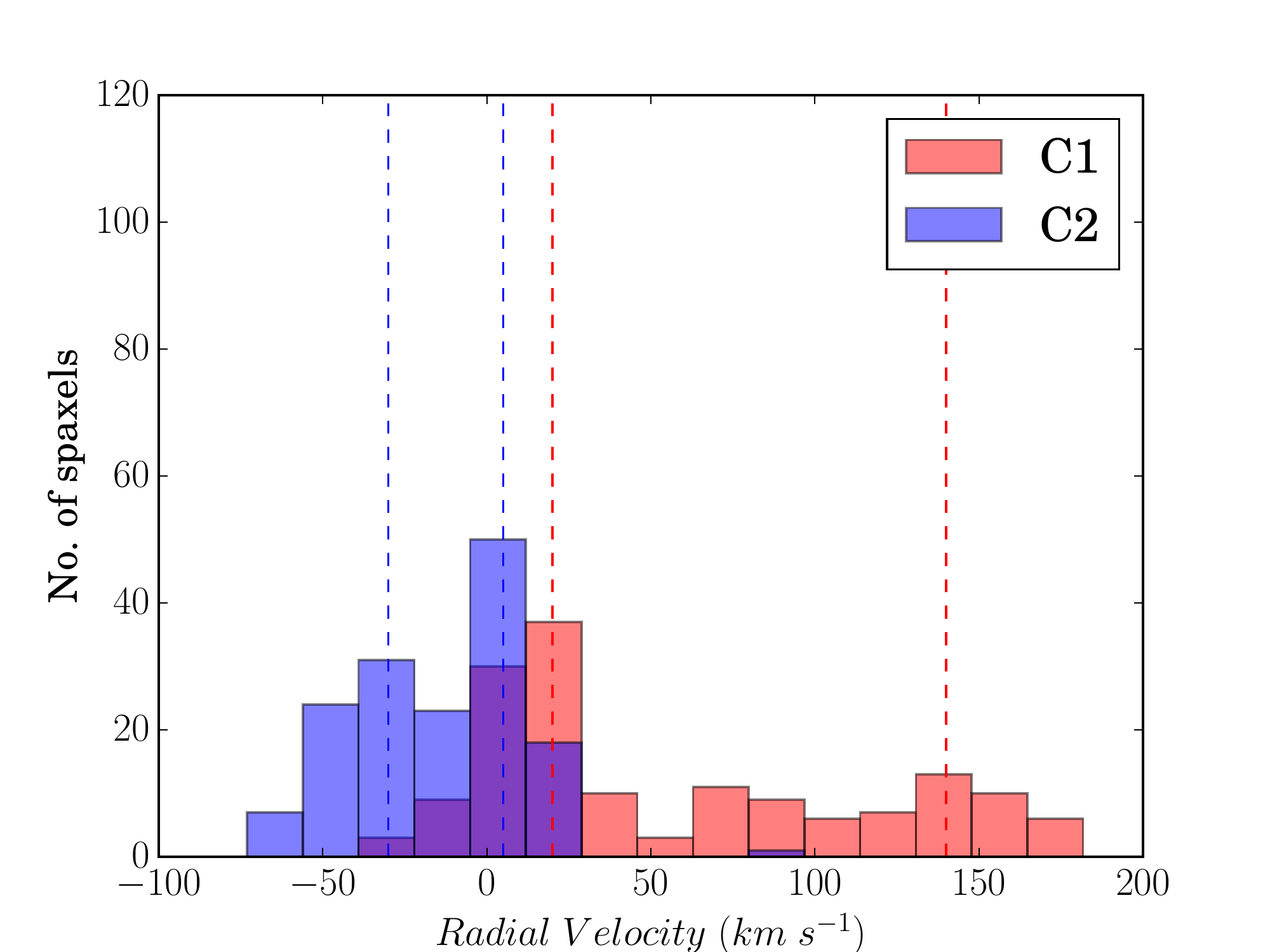}
 \includegraphics[width = 0.48\textwidth]{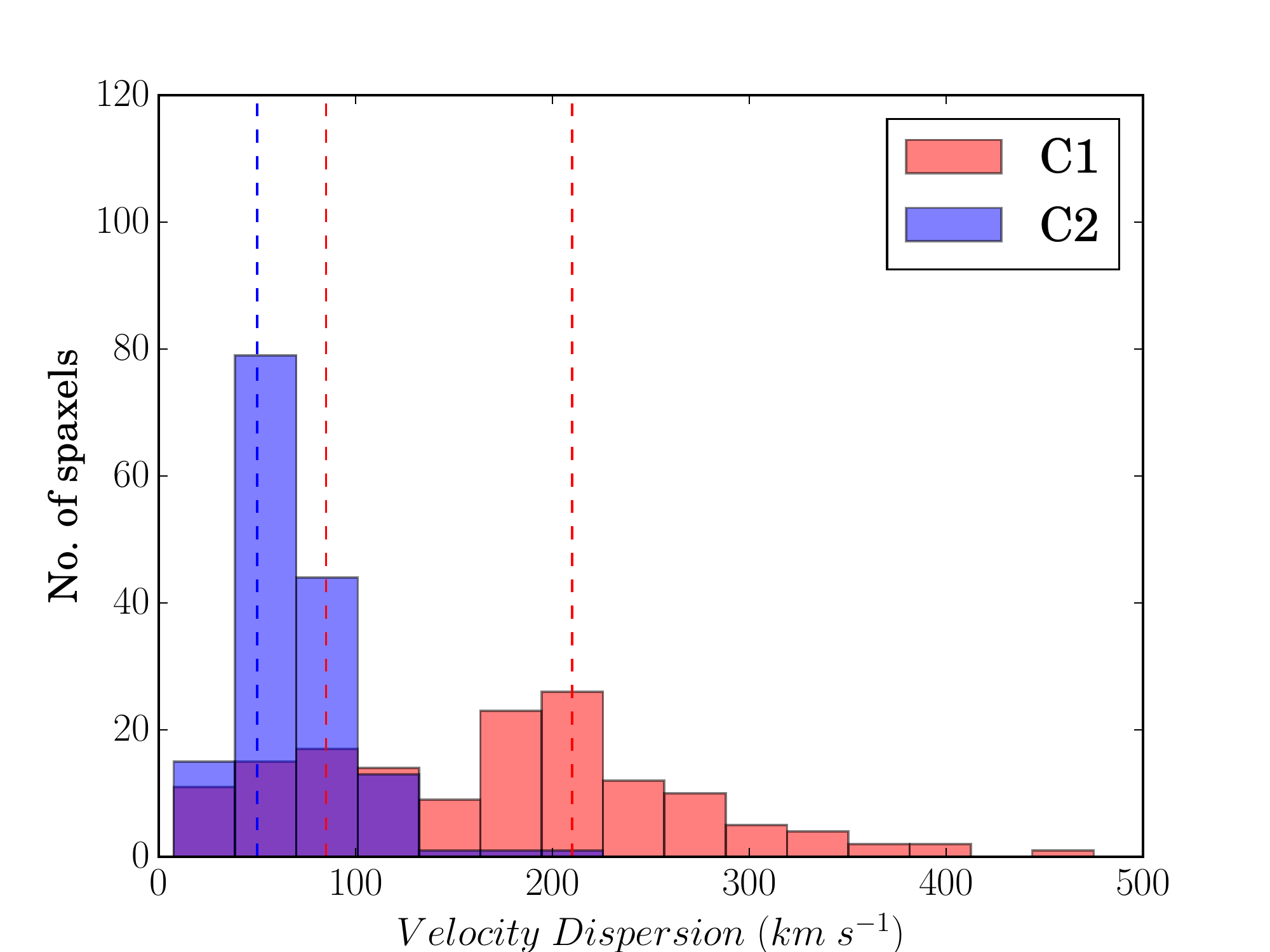}
\caption{The distributions of  radial velocity (left) and velocity dispersion (FWHM) (right) of the two kinematic components of H$\alpha$ emission line. C1: red, C2: blue for both distribution and the dashed vertical lines denote approximate modes for the distributions. Both C1 and C2 components show bimodal distribution in radial velocity  (left plot). The C1 component is probably bimodal while the C2 component follows a unimodal distribution in velocity dispersion (right plot).}
\label{distribution}
\end{figure*}

\section{Emission Line Kinematics}
\label{kinematics}

\indent We perform a detailed kinematic analysis of the central H \textsc{ii} region of the galaxy by separating different spectral components of the strong H$\alpha$ emission line. Figure \ref{velocity structure} shows different types of kinematic structure present at different locations in the H$\alpha$ map of our GMOS FOV. We used  an IDL-based curve-fitting software PAN (Peak ANalysis) to fit multiple spectral components to the H$\alpha$ emission line, which is based on $\chi^2$--minimisation. To determine the optimal number of spectral components, we experiment with single, double, and triple Gaussian fits. We found that a double Gaussian fit resulted in the value of $\chi^2$ significantly lower than that of the single Gaussian fit, while an attempt to fit the third component resulted in the Gaussian fits with either the FWHM of the third Gaussian less than the instrumental FWHM or unreasonable fits for a large number of spectra. Hence we decided that a double Gaussian provides the most suitable description of the H$\alpha$ emission line across the entire FOV. The double Gaussian component emission has a variety of structure throughout the FOV. The main emission region shows a narrow and a broad component, while the regions farther away show two separate as well as blended components which are best fit by two similar Gaussians.

\indent Figure \ref{spectral decomposition} shows the maps of the observed flux, radial velocity  and velocity dispersion of the two components of the H$\alpha$ emission line.  An attempt to separate the components on the basis of velocity dispersion leads to discontinuous/patchy maps.  We therefore assign broad spectral component to C1 and narrow spectral component to C2.  For the spaxels which show two spectral components of comparable velocity widths, we assign the component with higher radial velocity to C1 and the component with lower radial velocity to C2. The distributions of radial velocity  and  velocity dispersion (FWHM) of different kinematic components of H$\alpha$ are shown in Figure \ref{distribution}. The radial velocities take into account the barycentric correction (= $-$14.41 km s$^{-1}$) and are relative to the systemic velocity of 207 km s$^{-1}$ \citep{Schneider1992}. The FWHM values are corrected for the instrumental broadening of 1.7 \AA\hspace{0.01in}.

\indent The flux maps of H$\alpha$ (Figure \ref{spectral decomposition}, upper panel) show that the C1 component peaks in the central part of the main emission region (11 pc $\times$ 11 pc) and decreases by a factor of 2 in the surrounding $\sim$ 5.5 pc region, whereas the  C2 component is uniform over an area of 22 pc $\times$ 22 pc of the main emission region. Compared to C1, the C2 component shows a region of higher flux in the south-west of the main emission region. 

\indent The radial velocity maps of H$\alpha$ (Figure \ref{spectral decomposition}, middle panel) show rotating structures in both C1 and C2 components about an axis of rotation below the main emission region. The radial velocity of C1 varies between  $\sim$ $-$40 to 180 km s$^{-1}$. This component of the ionised gas has lower velocity in the main emission region (north-west) and, increases smoothly in the region farther away (south-east) - which could indicate a region of local shear stress between the layers of ionised gas at different velocities, that can lead to turbulence.  The C2 component shows a slower and smoother rotation compared to C1, with the radial velocity varying between $\sim$ $-$70 to 25 km s$^{-1}$. This component is blueshifted in the main emission region (north-west) and redshifted in the region below it (south-east). Both C1 and C2 components show bimodal distribution (Figure \ref{distribution}, left panel) in radial velocity.  

\indent From FWHM maps of H$\alpha$ (Figure \ref{spectral decomposition}, bottom panel) and the FWHM distribution (Figure \ref{distribution}, right plot), we find that the C1 component  spans a wide range of velocity dispersion varying between $\sim$ 10 to 400 km s$^{-1}$; while the C2 component is mostly narrow varying between $\sim$ 10 to 200 km s$^{-1}$ with a mean value of $\sim$ 70 km s$^{-1}$. The C1 component is probably bimodal while the C2 component follows a unimodal distribution in velocity dispersion (Figure \ref{distribution}, right plot).
 

\section{ionisation Conditions}
\label{ionisation}

\begin{figure*}
  \centering
 \includegraphics[width = 0.44\textwidth]{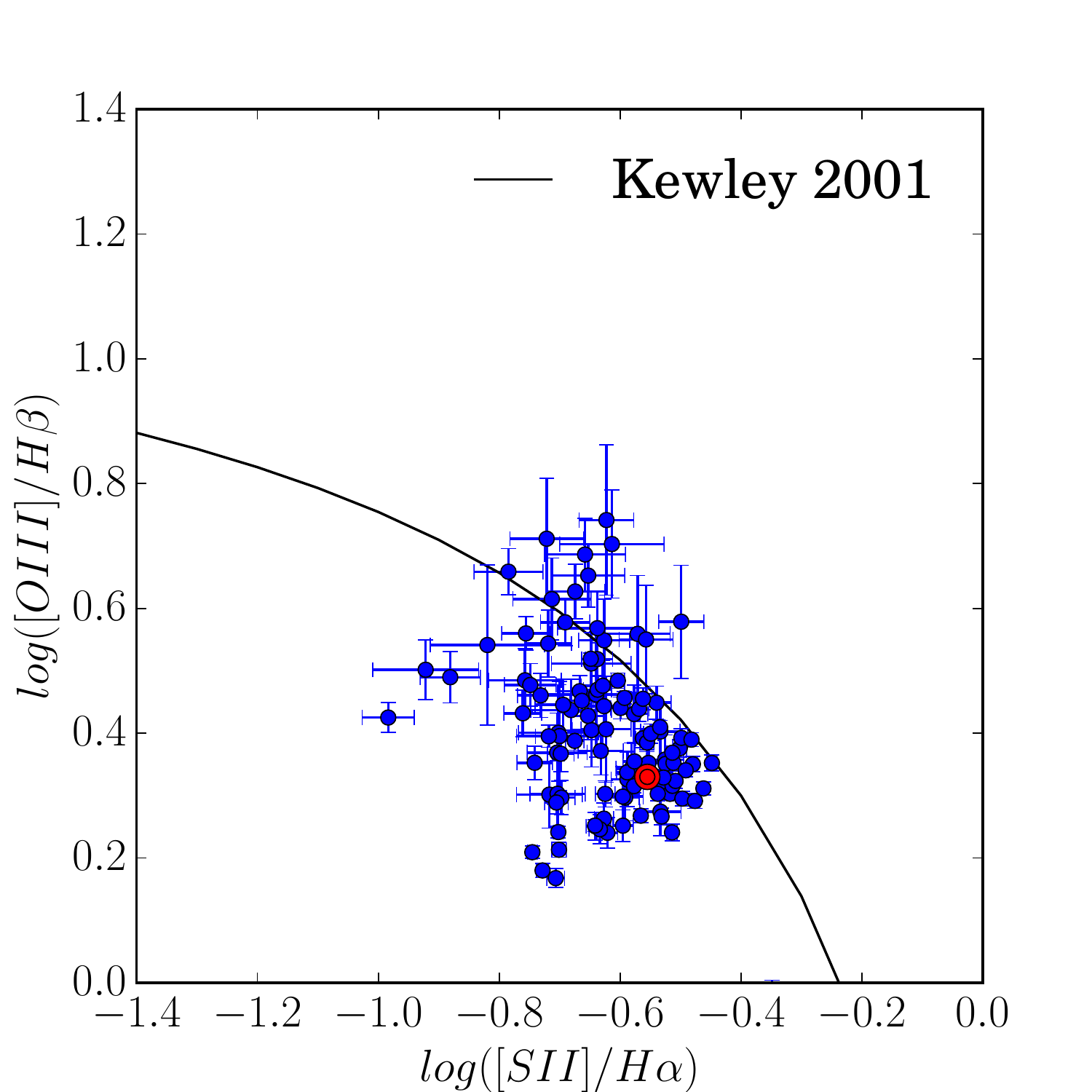}
 \includegraphics[width = 0.44\textwidth]{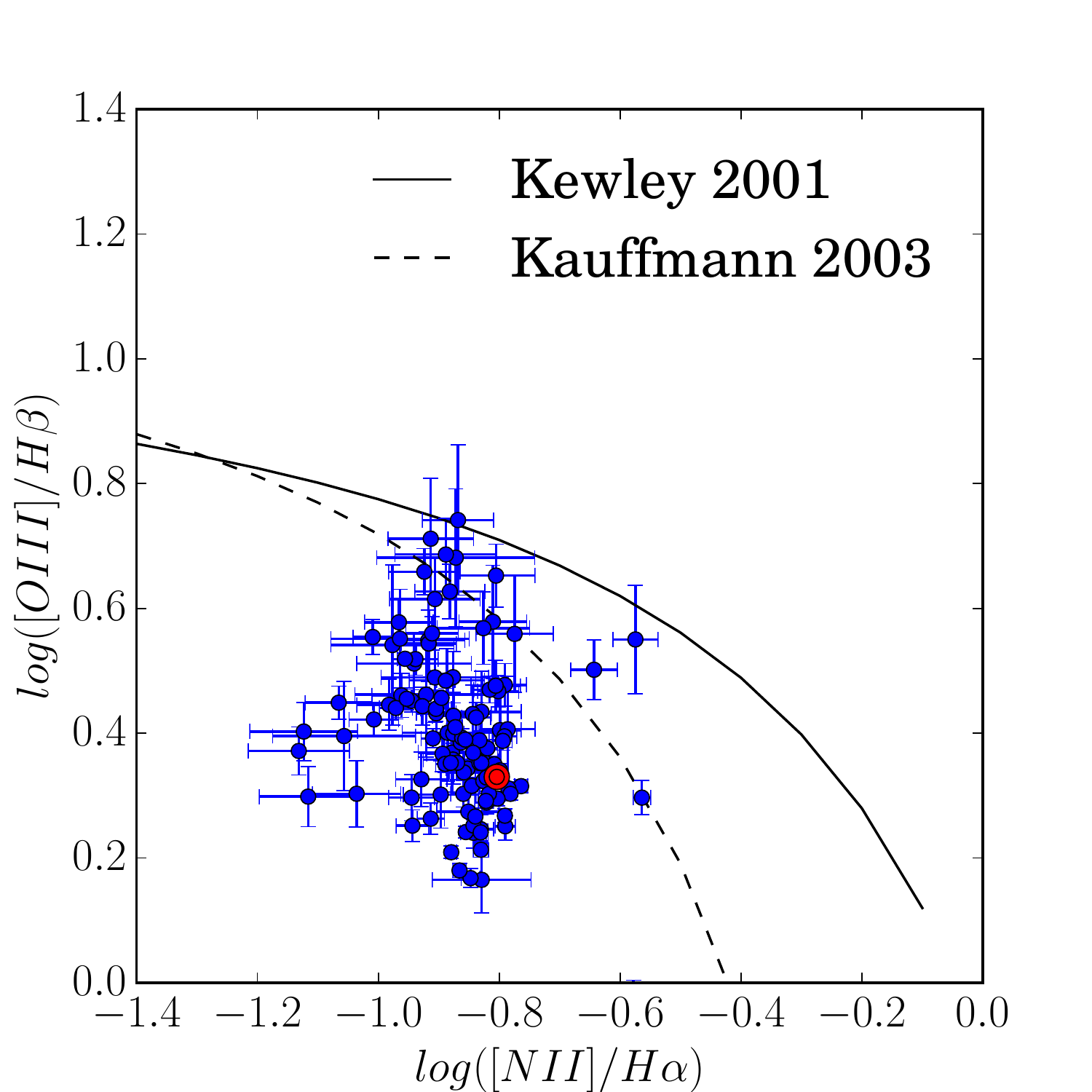}
 \caption{Emission line ratio diagnostic diagrams: [O \textsc{iii}]/H$\beta$ versus [S \textsc{ii}]/H$\alpha$ (left),  and [O \textsc{iii}]/H$\beta$ versus [N \textsc{ii}]/H$\alpha$ (right).  Black solid curve and dashed curve represent the theoretical maximum starburst line from \citet{Kewley2001} and \citet{Kauffmann2003} respectively, showing a classification based on excitation mechanisms. The points lying below and to the left of the Kewley line are those objects whose emission line ratios can be explained by the photoionisation by massive stars while the points above this line are those objects where some other source of ionisation is needed to explain their line ratios. The outliers above the Kewley line are the $\sim$ 2--$\sigma$ events but there are enough of them to argue for an additional mechanism. Red dot (with small error bars) denotes the corresponding ratios of the summed spectrum of the main emission region shown by the black square box in all maps.}
 \label{BPT plots}
 \end{figure*}

\begin{figure*}
\centering
\includegraphics[width = 0.48\textwidth]{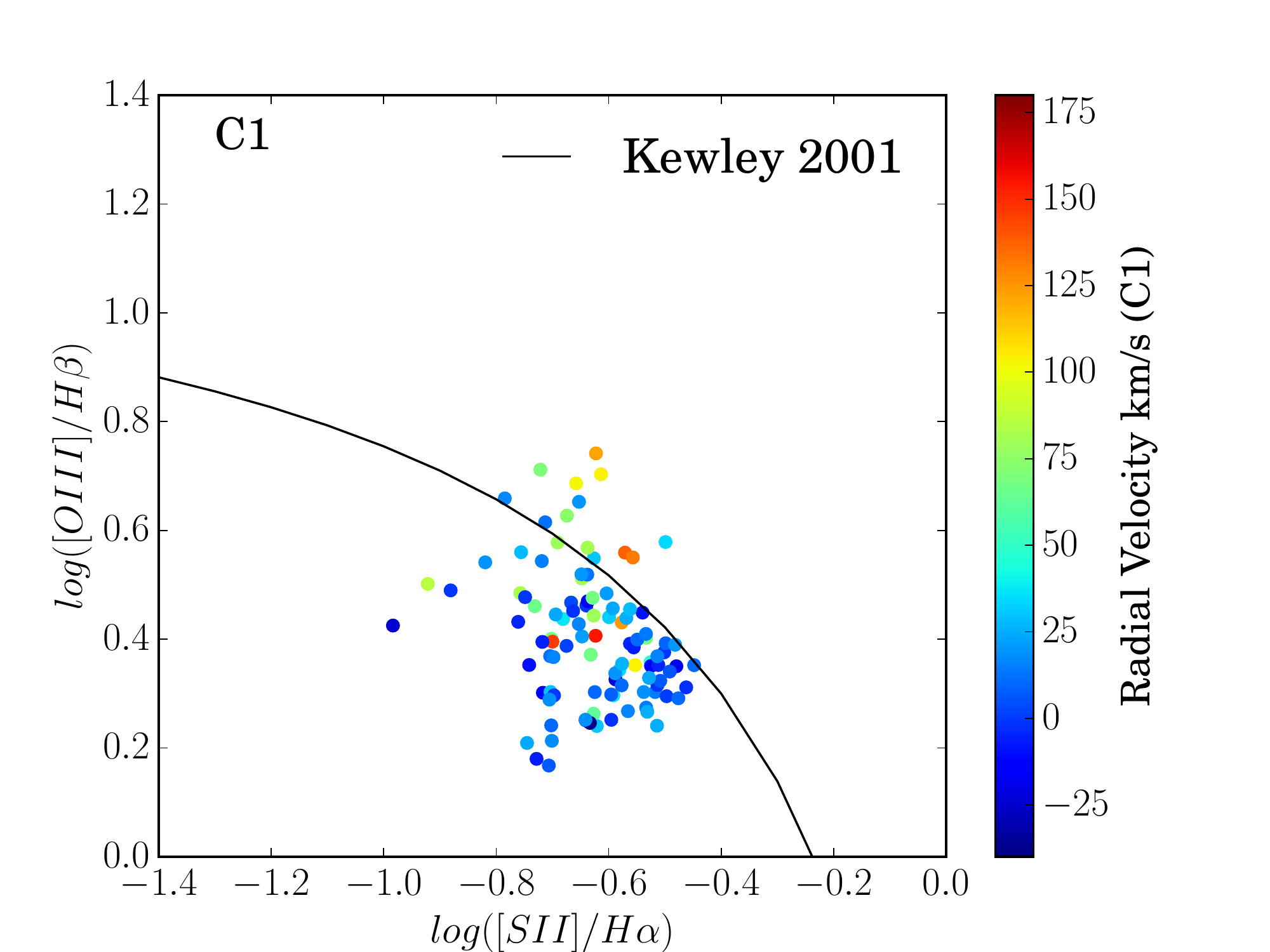}
\includegraphics[width = 0.48\textwidth]{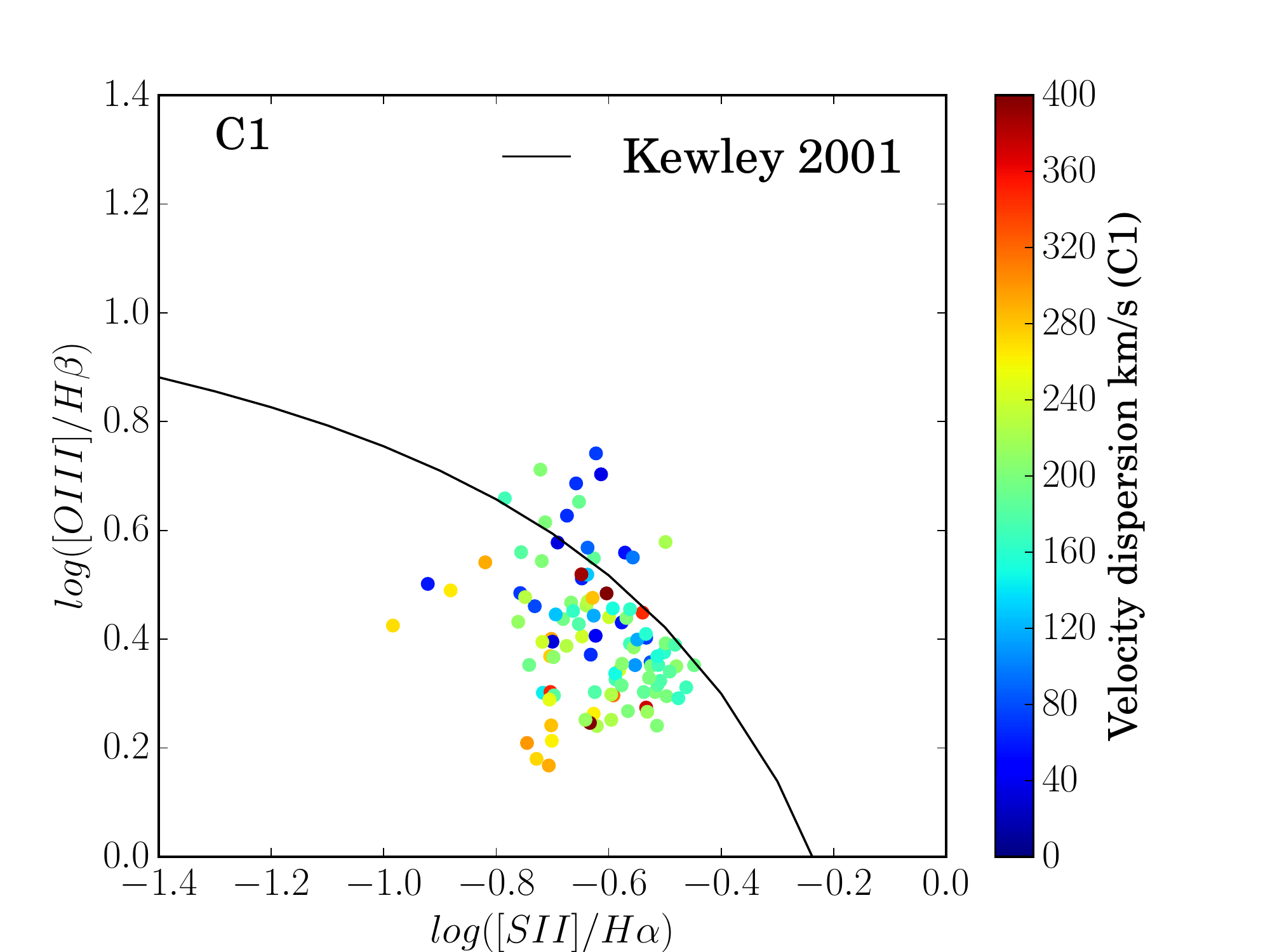}
\includegraphics[width = 0.48\textwidth]{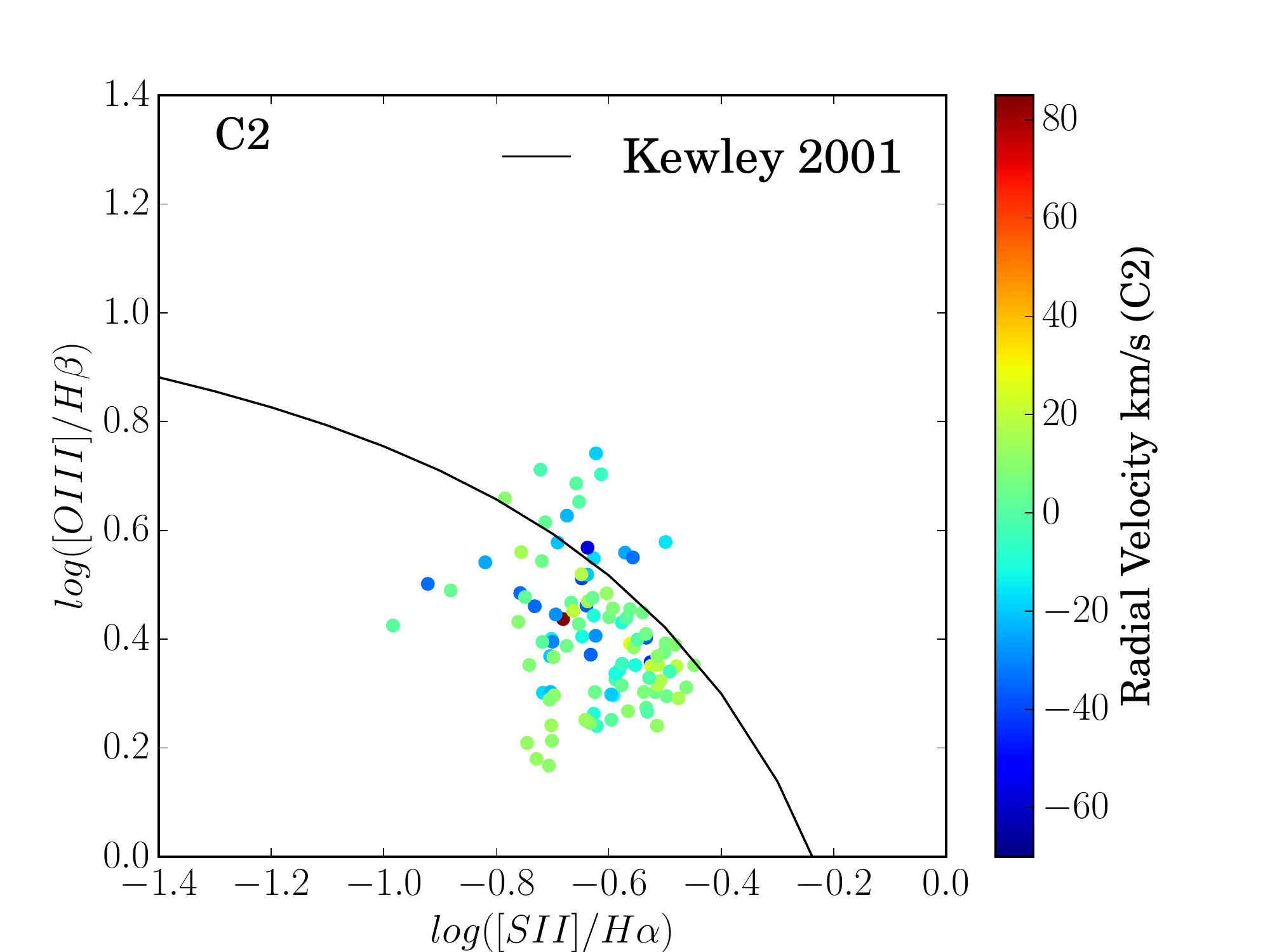}
\includegraphics[width = 0.48\textwidth]{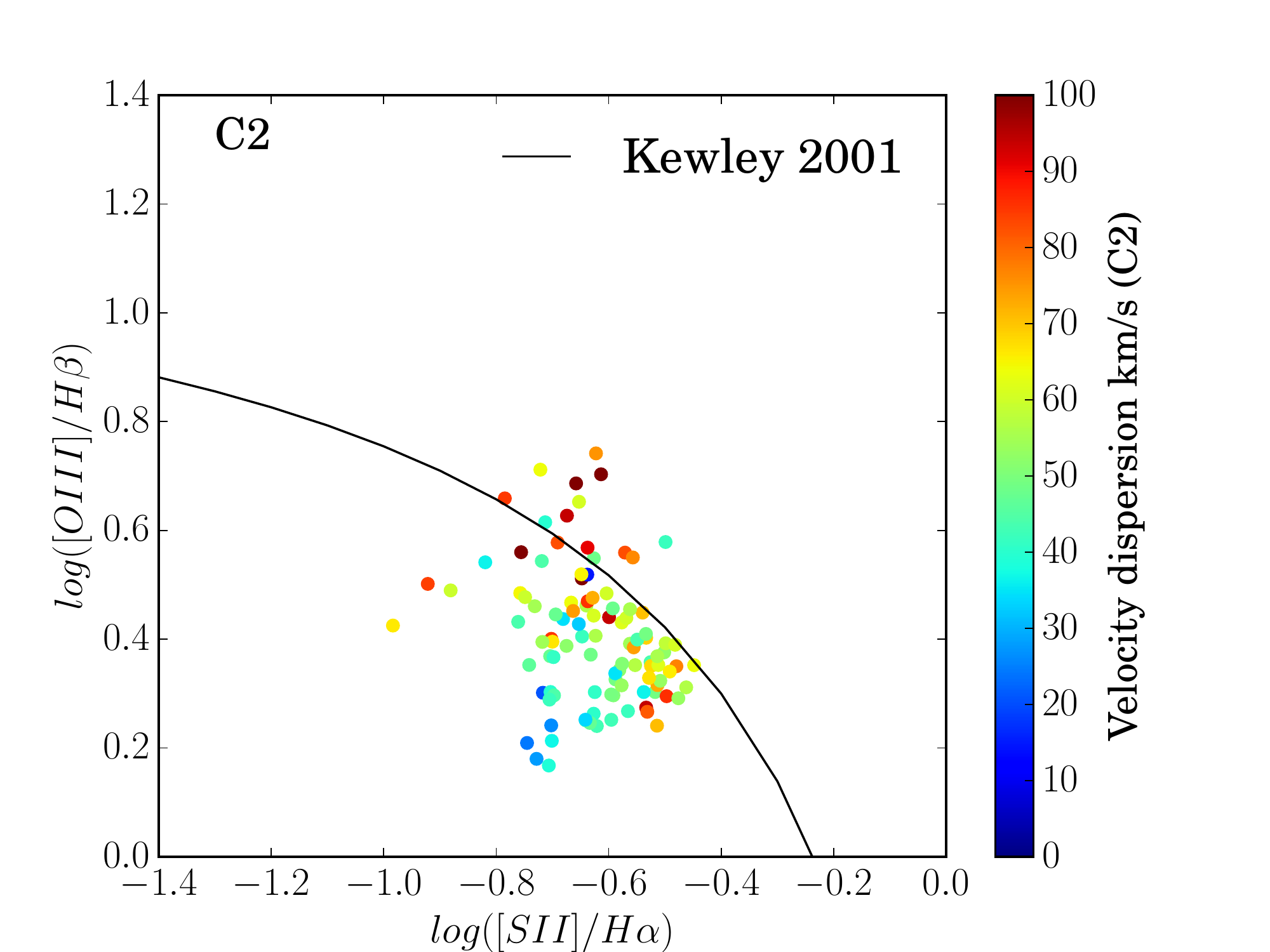}
\caption{[O \textsc{iii}]/H$\beta$ versus S \textsc{ii}]/H$\alpha$ diagnostic plots, where points are colour-coded with respect to radial velocity (left) and velocity dispersion (right) of the two spectral components of H$\alpha$ - C1 (upper panel) and C2 (lower panel). We do not find any obvious correlation/dependence between the points on the [S \textsc{ii}] diagnostic plot and radial velocity or velocity dispersion.}
\label{BPT velocity}
\end{figure*}

\begin{figure*}
 \centering
\includegraphics[width = 0.48\textwidth]{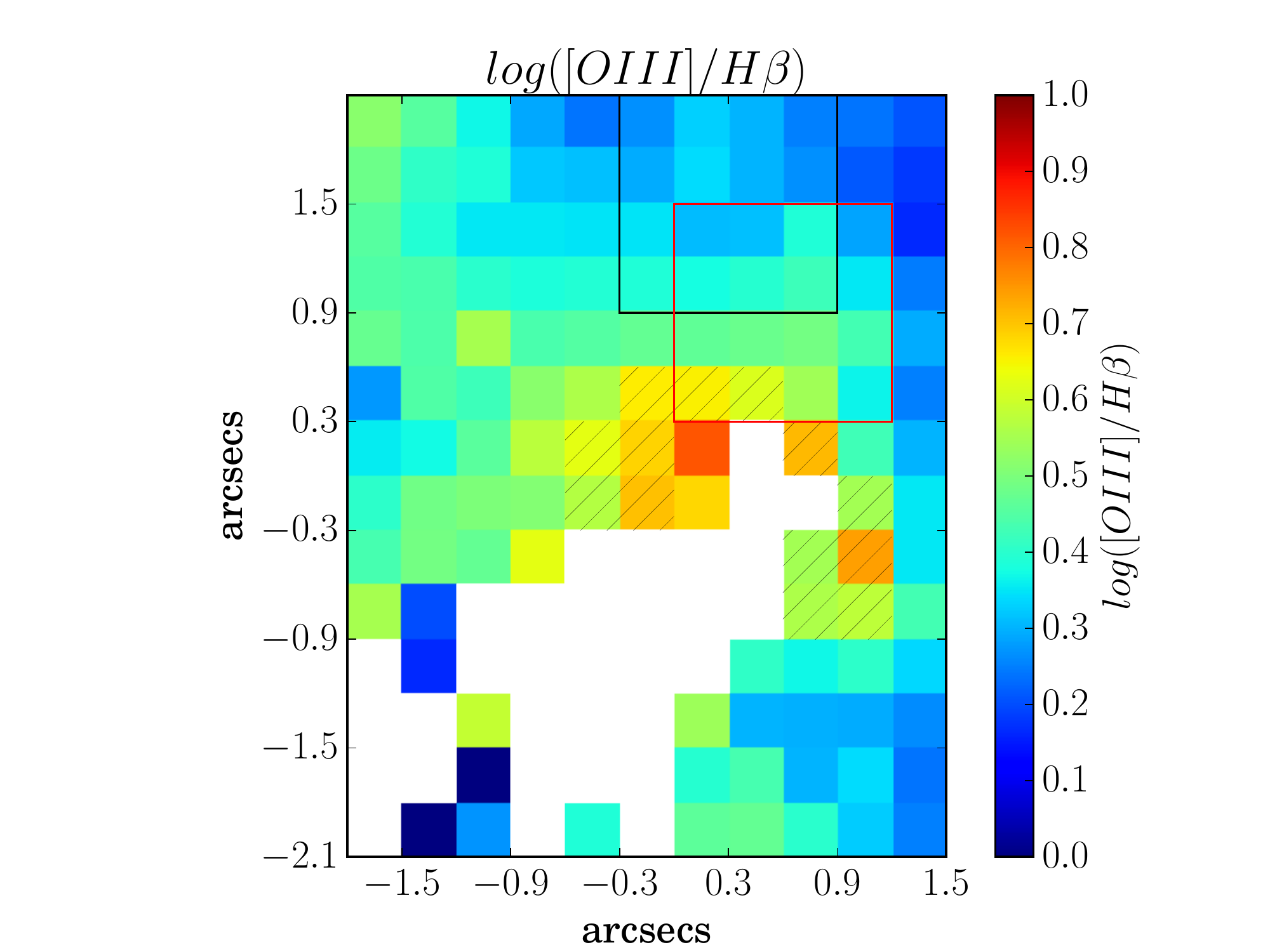}
\includegraphics[width = 0.48\textwidth]{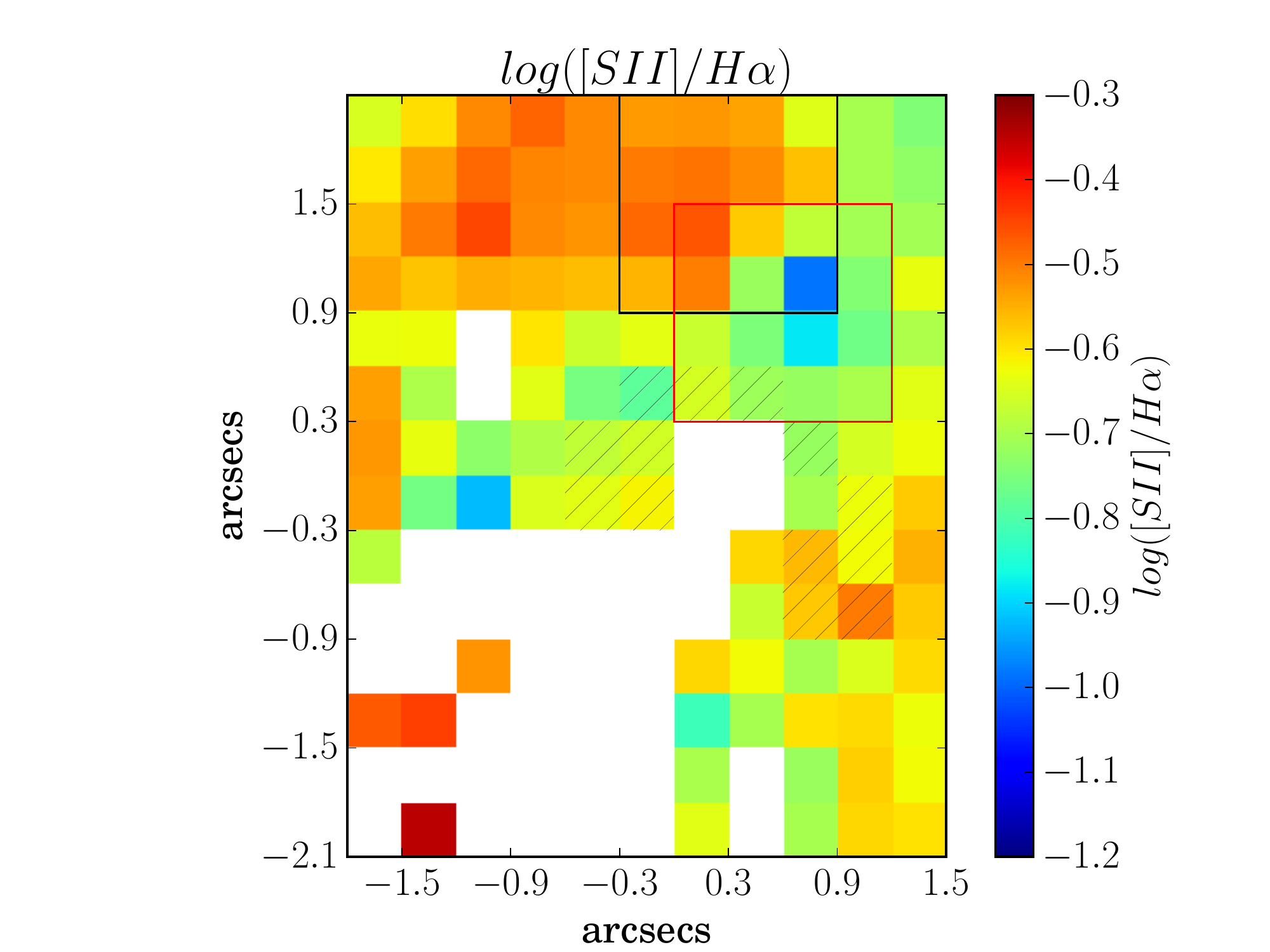}
\includegraphics[width = 0.48\textwidth]{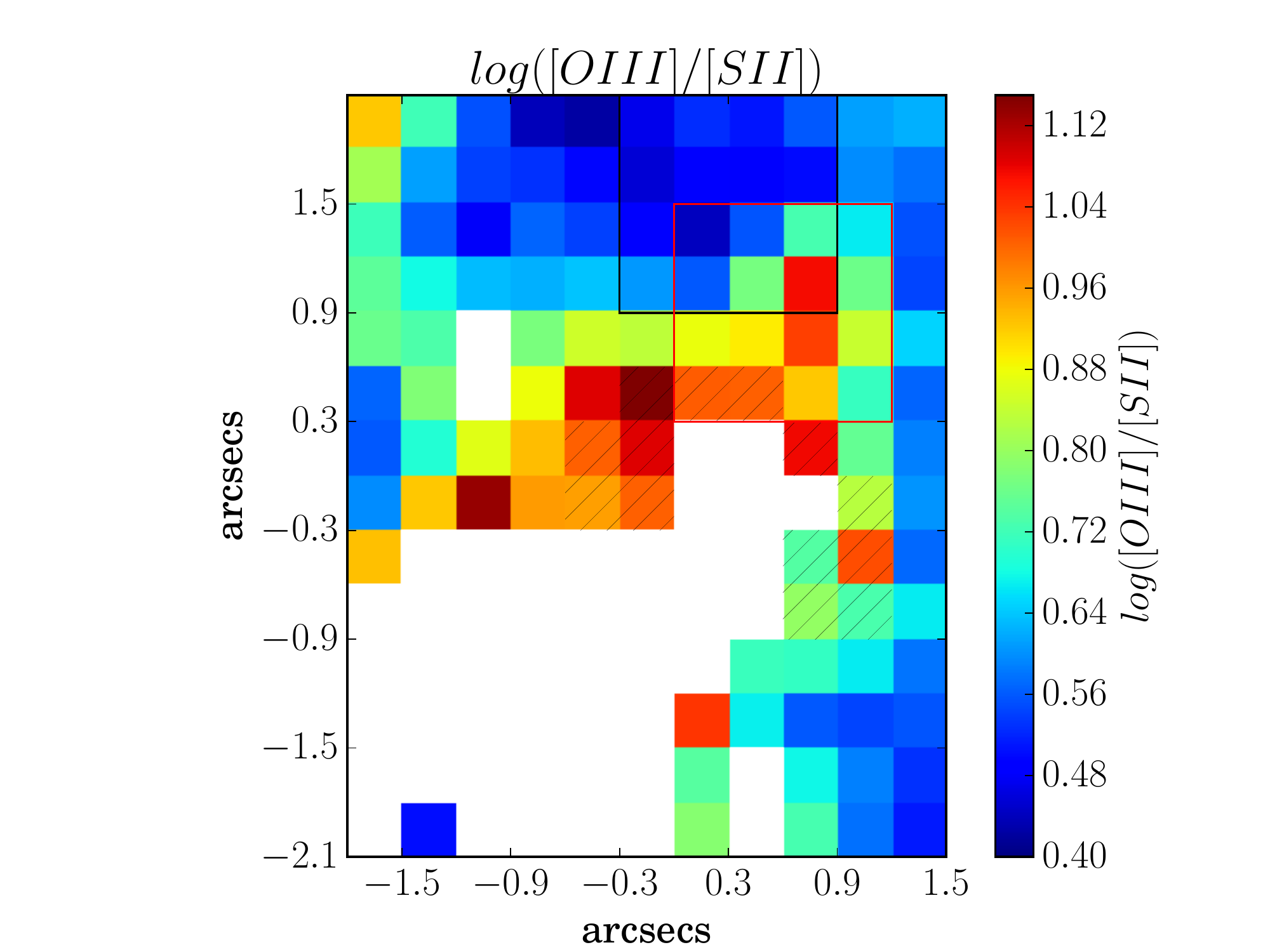}
\caption{Emission line ratio maps of [S \textsc{ii}]/H$\alpha$, [O \textsc{iii}]/H$\beta$ and [O \textsc{iii}]/[S \textsc{ii}]. The hatched region corresponds to the spaxels with line-ratios above the theoretical Kewley line in the [S \textsc{ii}] diagnostic plot (Figure \ref{BPT plots}, left). The black and red square box show the main emission region and the continuum region respectively for reference. The white spaxels correspond to the spaxels in which emission/flux had S/N $<$ 3. The shape of the hatched region correspond to the presence of highly ionised gas along the shock ionised front.}
 \label{overlap ratio1}
 \end{figure*}

\subsection{BPT diagrams}
\label{BPT diagram}

\indent In order to understand different excitation or ionisation mechanisms responsible for the kinematic components observed in Section \ref{kinematics}, we use the classical emission line diagnostic diagram by \citet{Baldwin1981}. Such diagrams are commonly known as the BPT diagrams which we show in Figure \ref{BPT plots} ([O \textsc{iii}]/H$\beta$ versus [S \textsc{ii}]/H$\alpha$ (left),  and [O \textsc{iii}]/H$\beta$ versus [N \textsc{ii}]/H$\alpha$ (right)). Note here that the line fluxes used to calculate the line ratios in the BPT diagram are not corrected for reddening due to their closeness in wavelength. The ``Kewley line" \citep{Kewley2001} indicating the theoretical maximum starburst line is over-plotted  as solid black curve on these diagrams. The emission line ratios of the region lying below and to the left of the Kewley line,  can be explained by the photoionisation of gas by massive stars while the emission line ratio in the region above this line is due to non-photoionised emission. The red dot denotes the line ratios derived from the summed spectrum fluxes of the main emission region (Tables \ref{properties}) shown by the black square box in all maps. As expected, the red dot (with very small error bars) lies well below the Kewley line on both diagnostic plots as it corresponds to a photoionised H \textsc{ii} region. The blue points with error bars denote the line-ratios and their uncertainties respectively for individual spaxels. In the [N \textsc{ii}] diagnostic plot (Figure \ref{BPT plots}, right) all the points lie below the Kewley line, indicating  photoionisation as the primary ionisation mechanism even in the region surrounding the star-forming region. This is an interesting observation because the gas surrounding a star-forming region may or may not be photoionised, and can be ionised by non-photoionising sources (e.g. mechanical shocks by supernovae,  stellar winds, expanding H \textsc{ii} regions). On the [N \textsc{ii}]/H$\alpha$ diagnostic plot, we also plot the line derived by \citet{Kauffmann2003} (referred to as the ``Kauffmann line") based on the SDSS spectra of 55 757 galaxies. The region enclosed between the Kewley line and the Kauffmann line is thought to be ionised either by photons or by shocks. On the [S \textsc{ii}]/H$\alpha$ diagnostic plot (Figure \ref{BPT plots}, left), we find points lying both below and above the Kewley line, which indicates that the surroundings of the H \textsc{ii} region may not be primarily photoionised but may also be shock ionised.  

\indent The identification of shocks and turbulence in the surrounding of a star-forming region is important because shocks can have a strong impact on the evolution of gaseous clouds which are the sites of star-formation, and hence can affect the star formation rate and efficiency in the star-forming regions. Hence, we further investigate different ionisation mechanisms depicted in the [S \textsc{ii}] diagnostic plot by using the information on radial velocity and velocity dispersion obtained from multi-component line fitting to the strong H$\alpha$ emission line. Figure \ref{BPT velocity} shows this analysis where points on the [S \textsc{ii}] diagnostic plot are colour-coded with respect to radial velocity (left) and velocity dispersion (right) of the two spectral components of H$\alpha$ - C1 (upper panel) and C2 (lower panel).  The points above the Kewley line show velocity widths $\sim$ 50--100 km s$^{-1}$ for both of the components (Figure \ref{BPT velocity}, right plots), which suggests that the gas ionised by shock excitation is primarily cool. The diagnostic plot corresponding to the velocity width of the C2 component (bottom right plot) shows that the velocity dispersion is slightly increasing as we go above the Kewley line, which suggests the shock ionised gas might be broadened. However we do not observe this behaviour in the broad C1 component of the gas (upper right plot). Hence overall we do not find any obvious correlation/dependence between the position of the points on the [S \textsc{ii}]/H$\alpha$ diagnostic plot and  radial velocity or velocity dispersion.

\subsection{Emission line ratio maps}
\label{line ratio maps}
\indent Interestingly, the points above the Kewley line on the [S \textsc{ii}] diagnostic plot (Figure \ref{BPT plots}) show higher [O \textsc{iii}]/H$\beta$ rather than [S \textsc{ii}]/H$\alpha$, suggesting highly ionised gas ([O \textsc{iii}]) along the shock front. We investigate this further through the line ratio maps of [O \textsc{iii}]/H$\beta$, [S \textsc{ii}]/H$\alpha$ and [O \textsc{iii}]/[S \textsc{ii}] in Figure \ref{overlap ratio1}. Comparing doubly- and singly-ionised gas of the same species (e.g. [O \textsc{iii}]/[O \textsc{ii}]) would give a better estimate of the relative amount of gas in the two ionisation states. However we instead use [O \textsc{iii}]/[S \textsc{ii}] because [S \textsc{ii}] map has larger number of spaxels with sufficient S/N ($>$  3) than [O \textsc{ii}] (Figure \ref{observed flux}), and [S \textsc{ii}] has similar ionisation potential as that of [O \textsc{ii}]. In the emission line maps in Figure \ref{overlap ratio1}, the hatched regions correspond to the spaxels with line-ratios above the theoretical Kewley line in the [S \textsc{ii}] diagnostic plot (Figure \ref{BPT plots}, left). Overall, the regions appear to be bow-shaped. The cause of this bow-shock remains unknown however, since there are no stellar clusters on the south-east of this region which could lead to supersonic local random motion and create the apparent bow-shock. The continuum map instead shows the presence of the stellar cluster in the north of the shock-ionised region. This suggests that the gas is ionised by the stellar winds originating from this stellar cluster.  It is also possible that the shock-front has been triggered by the recent interaction of NGC 4449 with DDO 125 present in the south of this galaxy \citep{Hunter1998} or due to a tidal interaction with another dwarf as evidenced by tidal stellar stream in the south-east \citep{Martinez-Delgado2012} of this galaxy. The spectra of shock-ionised gas are generally characterised by double Gaussians. Figure \ref{velocity structure} shows that the spectra corresponding to the shock-ionised region on the H$\alpha$ emission line map consist of these expected double Gaussians of equal velocity widths  and are blended.

\begin{figure}
\centering
\includegraphics[width = 0.48\textwidth]{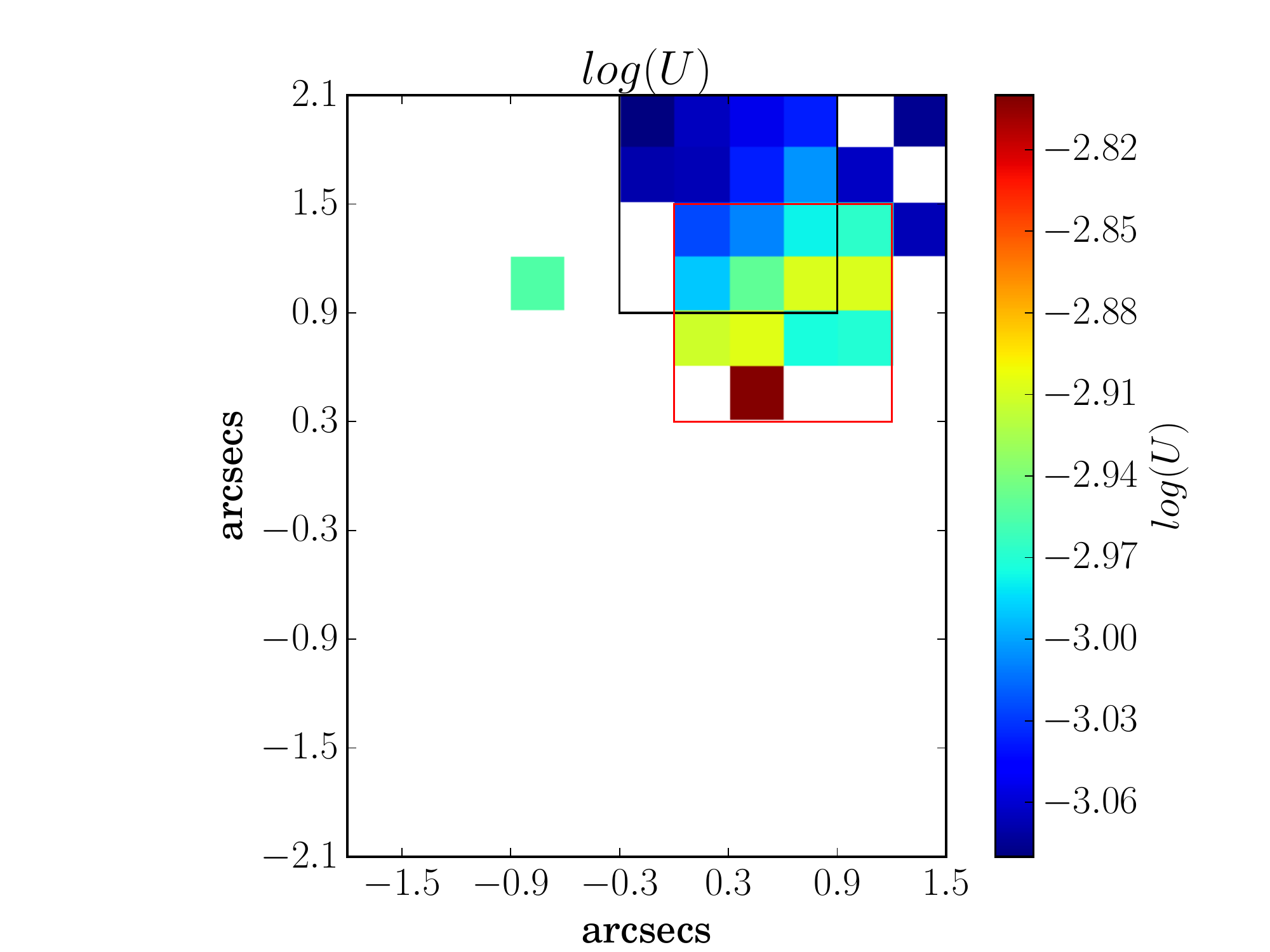}
\caption{Ionisation parameter $\mathcal{U} \equiv q/c$ as mapped from the diagnostic method of \citet{Kewley2002} based on the ratio of [O \textsc{iii}] $\lambda$5007  and [O \textsc{ii}] $\lambda\lambda$3726, 3729. The black and red square box show the main emission region and the continuum region for reference respectively. White spaxels correspond to the spaxels in which emission/flux had S/N $<$ 3.}
\label{Uparam}
\end{figure}

\subsection{Ionisation Parameter}

\indent The ionisation state of a medium is characterised by the ionisation parameter ($q$), defined as the ratio of the number of particles in the ionised state to that in the non-ionised state. For the photoionised gas, it is simply the ratio of the surface density of ionising photon flux  (radiation pressure) to the local neutral gas number density (gaseous pressure), and is interpreted as the maximum velocity at which the local radiation field can drive an ionisation front \citep{Kewley2002}. It is generally expressed as a dimensionless parameter $\mathcal{U} \equiv q/c$ where c is the speed of light. Using the ionisation parameter diagnostic devised by \citet{Kewley2002} which relates O32 (flux ratio of [O \textsc{iii}] $\lambda$5007  to [O \textsc{ii}] $\lambda\lambda$3726, 3729) and log ($q$) with the metallicity-dependent constants, we map log $\mathcal{U}$ (Figure \ref{Uparam}).  Note here that in order to calculate $\mathcal{U}$,  we use the metallicity map derived from direct T$_e$ method which is the most robust metallicity estimator as we discuss later in Section \ref{chemistry}. We find that log $\mathcal{U}$ is highest at the peak of the stellar continuum (see first plot of the Figure \ref{observed flux}) and decreases radially outwards. We expect $\mathcal{U}$ to be highest at the inner edge of the ionised region and falling to zero at the outer edge in case of full depletion of ionising flux \citep{Nicholls2013}, i.e. when the source of ionising radiation (indicated by the continuum) is aligned with respect to the ionised gas. In our case, the peak of the continuum is not aligned with the peak of the main emission region. Hence, we do not find the expected decrease from the inner to outer region of the H \textsc{ii} region. We find one spaxel on the outer edge of the continuum which has unusually high log $\mathcal{U}$ = $-$ 2.8 and also high error bar 0.19 compared to the standard error ($\sim$ 0.01) of measurement on log $\mathcal{U}$. A variation of $\sim$ $-$3.08 to $-$2.80 is observed across the region with S/N > 3, with average value of  log $\mathcal{U}$ = $-$3.00$\pm$0.07.


\begin{table*}
\centering
\caption{Metallicities derived using direct and indirect methods for integrated spectrum of the bright emission region as well as on spaxel-by-spaxel basis.}
\label{integrated Z}
\begin{tabular}{@{}ccccc@{}}
\toprule
Diagnostic      & R$_{23}$      & N2            & O3N2          & Direct         \\ \midrule
& & Integrated Spectrum Abundance  &\\ \midrule
12 + log(O/H)$^a$   & 8.16 $\pm$ 0.15 & 8.39 $\pm$ 0.18 & 8.37 $\pm$ 0.14 & 7.88 $\pm$ 0.14 \\
         Z (Z$_{\odot}$) & 0.29 $^{+0.12}_{-0.08}$          & 0.50 $^{+0.27}_{-0.17}$          & 0.48 $^{+0.19}_{-0.13}$          & 0.15 $^{+0.06}_{-0.04}$\\ 
          \midrule
& & Spatially-resolved Abundance & \\ \midrule
Mean metallicity$^b$ & 8.17 $\pm$ 0.15&8.35 $\pm$ 0.06 &8.33 $\pm$ 0.06 &7.89 $\pm$ 0.08 \\
Maximum deviation $\Delta$ & 0.58& 0.35& 0.34& 0.32 \\ \midrule
Calibration uncertainty &0.15 & 0.18& 0.14& --\\ \bottomrule
\end{tabular}

Notes: $^a$: The quoted error on each metallicity estimate takes into account both the statistical error (in flux measurement) and the calibration error associated with each of the empirical calibrators (given in the last row). Section \ref{direct} details the error estimate on the direct method metallicity. $^b$: Mean metallicity along with standard deviation across the FOV.

\end{table*}

\begin{figure}
 \centering
\includegraphics[width = 0.48\textwidth]{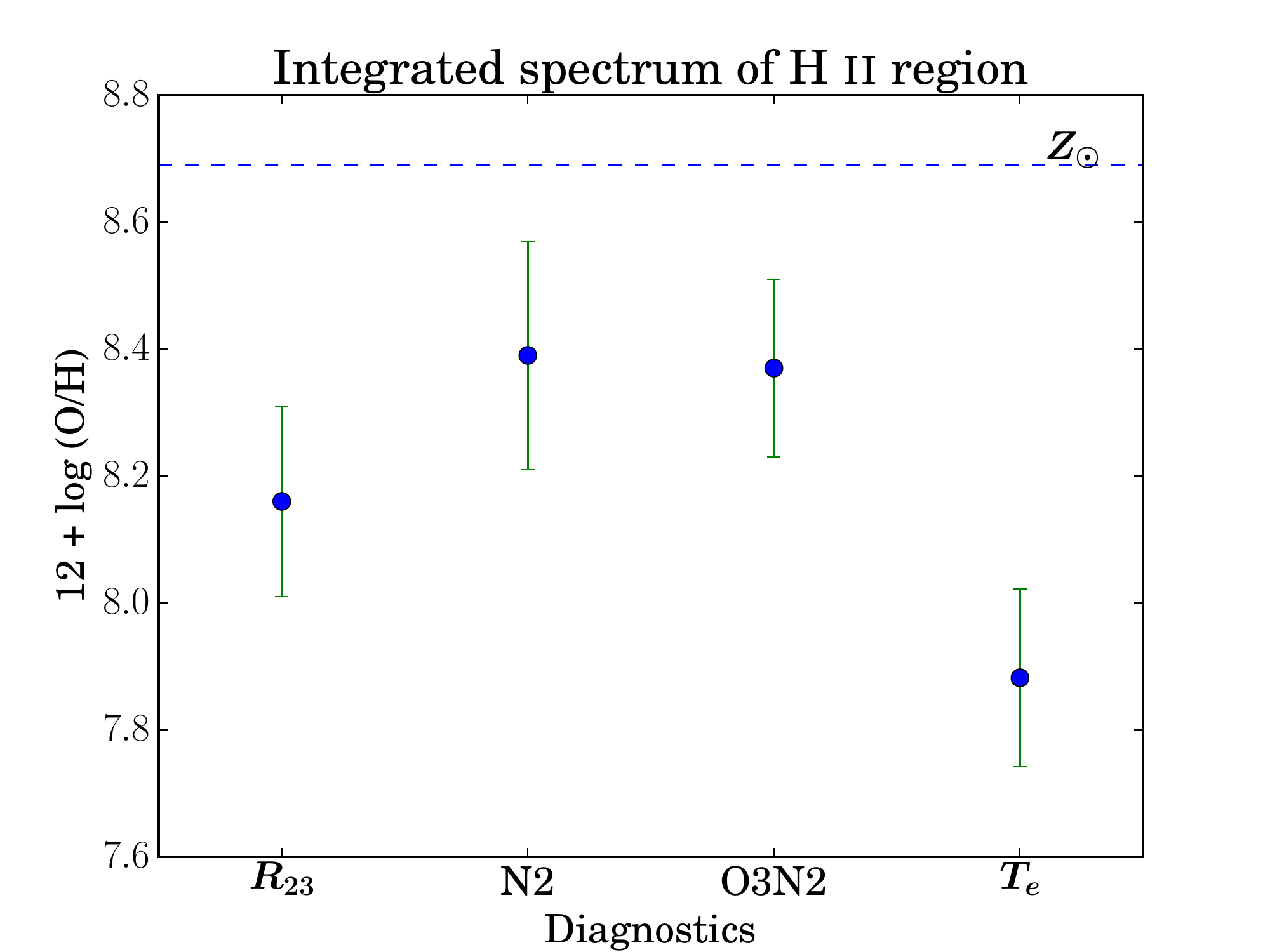}
\caption{Comparison of metallicities derived from the integrated spectrum of the main emission region using direct and indirect methods. The error bar on each indirect metallicity estimates takes into account both the statistical error (in flux measurement) and the calibration error associated with each of the empirical calibrators. See Section \ref{direct} for the error estimate on the direct-method (T$_e$) metallicity. The horizontal blue dashed line corresponds to the solar abundance of 12 + log(O/H) $\sim$ 8.69 \citep{Asplund2009}.}
\label{summary Z}
\end{figure}

  \begin{figure*}
 \centering
\includegraphics[width = 0.48\textwidth]{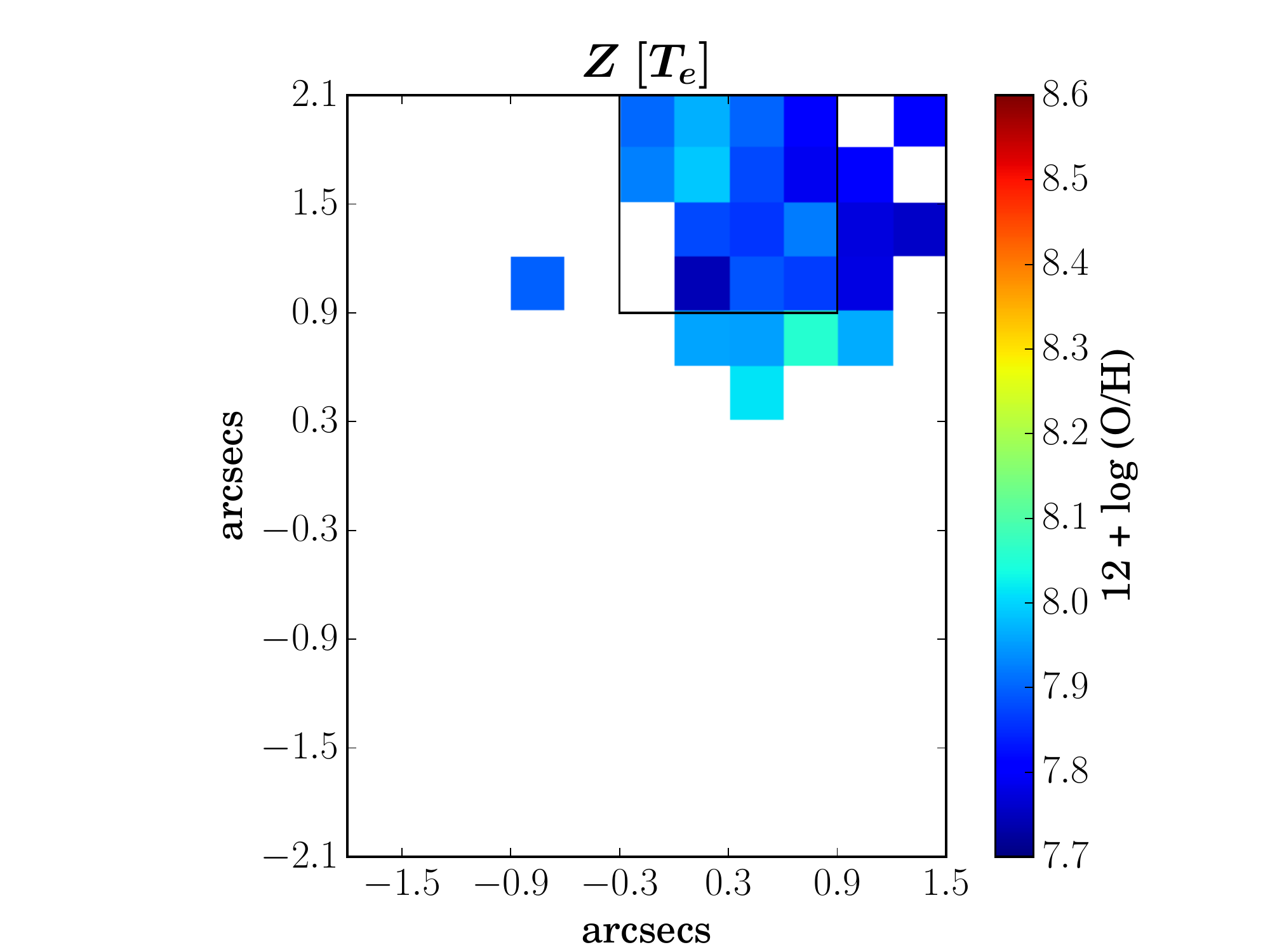}
\includegraphics[width = 0.48\textwidth]{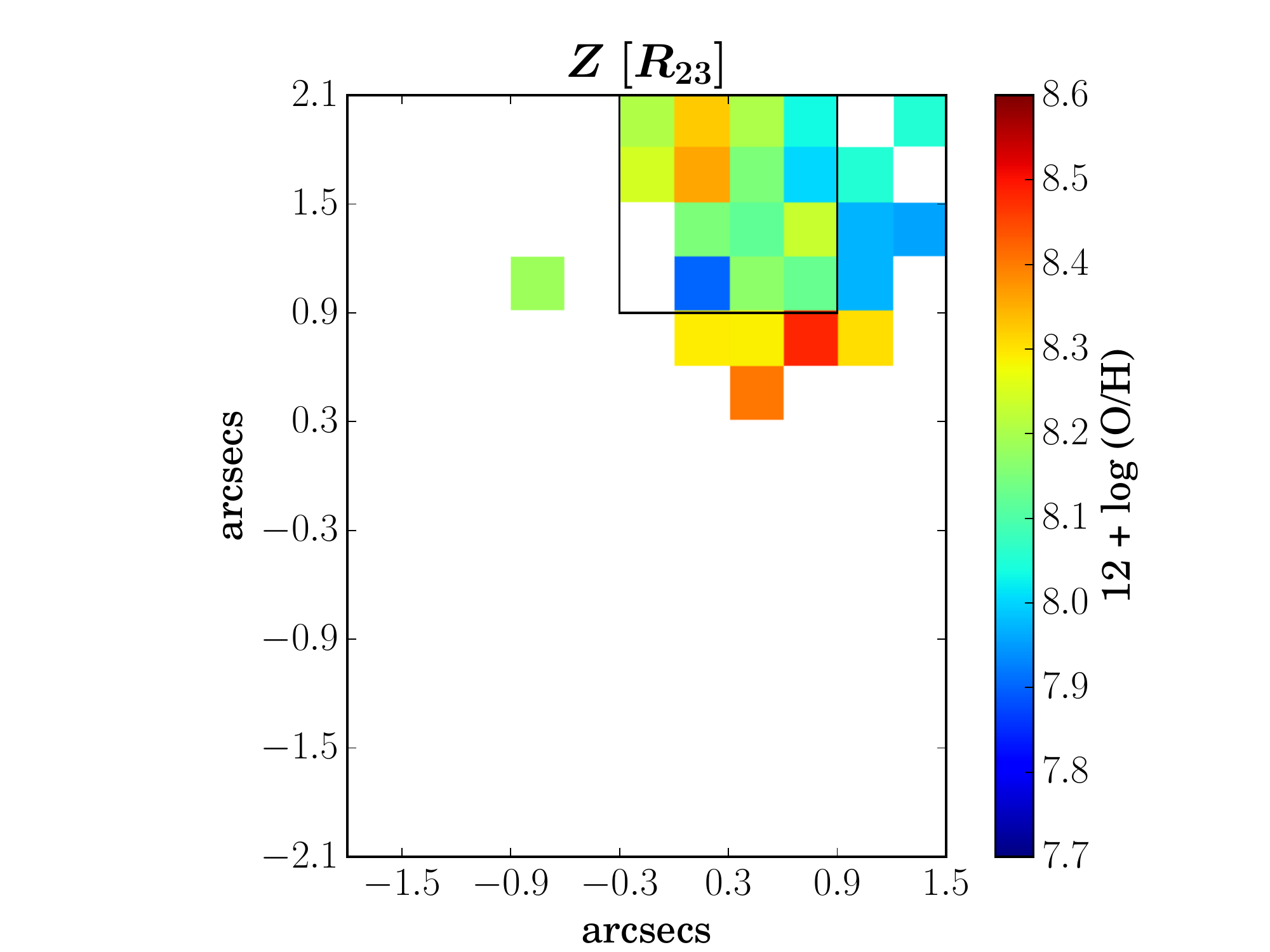}
\includegraphics[width = 0.48\textwidth]{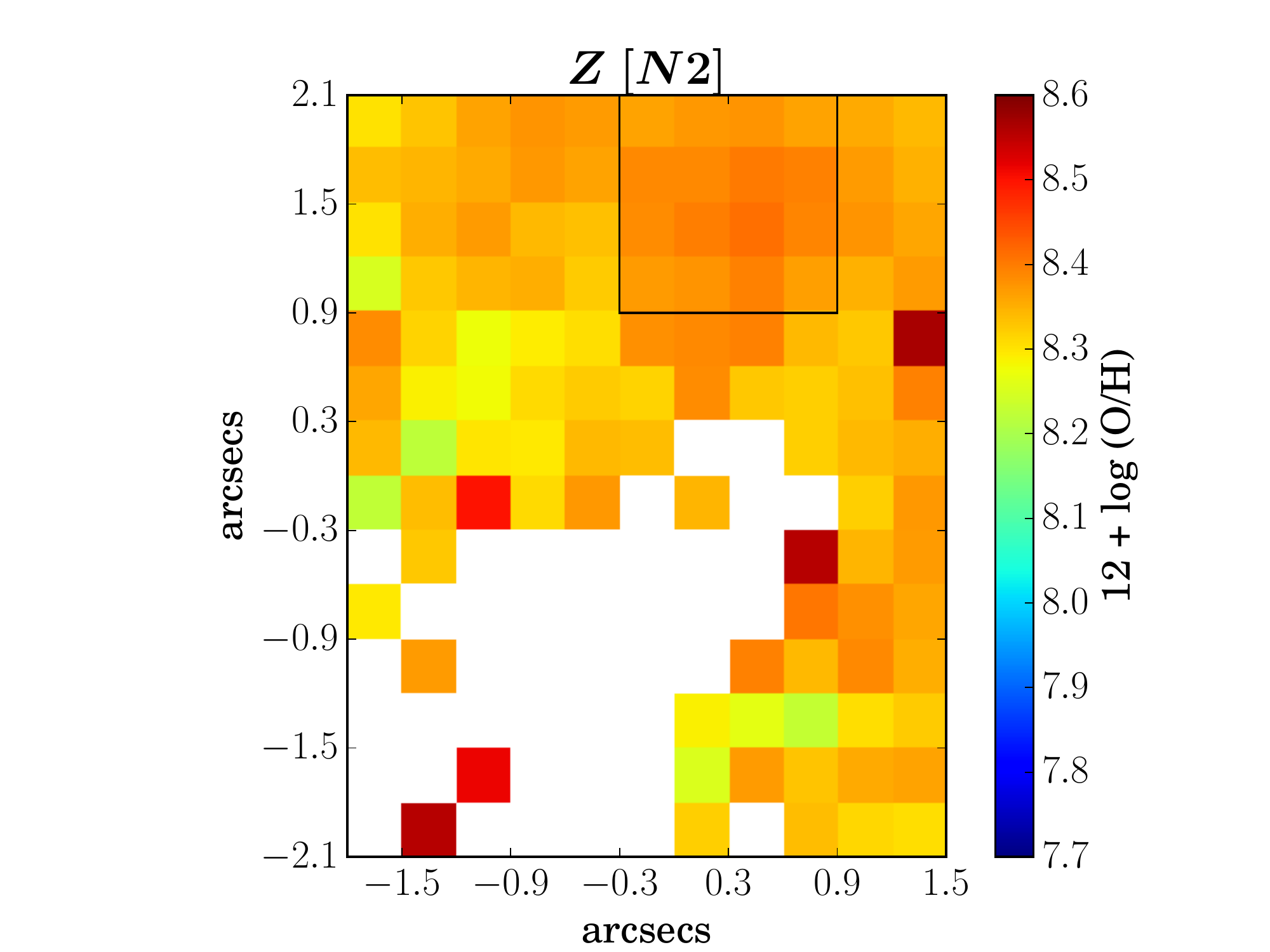}
\includegraphics[width = 0.48\textwidth]{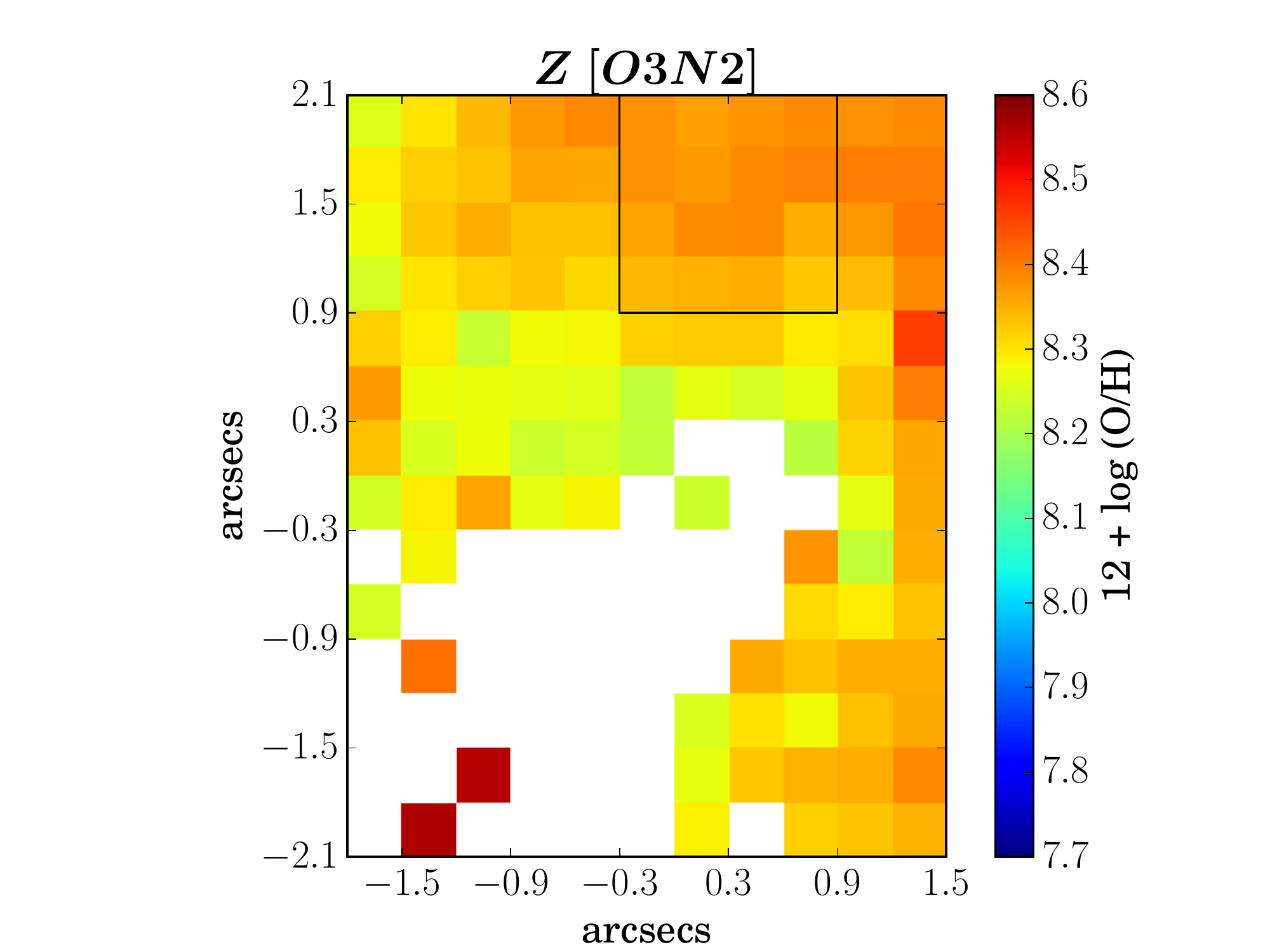}
\caption{Metallicity maps derived from the direct Te method (Z [T$_e$]) and indirect methods (Z [R$_{23}$], Z [N2] and Z [O3N2]). The scales of all maps are adjusted to the same limits so that the obvious offsets between metallicities from different methods and  and relative variation within maps are clear. The black square box shows the main emission region and white spaxels correspond to the spaxels in which emission line fluxes had S/N $<$ 3.}
\label{all Zs}
\end{figure*}

 \begin{figure*}
 \centering
\includegraphics[width = 0.48\textwidth]{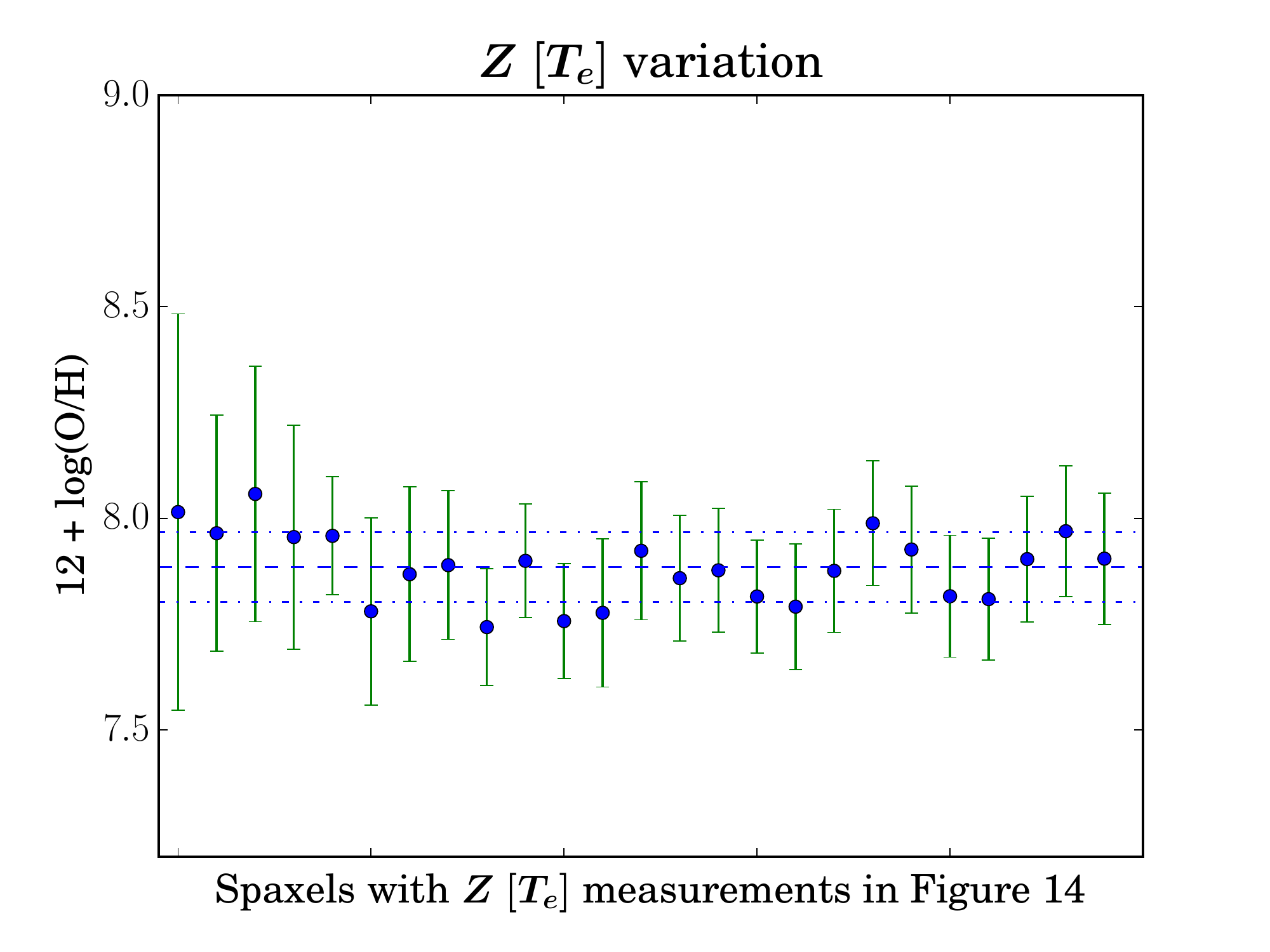}
\includegraphics[width = 0.48\textwidth]{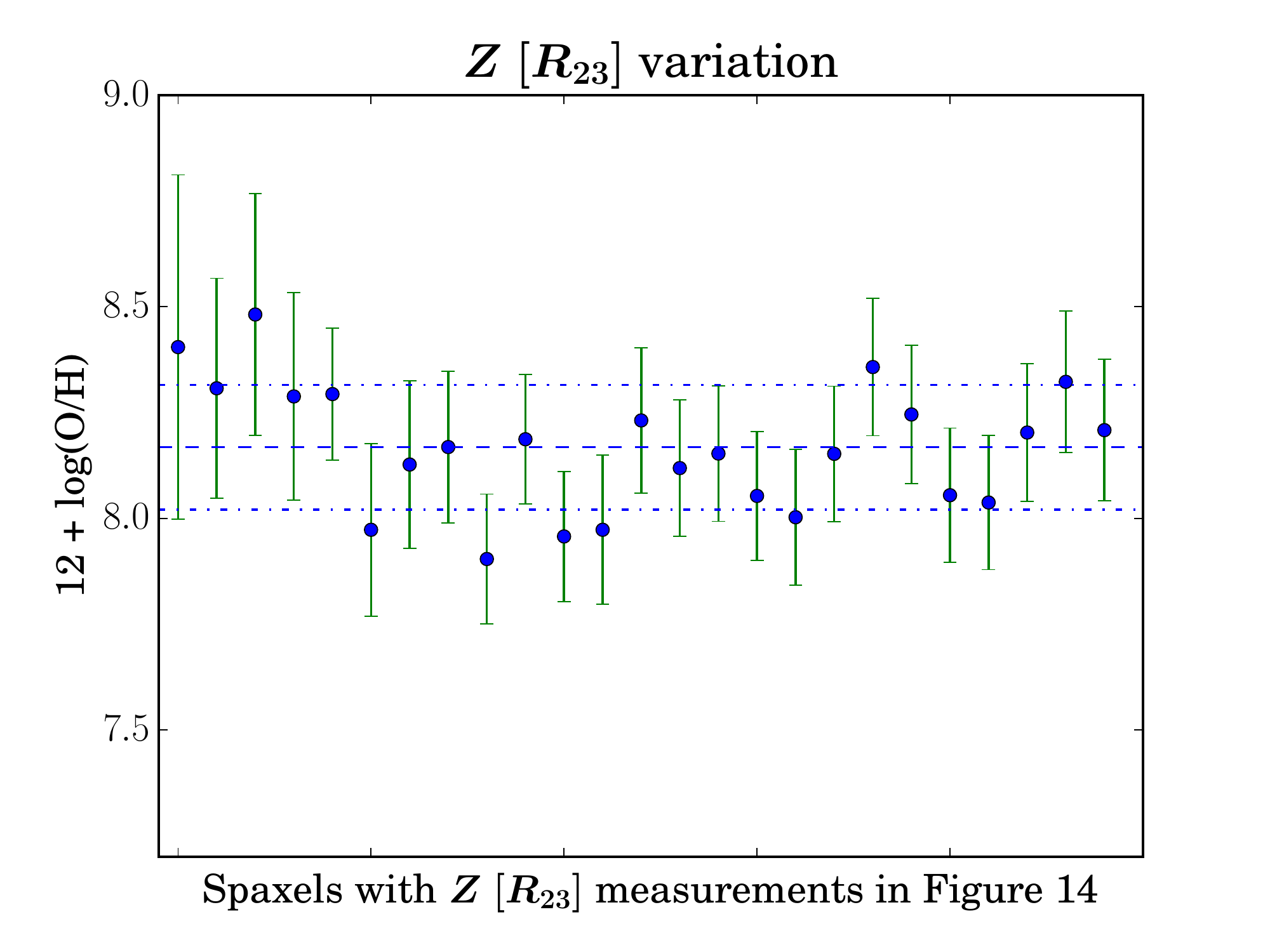}

\caption{Scatter plots of metallicity values in each spaxel of Z [T$_e$] (left) and Z [R$_{23}$] (right) maps. Section \ref{direct} details error estimate on the direct-method (T$_e$) metallicity. The error bar on each metallicity value Z [R$_{23}$] in the right plot also includes calibration error of 0.15 dex along with the propagated error on the flux measurement. The dashed and dotted line in each of the plots indicate the mean and standard deviation of all metallicities in the corresponding metallicity maps. In both the plots, we find that all the individual metallicity values with error bars lie within the 1-$\sigma$ standard deviation of the mean metallicity value, indicating chemical homogeneity throughout the regions.}
\label{Z scatter}
\end{figure*}

\section{Chemical Abundance Analysis}
\label{chemistry} 

\subsection{Integrated spectrum chemical abundances}
\label{integrated abundance}
\indent  In this section, we present a chemical abundance analysis of the integrated spectrum of the main emission region (Figure \ref{spectra}, upper two panels) based on direct and indirect metallicity diagnostic methods. The direct method involves the measurement of the abundance $directly$ from the electron temperature T$_e$ combined with the relative emission flux of collisional-to-Balmer emission lines (Section \ref{direct}), while the indirect method involves the ratio of strong emission lines (Section \ref{indirect}). Since oxygen is the most prominent heavy element generally observed in the optical spectrum in the form of O$^0$,  O$^+$,  O$^{2+}$ and  O$^{3+}$, it is used as a proxy for total metallicity. Our analysis primarily includes oxygen abundance determination, however we also briefly mention nitrogen-to-oxygen abundance ratios. 

\subsubsection{Electron Temperature and Density}
\label{temden}
\indent The electron temperature T$_e$ is derived from the deredenned [O \textsc{iii}] line ratio, [O \textsc{iii}]($\lambda$5007+$\lambda$4959)/$\lambda$4363 and the calibrations based on the MAPPINGS IV photoionisation code which uses up-to-date atomic data \citep{Nicholls2013}. Unfortunately, the S/N ratio of [O \textsc{iii}] $\lambda$4363 and the [S \textsc{ii}] doublet in the individual spectrum across the FOV was too low on an individual spaxel basis to map the spatially-resolved  electron temperature T$_e$ and electron density N$_e$.  We can however calculate the  T$_e$ and  N$_e$ from the integrated spectrum of the main emission region which has S/N > 3 for [O \textsc{iii}] $\lambda$4363 and the [S \textsc{ii}] doublet.  
From the integrated spectrum of the main emission region, we obtain an electron temperature of $\sim$ 15100$\pm$1700 K. Using the derived T$_e$ value and the [S\textsc{ii}] doublet ratio $\lambda6717/\lambda6731$ from the integrated spectrum of the main emission region, we compute an electron density N$_e$ of $\sim$ 316$\pm$47 cm$^{-3}$ for the main emission region. 

\subsubsection{Direct method}
\label{direct}
 \indent  To calculate the total elemental abundance, ionic abundance ratios  are calculated using the electron temperature of the ionisation zone dominated by the corresponding ion.  We estimate O$^+$/H$^+$ and  O$^{2+}$/H$^+$ using T$_e$ ([O $\textsc{ii}$]) and T$_e$ ([O $\textsc{iii}$])  respectively using the formulations of \citet{Nicholls2014}. Using the empirical relation between T$_{e}$([O  \textsc{ii}]) and  T$_e$ [O $\textsc{iii}$] given in \citet{Lopez2012b}, we calculate T$_{e}$([O  \textsc{ii}]) $\sim$ 12300 $\pm$ 900 K. The O$^+$/H$^+$ and  O$^{2+}$/H$^+$ are finally summed to calculate the elemental O/H. We derive a metallicity 12 + log(O/H) = 7.88 $\pm$ 0.14 $\sim$ 0.15 $^{+0.06}_{-0.04}$ Z$_{\odot}$ from the integrated spectrum of the main emission region. Note here that the error estimate on  T$_{e}$([O  \textsc{ii}]) obtained by the error propagation on flux measurement error is underestimated because of the shape of the empirical model of \citet{Lopez2012b} at the required T$_{e}$([O  \textsc{ii}]). Hence we estimated the error on T$_{e}$([O  \textsc{ii}]) through a Monte Carlo simulation (100,000 samples) by using Gaussian distribution of T$_{e}$([O  \textsc{iii}]) with a sigma (width) equal to the error on T$_{e}$([O  \textsc{iii}]). This error is hence propagated to obtain error on metallicity measurements.  

\indent Assuming the same T$_e$ for [O \textsc{ii}] and [N \textsc{ii}], and that N$^+$/O$^+$ = N/O, we use the formulation from \citet{Nicholls2014} and derive N/O = $-$1.03 $\pm$ 0.07.
  \subsubsection{Indirect methods}
  \label{indirect}
 \indent Although the direct method is deemed the most reliable to estimate chemical abundances, it is sometimes difficult to calculate because it requires the detection of usually weak auroral lines (e.g. [O \textsc{iii}] $\lambda$4363, [N \textsc{ii}] $\lambda$5755). This has to led to the development of theoretical and empirical calibrations of metallicity based on ratios of strong lines \citep[see][for a review]{Kewley2008}. We estimate metallicity from the following three widely used methods:
  
  \begin{enumerate}
  
\item \indent The \textit{R$_{23}$} parameter was first defined by \citet{Pagel1979} as R$_{23}$ = ([O $\textsc{ii}$] $\lambda\lambda$3727,29 + [O $\textsc{iii}$] $\lambda\lambda$4959,5007)/H$\beta$. Since oxygen is one of the principle coolants in H \textsc{ii} nebulae, R$_{23}$ is in turn sensitive to the oxygen abundance. However, it has some drawbacks and requires caution. One of the main problems is the degeneracy of R$_{23}$ with abundance, i.e. a single value of R$_{23}$ corresponds to two values of the O abundance. From our abundance measurement using the direct method (Section \ref{direct}), we can break the degeneracy and decide to use the calibrations corresponding to the lower-metallicity branch. The second problem arises due to the dependence of R$_{23}$ on the ionisation parameter.  Hence, of the many calibrations available \citep{Zaritsky1994, Kewley2002, Kobulnicky2004}, we use the metal-poor (lower) branch calibration from \citet{Kobulnicky1999} which takes  into account of the effect of the ionisation parameter and is based on a set of H \textsc{ii} region models of  \citet{McGaugh1991} using the photoionisation code Cloudy \citep{Ferland1998}. From the integrated spectrum of the main emission region, we find 12+log(O/H) = 8.16 $\pm$ 0.02 (statistical) $\pm$ 0.15 (calibration).

\item \indent The \textit{N2} parameter was initially defined as N2 = log ([N \textsc{ii}] $\lambda$6584/H$\alpha$) and proposed as an abundance estimator for low-metallicity galaxies by \citet{Denicolo2002}. Since the two emission lines involved in the N2 parameter are very close to each other spectrally, the use of this parameter requires neither reddenning correction nor flux calibration. We use the calibration relating N2 and metallicity from \citet{Pettini2004} and find 12 + log(O/H) = 8.39 $\pm$ 0.01 (statistical) $\pm$ 0.18 (calibration).

\item \indent The \textit{O3N2} parameter was  defined as O3N2 = log( [O\textsc{iii}] $\lambda$5007/H$\beta$ / [N \textsc{ii}] $\lambda$6584/H$\alpha$) and introduced as an abundance estimator by \citet{Pettini2004}. Using their calibration, we find 12 + log(O/H) = 8.37 $\pm$ 0.01 (statistical) $\pm$ 0.14 (calibration) for the integrated spectrum of the main emission region.

\end{enumerate}
 \indent Table \ref{integrated Z} and Figure \ref{summary Z} summarise our metallicity estimates of the main emission region from all the methods described above. The error on each of the metallicity estimates takes into account both the statistical error (i.e. in flux measurement) and the calibration error associated with each of the empirical calibrators. We find that the metallicity estimates using all indirect methods agree with each other within the errors, and all of them are at significant offsets (0.3--0.5 dex) from the metallicity obtained using the direct method. The observed offset is consistent with other works \citep[e.g.][]{Kewley2008} which show that the theoretical and empirical calibrations produce oxygen abundances higher than those derived from the direct method. 
  
\subsection{Spatially-resolved chemical abundances}  
  \label{spatially-resolved abundance}

\indent We create the metallicity maps of the FOV using the three indirect diagnostic methods described in Section \ref{integrated abundance}. We cannot map metallicity from the direct method due to poor S/N in the [O \textsc{iii}] $\lambda$4363 line required to map T$_e$. Hence, we assume a constant temperature H \textsc{ii} region and map Z [T$_e$] using the constant T$_e$ of the integrated spectrum of the main emission region and the maps of relative intensities of  collisional-to-Balmer emission lines (i.e. O$^+$/H$^+$ and O$^{2+}$/H$^+$).  Figure \ref{all Zs} presents the metallicity maps obtained from the four methods. Scales of all maps are adjusted to the same limits so that the relative offsets between metallicities from different methods and variations across the FOV are clear.  Due to the low S/N of [O \textsc{ii}] $\lambda\lambda$3727,29 (see Figure \ref{observed flux}), the Z [T$_e$] and Z [R$_{23}$] maps only show the values for the main emission region and $\sim$ 5--10 pc area surrounding it. The metallicity maps obtained from N2 and O3N2 parameters (Z [N2] and Z [O3N2] respectively) cover larger area and are mainly limited by the low S/N of  [N \textsc{ii}] $\lambda$6584 in the low-brightness region to the south-east of the H \textsc{ii} region. Large offsets between metallicity maps obtained from indirect method and direct method are clearly seen from the colour scale. Table \ref{integrated Z} presents the mean metallicities, maximum deviation within each metallicity map, along with the calibration uncertainty associated with each indirect method employed. The quoted error on the mean is the 1-$\sigma$ standard deviation across each map. These values are particularly useful in identifying any chemical variation of the ionised gas in the star-forming region and its surroundings. We find that the standard deviation ($\sigma$) and maximum variations ($\Delta$) are comparable in Z [T$_e$], Z [N2] and Z [O3N2] ($\sigma\sim$ 0.07 dex and $\Delta\sim$ 0.3 dex) while they are a factor of 2 higher in Z [R$_{23}$] ($\sigma\sim$ 0.15 dex and $\Delta\sim$ 0.6 dex).


 \indent We further investigate the possible chemical variation in Figure \ref{Z scatter}, which shows scatter plots of metallicity values with error bar in each spaxel of Z [T$_e$] (left) and Z [R$_{23}$] (right) maps.  The error bar on each point in both of the plots includes the error on flux measurement as well as the calibration error. The dashed and dotted line in each of the plots indicate the mean and standard deviation of all metallicities in the corresponding metallicity maps. In both of the panels, we find that the individual metallicity values with error bars lie within the 1-$\sigma$ standard deviation of the mean metallicity value. We do not show the scatter plots for Z [O3N2] and Z [N2] because the calibration uncertainties on these metallicity calibrations errors are high ($\sim$0.14 dex and $\sim$0.18 dex on O3N2 and N2 respectively), due to which all metallicity values are within 1-$\sigma$ standard deviation of the mean metallicity value.  Hence we conclude that the ionised gas in the central H \textsc{ii} region in NGC 4449 and its surrounding  is chemically homogeneous.


\begin{figure}
\centering
 \includegraphics[width = 0.48\textwidth]{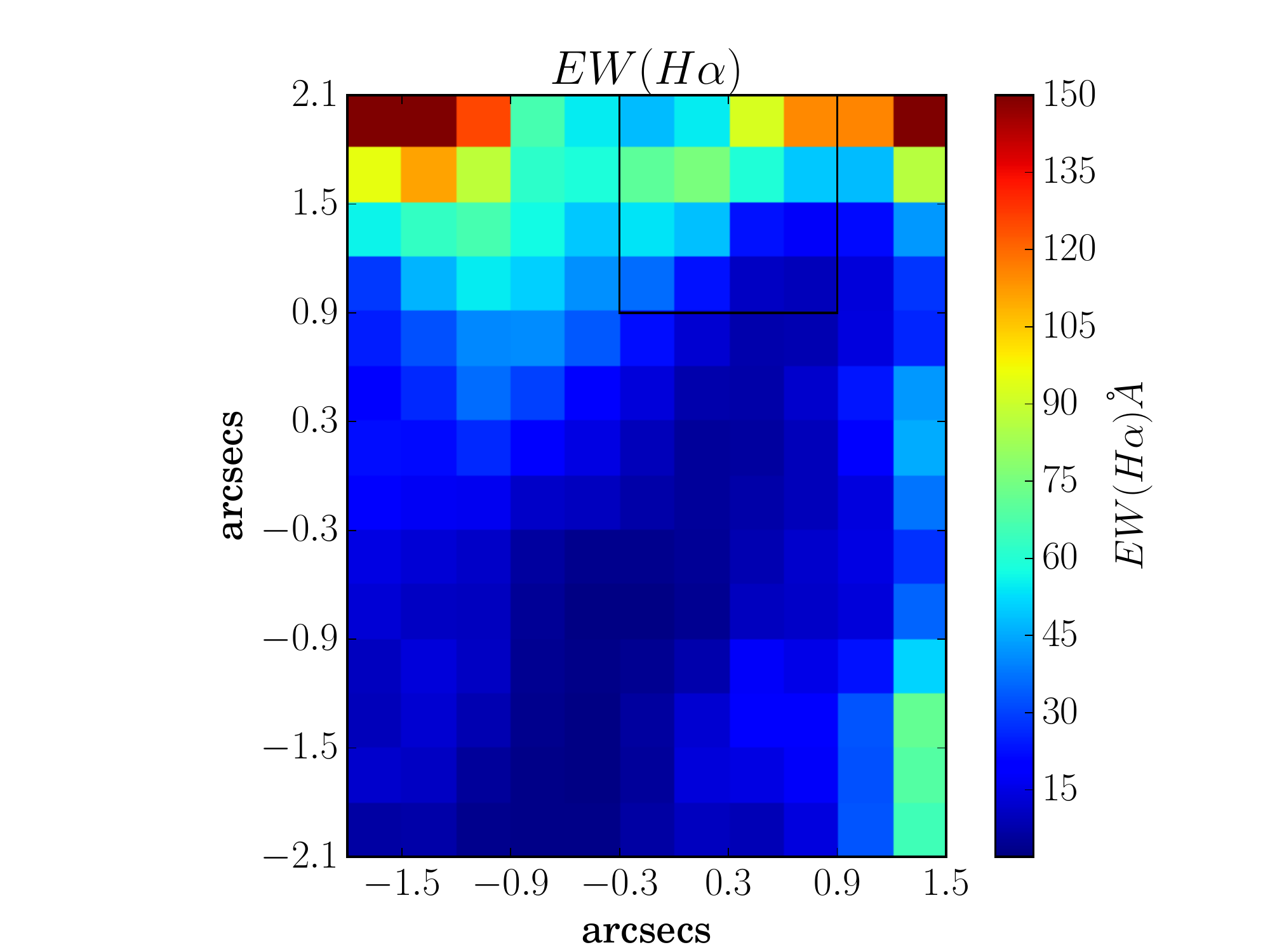}
\caption{Map of the equivalent width of H$\alpha$. The black square box shows the main emission region.}
\label{EW}
\end{figure}

\begin{figure}
\centering
\includegraphics[width = 0.48\textwidth]{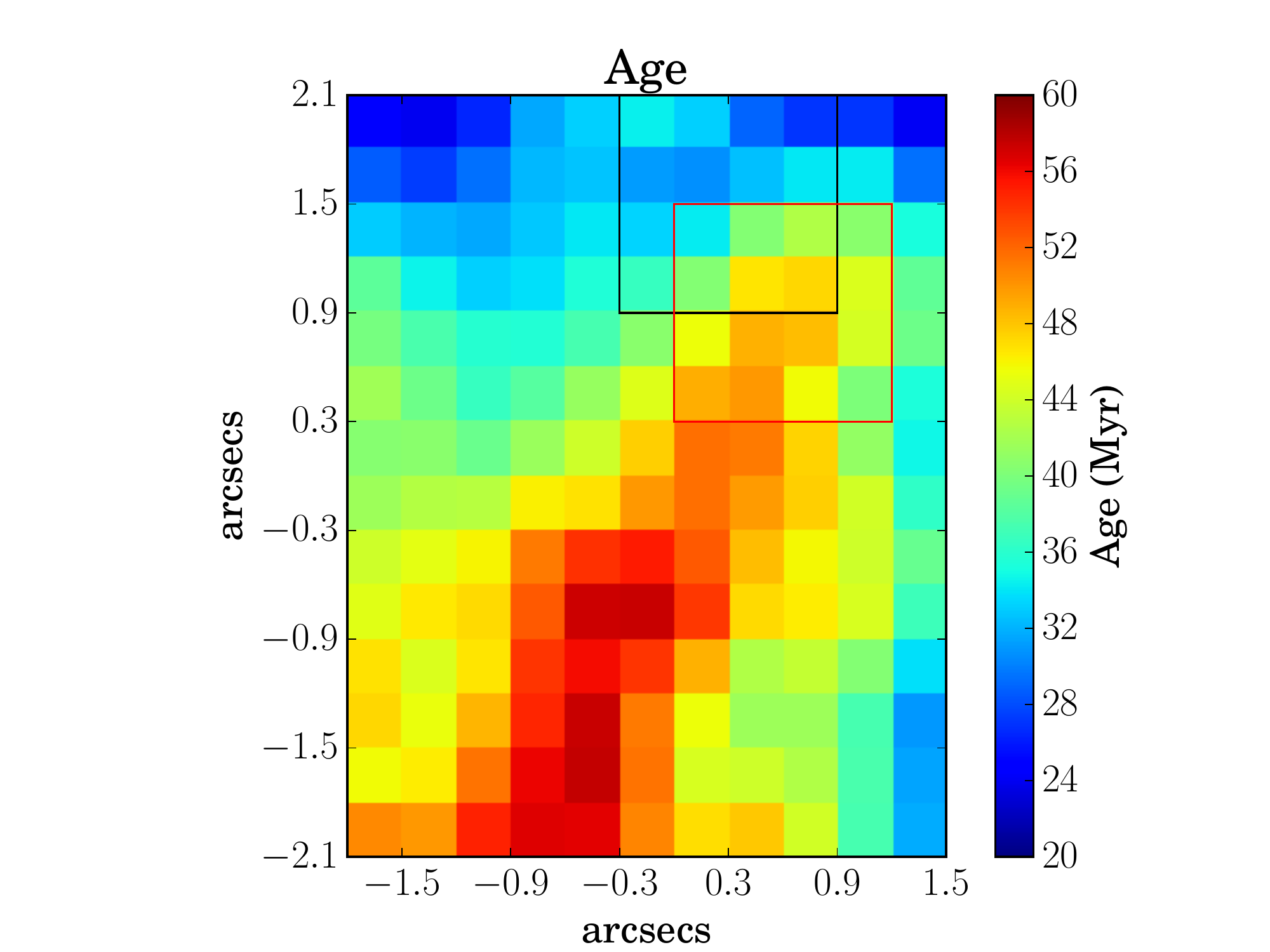}
\caption{Age map in Myr calculated from Starburst99 models at the ``constant'' metallicity Z = 0.15 Z$_{\odot}$ (calculated from the integrated spectrum of the main emission region). The black and red square box show the main emission region and the continuum region for reference respectively.}
\label{Age}
\end{figure}

 \begin{figure*}
 \centering
 \includegraphics[width = 0.48\textwidth]{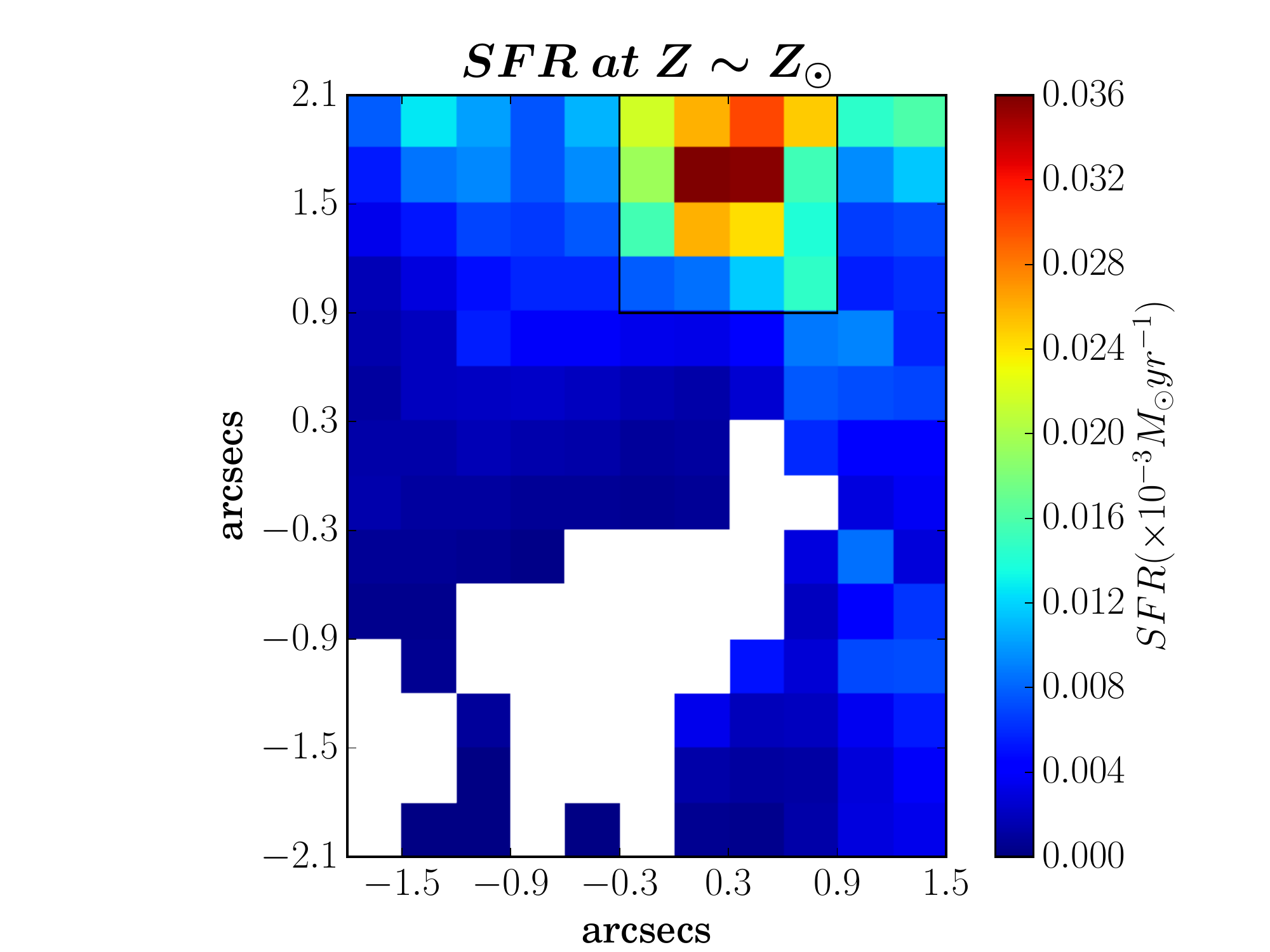}
\includegraphics[width = 0.48\textwidth]{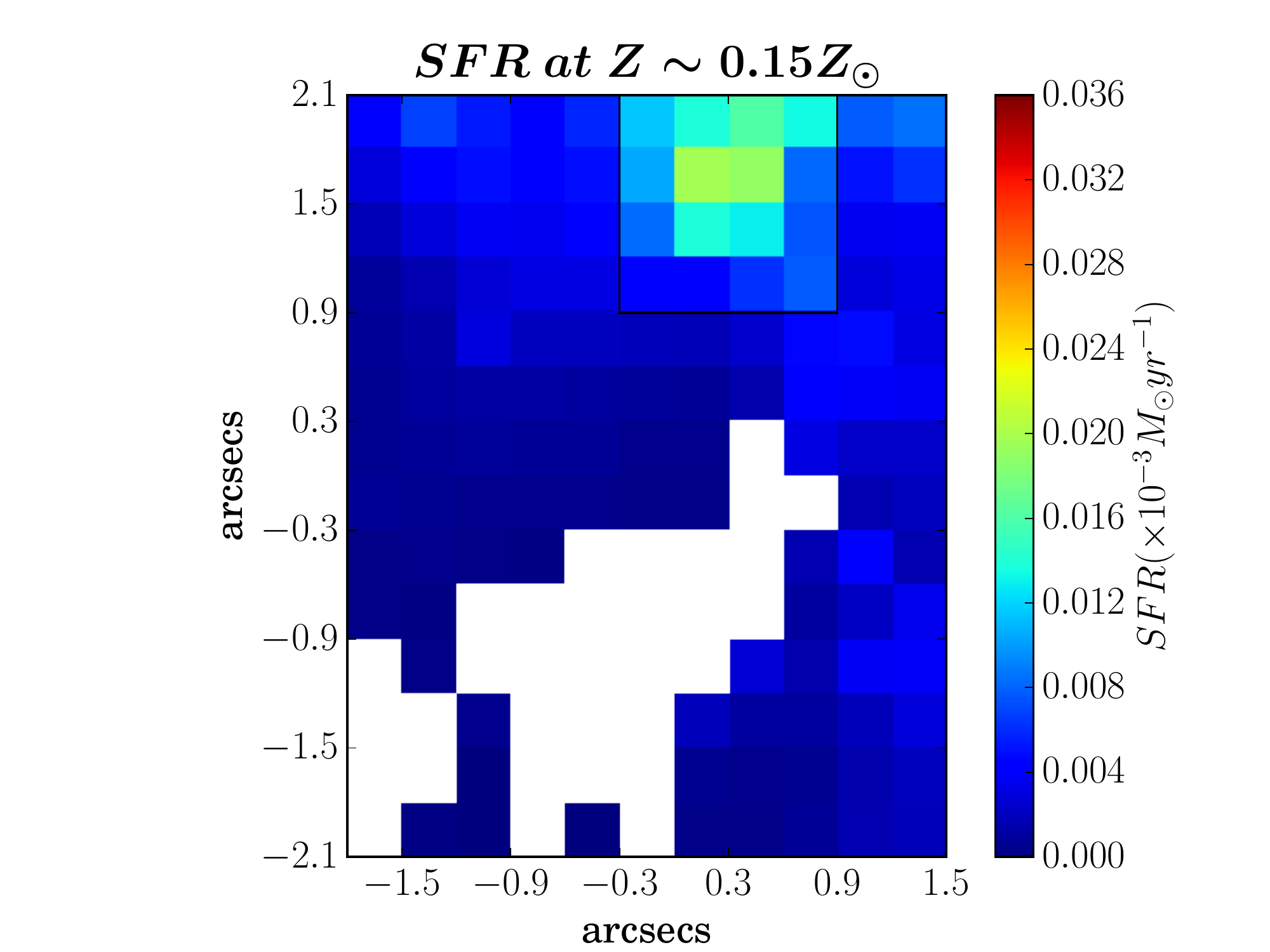}
\caption{SFR maps of the FOV in units of $\times$ 10$^{-3} $M$_{\odot}$ yr$^{-1}$ at Z$_{\odot}$ (left) and 0.15 Z$_{\odot}$ (right). The black square box shows the main emission region and white spaxels correspond to the spaxels in which emission line fluxes had S/N $<$ 3.}
\label{SFR}
\end{figure*}

\section{Stellar Properties}
\label{stars}

    
\subsection{Age of stellar population}


\indent The integrated spectra of  the main emission region  (Figure \ref{spectra}, upper two panels) as well as its surroundings (i.e. whole FOV - main emission region; lower two panels show Balmer emission which are indicative of young stellar systems (O and B stars) embedded in gas and absorption lines are characteristic of early-type (A-type) stars. None of these spectra show any obvious Wolf-Rayet (WR) features, despite previous reports of WR stars in other regions of NGC 4449 \citep{Milisavljevic2008, Sokal2015}.

\indent  To determine the age of the current ionising population, we first map the equivalent width (EW) of the strongest hydrogen recombination line H$\alpha$ as shown in Figure \ref{EW}. The EW values vary from 15--150 \AA \hspace{0.01in} across the entire FOV. Note here that the EWs of other recombination lines (H$\beta$, H$\gamma$) can also be used for age-dating, but these lines in our spectra also show broad photospheric absorption which can introduce considerable uncertainties on the age determined from their equivalent widths.

\indent We employ a metallicity-dependent age-dating method in which, we estimate the age of the currently ionising stellar population by comparing the measured EW(H$\alpha$) with those predicted by the evolutionary synthesis models of Starburst99 \citep{Leitherer1999} at a constant metallicity of 0.15 Z$_{\odot}$ across the chemically-homogeneous FOV (Section \ref{chemistry}). Each spatial pixel in the FOV has a size of 5.5 pc, which is smaller than the size of an individual giant H $\textsc{ii}$ region ($\sim$ 100--1000 pc). This is small enough that the free-fall time at this spatial scale is much less than the evolutionary timescale of the massive stars, and hence the star-formation activity can be described by an instantaneous starburst \citep{Leitherer1995}. For generating models we assume instantaneous star-formation with the epoch of the onset of star-formation as 0.01 Myr and the oldest age of the model as 100 Myr. We adopt a Salpeter-type initial mass function  (IMF) \citep{Salpeter1955} with the mass boundaries 1 M$_{\odot}$ and 100 M$_{\odot}$ and total stellar mass extent of 10$^6$ M$_{\odot}$ (the default mass). We choose  Geneva tracks with standard mass-loss rates over Padova tracks, though note here that a different choice of the evolutionary tracks produces a relatively small change in the predicted ages  \citep[$\sim$20--23\% as found by][]{James2010, James2013}. We use recommended more realistic models of Pauldrach/Hiller of expanding atmosphere. At the specified metallicity and IMF, the Starburst99 code first generates the mass-distribution of the stellar population. The stellar evolutionary track is then used to predict the EW(H$\alpha$) and its variation in specified time-steps of 0.1 Myr in the time-period between the zero-age main sequence and the end-point of the stellar evolution. These time steps are representative of the age at a given EW(H$\alpha$). If we take into account the stellar rotation in the starburst models, we find an insignificant increase of $\sim$ 3\% in the age. 

\indent Figure \ref{Age} shows the resulting age map which varies between 20--60 Myr. This result is quite interesting since it is longer than the ionisation lifetime of an H \textsc{ii} region  \citep[up to 5 Myr,][]{KennicuttEvans2012}. These results should be interpreted with caution because of the systematic uncertainties involved in modelling. For example, we do not take into account the internal attenuation due to dust present in the region of study. Moreover the observed continuum in a region is the combination of the nebular continuum and stellar continuum resulting from stellar population of different ages in the same region \citep[see for example,][]{Cantin2010}. Note here that at a given EW(H$\alpha$), the derived age will be even longer if the star-formation history is assumed to be continuous rather than instantaneous. This is because a continuous star-formation history will result in continuous production of younger stars as well as an increased fraction of older stars, which will decrease the EW(H$\alpha$) and increase the derived age. We find that the age of the main emission  region (black box) is relatively younger than the continuum region (red box). The spatial distribution of older stars with respect to the younger stellar population has approximately the same shape as the bow-shock region found earlier in Section \ref{line ratio maps}, and hence the older stellar population could be the possible cause of the shock.  

\indent The age of the stellar population in NGC 4449 has been estimated in various observational studies at global and spatially-resolved scales using different modelling techniques. For example, \citet{Karczewski2013} analysed the global optical spectrum of NGC 4449 by using the spectral fitting code \textsc{starlight} and found that only  1\% of the mass fraction of the stellar populations in NGC 4449 have age $<$ 10 Myr, 20--25\%  is  $\sim$ 100 Myr old and 60-75\%  of $>$ 1000 Myr. Spatially-resolved stellar mass-estimates of the entire galaxy are required to compare the ages derived in their work with ours. However, considering that the region of NGC 4449 studied in the present work is very bright (see Figure \ref{sdss}), young and massive stars must be present there, which is likely to correspond to the mass-bin of 20-25\% corresponding to the age of $\sim$ 100 Myr. This is in good agreement with the age range of 20--60 Myr obtained from our analysis. \citet{Boker2001} reports the detection of very deep Calcium triplet (CaT) in absorption based on their optical \'echlette spectra of the nucleus of NGC 4449 and hence predicts a younger cluster age between 5--20 Myr, which is in relatively good agreement with our results of the main emission region whose average age is found to be $\sim$ 30 $\pm$ 3 Myr. This age-estimate is consistent with the results of \citet{Gelatt2001} who report the detection of $^{12}$CO(2, 0) and $^{12}$CO(3, 1) absorption features in the near-infrared spectrum of  the central cluster of NGC 4449, indicative of the presence of cool stars (7--100 Myr). However our age-estimate is higher than the age-limit of 8--15 Myr derived in \citet{Gelatt2001}, obtained by comparing the colour of central region with cluster evolutionary models fitted to the clusters in the galaxy.  The discrepancy in the results is likely due to the metallicity Z = 0.4 Z$_{\odot}$ assumed in \citet{Gelatt2001} which is significantly higher than the metallicity (Z = 0.15 Z$_{\odot}$) of NGC 4449. We analyse our age-estimates in light of the star-formation history of NGC 4449 derived by \citet{McQuinn2010} using a colour-magnitude diagram of the stellar population in the entire galaxy. These authors report a lower average activity in the last 50 Myr than that in the last 4--10 Myr. This time bin is smaller than their inherent temporal resolution ($\sim$25--50  Myr) of star-formation history  and hence should be interpreted with caution. In the case where the galaxy has undergone starbursts in the last 10 Myr, we expect the stellar age of the corresponding star-forming regions to be younger. The distribution of stellar populations younger than $\sim$ 10 Myr is presented in Figure 20 of \citet{Annibali2008}, which shows that the majority of the younger stellar population does not reside in the central star-forming region. Moreover, \citet{Annibali2008} also note that over-crowding prevents the detection of the older stellar population in the central star-forming region. Thus the results of \citet{McQuinn2010}, combined with those of \citet{Annibali2008} are in agreement with the  average age-estimate of the central star-forming region found in our analysis. It may be possible that the central star-forming region is in a post-starburst regime.

 
 \subsection{Star Formation Rate}
  
\indent We map the SFR at Z = 0.15 Z$_{\odot}$ (Figure \ref{SFR}, right panel) using the following metallicity-dependent relation between dereddened luminosity of H$\alpha$, L(H$\alpha$) and SFR from \citet{Ly2016}: $log[\frac{SFR}{L(H\alpha)}] = - 41.34 + 0.39y + 0.127y^2$, where y = log(O/H) + 3.31\footnote{y = 0 is the solar oxygen abundance.}. This relation is derived from the Starburst99 models assuming a Padova stellar track and Chabrier IMF \citep{Chabrier2003}. We also map the SFR derived at the solar metallicity (Figure \ref{SFR}, left panel) for comparison using the L(H$\alpha$)--SFR relation: $SFR =  \frac{L(H\alpha)}{1.26 \times 10^{41} ergs s^{- 1}}$. This formula was initially calibrated  by \citet{Kennicutt1998} for Salpeter IMF and we multiplied it by a factor of 0.63 to convert it to Chabrier IMF.  The chosen IMF has more realistic distribution at lower stellar masses and matches the IMF assumption of  \citet{Ly2016}. Figure \ref{SFR} shows that at Z = Z$_{\odot}$, SFR varies from 0--0.04 $\times$ 10$^{-3}$ M$_{\odot}$ yr$^{-1}$ (left panel), while it varies from 0--0.02 $\times$ 10$^{-3}$ M$_{\odot}$ yr$^{-1}$ at Z = 0.15 Z$_{\odot}$ (right panel).  The spatially-resolved (spaxel-by-spaxel) values of SFR given here should be interpreted with caution because the SFR recipes fail at these scales because of the stochastic sampling of IMF. From the integrated spectrum of the main emission region, we find SFR = 0.328 $\pm$ 0.005  $\times$10$^{-3}$ M$_{\odot}$ yr$^{-1}$ at solar metallicity and SFR = 0.176 $\pm$ 0.003 $\times$ 10$^{-3}$ M$_{\odot}$ yr$^{-1}$ at Z = 0.15 Z$_{\odot}$, which correspond to the SFR surface density (SFRD) of 0.67 M $_{\odot}$ yr$^{-1}$ kpc$^{-2}$ and 0.34 M $_{\odot}$ yr$^{-1}$ kpc$^{-2}$ respectively. We find a good agreement between our SFRD at solar metallicity  ($\sim$ 0.67 M $_{\odot}$ yr$^{-1}$ kpc$^{-2}$) to that found by \citet[][$\sim$ 0.76 M $_{\odot}$ yr$^{-1}$ kpc$^{-2}$]{Buckalew2005}\footnote{The SFRD value takes into account the IMF conversion from Salpeter to Chabrier.}. For both the integrated spectrum of the main emission region and the spatially-resolved data, we find that the SFR at solar metallicity is higher than SFR at sub-solar metallicity in the H \textsc{ii} region and its surroundings. This is because the atmosphere of the metal-poor O stars are less-blanketed at sub-solar metallicity which results in an increased escape fraction for the ionising photons and hence requires a lower SFR. 


\indent  The SFR of the central star-forming region in NGC 4449 both at solar and sub-solar metallicity, is comparable to the SFR of star-forming regions in other BCDs, e.g. \citet{Buckalew2005} reports SFR of 0.208 $\times$ 10$^{-3}$ M$_{\odot}$ yr$^{-1}$ and 1.53 $\times$ 10$^{-3}$ M$_{\odot}$ yr$^{-1}$ for two different star-forming regions in Mrk 178, a BCD found at approximately the same distance ($\sim$ 4.5 Mpc) and has similar metallicity (12 + log(O/H) =  7.82) as NGC 4449.  The literature suggests that the SFRs of BCDs might span a range of 10$^{-3}$ to 10$^2$ M$_{\odot}$ yr$^{-1}$, with the mean SFR varying from 0.1 M$_{\odot}$ yr$^{-1}$ to 10 M$_{\odot}$ yr$^{-1}$ \citep{Fanelli1988, Sage1992, Hopkins2002}.  

\section{Discussion: Possibility of chemical inhomogeneities}
\label{discussion}

\indent Comparing the metallicity values of the integrated spectrum of the main emission region and the average values of all the metallicity maps in Table \ref{integrated Z}, we find that the global values agree very well with the spatially-resolved values irrespective of the metallicity diagnostic used.  For both integrated spectra and spatially-resolved analysis, we find that the indirect metallicity values (Table \ref{integrated Z}) are offset by about 0.3--0.5 dex from the direct method metallicity (12 + log(O/H) = 7.88 $\pm$ 0.14) estimates. This offset has been observed before for different data sets \citep[e.g.][]{Kewley2001, Sanchez2011} and depends on many factors including the assumptions on the photoionization models employed for calibrating the indirect diagnostics. 

\begin{figure}
\centering
\includegraphics[width = 0.45\textwidth]{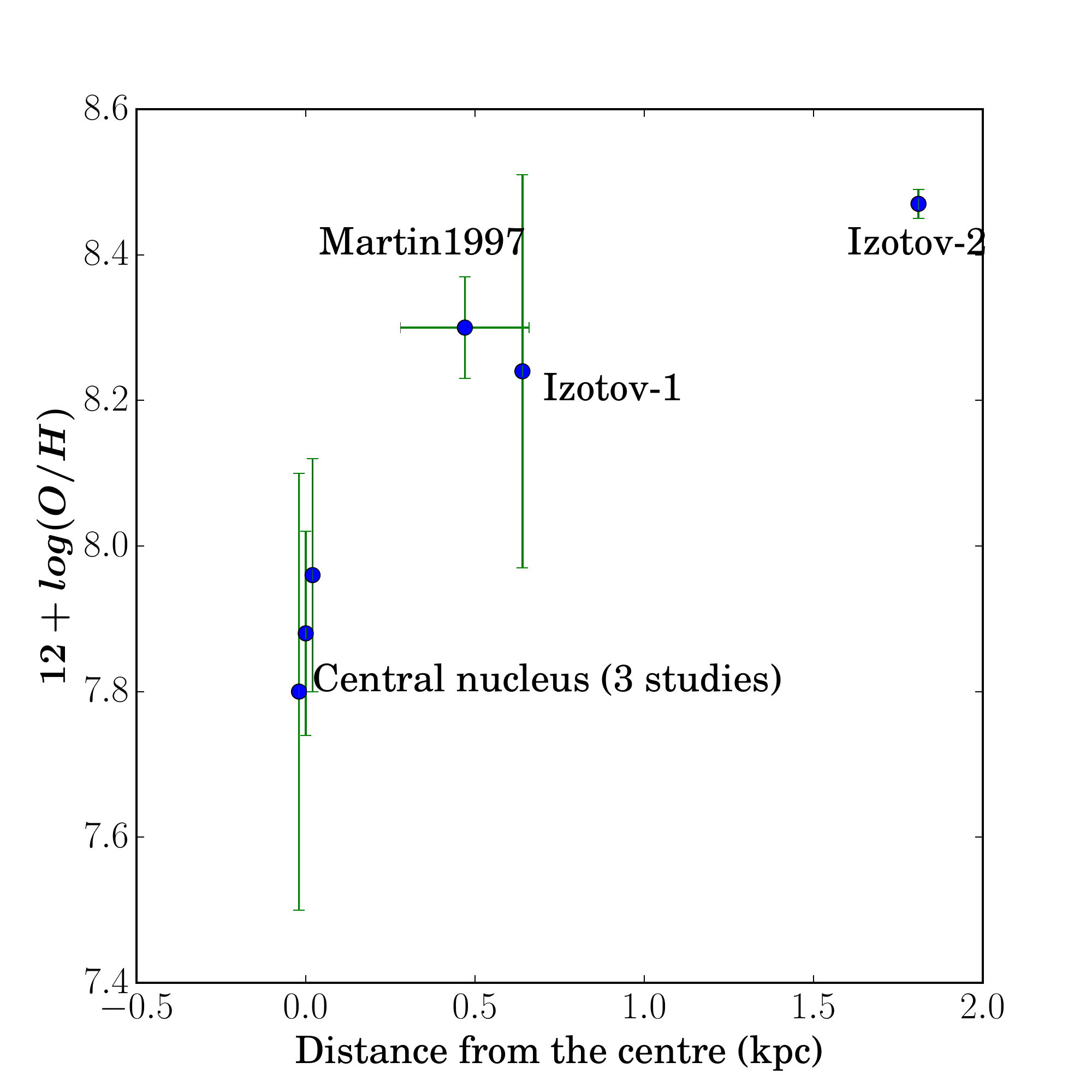}
\caption{Variation of metallicity estimates with respect to distance from the centre. ``Central nucleus (3 studies)" refers to the central H $\textsc{ii}$ region (RA, Dec: 187.046567, +44.0934) from present study (7.88 $\pm$ 0.14),  \citet[][7.8 $\pm$ 0.3]{Boker2001} and \citet[][7.96 $\pm$ 0.16]{Buckalew2005}. Note here that the two data points (excluding ours) corresponding to the central nucleus have been shifted on the x-axis to avoid overlap at 0 kpc. ``Martin1997" refers to the average metallicity of two long-slit spectroscopic observations from \citet{Martin1997} at two different locations from the central star-forming region.  ``Izotov-1" and ``Izotov-2" refer to the other two pointings in NGC 4449 ((RA,Dec): (187.03861, +44.08556) and (187.05776, +44.11956)) from \citet{Izotov2006}. }
\label{Z literature}
\end{figure}


Figure \ref{Z literature} shows the variation of metallicity estimates with respect to distance from the centre, and includes metallicity estimate from the integrated spectrum of the central star-forming region from this study as well as from the literature for different regions. In this figure the ``Central nucleus (3 studies)" refers to the central H $\textsc{ii}$ region (RA, Dec: 187.046567, +44.0934) from the present study (7.88 $\pm$ 0.14, direct method),  \citet[][7.8 $\pm$ 0.3]{Boker2001} and \citet[][7.96 $\pm$ 0.16\footnote{We estimated this metallicity from the flux measurements given in \citet{Buckalew2005}.}]{Buckalew2005}.  ``Martin1997" refers to the average metallicity of two long-slit spectroscopic observations from \citet{Martin1997} at two different locations from the central star-forming region. Here we have approximated distance and the error bar as the mean and the standard deviation of the distances between the central star-forming region and the mid-points of the two long-slit position. Note here that the metallicity measurements of \citet{Martin1997} are ambiguous with respect to the number of slits (two or three) used in the measurement. The trend in the plot is unaffected even if we assume three instead of two slits for measuring distance, hence our conclusion remains the same. ``Izotov-1" and ``Izotov-2" refer to the other two pointings in NGC 4449  in the north and south-west of the centre respectively. This plot suggests an overall increase in metallicity with respect to the distance from the centre.

\indent The above comparison indicates that chemical inhomogeneities do exist across the NGC 4449.  Metallicity inhomogeneities have been observed before in other galaxies. \citet{Sanchez2014} observed low metallicity inner regions with enhanced star-formation activity in two out of five dwarf star-forming galaxies with measurable metallicities. \citet{Sanchez2015} found low-metallicity starburst regions in nine out of ten extremely-metal poor galaxies. \citet{Elmegreen2016} reported a drop in metallicity at the star-forming head in the tadpole galaxy Kiso 5639. All these previous works interpret their finding of metallicity drops in the star-forming regions as a result of accretion of metal-poor gas. However, other mechanisms such as gas outflows, supernova blowout and fallback can also lead to chemical inhomogeneities as are discussed by \citet{Werk2010} in the case of BCD NGC 2915.  In the following sections we discuss alternative possible mechanisms that may yield a more metal-poor central star-forming region. 
 
\subsection{Metal-poor gas accretion from merger-event or cosmic web}

\indent Low metallicity at low-redshift can be caused by the accretion of either low-metallicity gas from a merger event or the pristine gas from the cosmic web mixed with the metal-rich gas of the galaxy.  The accretion of gas onto the galaxy will not only dilute the metallicity but also may increase the star-formation rate. 

\indent To investigate the scenario of increased star-formation and decreased metallicity caused by the accretion of gas in NGC 4449, we compared the H$\alpha$ fluxes (from the HST narrow band F658N image) in the central star-forming region as well as the two pointings of \citet{Izotov2006} (e.g. see Figure \ref{Z literature}). We find that the central star-forming region is the brightest of all the three star-forming regions\footnote{Our flux estimates take into account the attenuation correction.}, which indicates that star-formation is enhanced in the centre compared to other regions in the galaxy. Moreover, the central star-forming region also shows an increased concentration of atomic gas as revealed by the H $\textsc{i}$ map of this galaxy obtained from Very Large Array (VLA) in \citet{Hunter1998}, indicating the possible inflow of gas in the centre. 


\indent Merger events can cause inflow of gas in a galaxy, particularly in very dense environments, as demonstrated by numerical simulations. For example, \citet{Bekki2008} shows that a merger event between gas-rich dwarf galaxies can lead to the formation of a BCD, leading to the transfer of metal-poor gas from the outer parts of the merger progenitors to the centre of BCDs. NGC 4449 is quite likely currently interacting with a nearby dwarf galaxy \citep{Martinez-Delgado2012} and has potentially undergone interactions in the past \citep{Hunter1998, TheisKohle2001}. The low-metallicity gas in the central star-forming region could well be from a now disrupted gas-rich satellite.


\indent Cosmic-web gas can also feed galaxies with low-metallicity gas, sustaining the star-formation and creating chemical inhomogeneities, as suggested by cosmological simulations \citep{Ceverino2016} and several observational studies.  An inverse metallicity gradient is observed in three rotationally-supported star-forming galaxies at z=3 by \citet{Cresci2010}, who conclude that the low-metallicity centre in these galaxies is the result of dilution by accreted primordial gas \citep[``Cold flow model"][]{Dekel2009}. Our finding of low-metallicity in the central star-forming region of NGC 4449 is similar to the finding of \citet{Sanchez2015} who reports localised starbursts associated with low-metallicity gas in extremely metal-poor galaxies in the local Universe and suggests its origin to be the infall of low-metallicity cosmic clouds. However, the infall of gas may not always lead to chemical inhomogeneities as is the case in BCD NGC 5253, where a flat metallicity profile is observed \citep{Westmoquette2013} in spite of an infall of low-metallicity H \textsc{i} gas \citep{Lopez2012a}. Hence accretion of cosmic gas may or may not be the cause of the observed low-metallicity in the central star-forming region in NGC 4449.

\subsection{Metal-rich gas loss via  supernovae-driven blowout or galactic winds}

\indent High-metallicity gas can be easily lost from low-mass galaxies like NGC 4449 because of the shallow gravitational potential.
The gas loss can take place via supernova explosions  \citep{MacLow1999}, galactic winds generated by supernovae \citep{Tremonti2004}, or energy-driven outflows \citep{Dave2013} in dwarf galaxies. 


\indent Although supernova remnant candidates have been proposed for NGC 4449 \citep{Summers2003},  no supernova explosion/remnant or outflows are reported near the central star-forming region of NGC 4449.  The closest suggested supernova remnant is about 1$\arcmin$  north of the central star-forming region \citep{Seaquist1978, Kirshner1980}, with size $>$ 1.2 pc, age $<$ 100 yr and expansion velocity of 6000 km s$^{-1}$ \citep{Milisavljevic2008}. With this expansion velocity, this supernova is unlikely to affect the  central star-forming region at a distance of 1.1 kpc from the explosion. However, the age of the stellar populations of the central star-forming region are $\sim$ 20--60 Myr, which is old enough to harbour earlier core-collapse supernovae and hence cannot be ruled out as a possible cause of metal-rich gas loss leading to low-metallicity in the centre of NGC 4449.
 

\indent Galactic superwinds can occur when supernova explosions occur in galaxies with a high local star formation rate, which may trigger chimneys ejecting the metals out of the galaxy. Similarly, stellar winds from massive stars can also lead to metal-enriched gas loss. Gas loss in NGC 4449 has been predicted based on the H$\alpha$ morphology, but outflows coud also be prevented by the low-density neutral gas around the galaxy \citep{Ott2005}. This scenario is somewhat similar to BCD NGC 5253, where the unusually high densities and  pressures  in the supernebula and the surrounding medium stalls the ionised gas outflow from the supernebula cluster winds \citep{Westmoquette2013}. However, galactic winds may cause outflows in BCDs as suggested by various observational studies, e.g.  \citet{Meurer1992} reports an outflow H$\alpha$ morphology in the nucleated BCD NGC 1705, suggesting that stellar winds and supernova from recent cluster-formation in the nucleus power the outflow. Large scale outflows has been reported in another BCD, Mrk 71 possibly caused by the winds from massive stars \citep{Roy1991, James2016}. Thus, galactic winds can transport metal-rich gas out of galaxies depending on the environment, and hence cannot be ruled out as a possible cause of the chemical inhomogeneity in NGC 4449.


\indent 

\indent \textbf{Alternative Mechanisms:} Two possible scenarios are presented in the above discussion where the transport of gas in and out of the galaxy can lead to metallicity inhomogeneities.  It is also possible that the metal-rich star-forming region on the outskirts of the galaxy could have been formed from the pre-enriched gas accreted from a past-merger event, for example from DDO 125. There can also be another scenario where low-metallicity gas is formed in-situ and has remained un-enriched within the galaxy. In this case, supernova explosions may eject the metal-enriched gas into the galactic halo which may rain back to the outskirts of the galaxy causing a higher metallicity in the outer regions without any gas loss.  Such a scenario is proposed for BCD NGC 2915 by \citet{Werk2010} where they find the outer disk to be overabundant and the central star-forming core to be underabundant for the gas fraction in the respective regions. The low-metallicity central star-forming region in NGC 4449 can possibly be explained by this alternative mechanism without invoking any accretion or loss of gas from the galaxy.


\indent 

\indent A spatially-resolved kinematic and chemical abundance study complemented by a detailed star-formation study of NGC 4449 is required to determine the cause of observed low-metallicity central star-forming region in this galaxy. This can be achieved by carrying out a spatially-resolved spectroscopic (IFU) study of the entire galaxy, along with the central star-forming region. The Keck Cosmic Web Imager (KCWI) would be an excellent instrument available in the northern hemisphere for this type of study because of its large FOV and full wavelength coverage, which allow mapping important emission lines ([O \textsc{ii}] $\lambda\lambda$3727, 3729) in the blue-end of the optical spectrum. This can be complemented by multiwavelength data, particularly H \textsc{i} which provide signatures of cold-flow streams (if present) and  CO (e.g. using Very Large Baseline Array), which will trace the cool gas fuelling star-formation. This would allow us to confirm the chemical inhomogeneity across the galaxy from a single study and also determine its cause by mapping the gas kinematics and signatures of gas flows - in, out and within the galaxy.

\section{Summary}
\label{summary}
\indent We have carried out a spatially-resolved study of the central H \textsc{ii} region in the blue compact dwarf galaxy NGC 4449 using GMOS-N IFS data. We summarise our main results with respect to the questions we posed in the introduction.

\indent (a) Do chemical inhomogeneities exist at the parsec-scale in this star-forming region? In order to investigate the possibility of chemical inhomogeneity across the central star-forming region, metallicity maps of the entire region of study were created using both direct and indirect methods. The region under study is chemically homogeneous. The average values of metallicity across the FOV agreed with the values obtained from the integrated spectrum of the main emission region for each diagnostic (T$_e$, R$_{23}$, N2 and O3N2). We found a metallicity of 12 + log(O/H) = 7.88 $\pm$ 0.14 from the integrated spectrum of the main emission region from the direct T$_e$ method. Comparison with direct-method metallicity measurements (from the literature) of other regions in this galaxy show a radial increase in metallicity and hence signatures of chemical inhomogeneity at kiloparsec-scales.

\indent (b) What are the possible ionisation mechanisms at play in the gas surrounding this H \textsc{ii} region? A complex velocity structure of ionised gas  is found in the central H \textsc{ii} region of NGC 4449 and its surroundings. This includes distinct and blended shell structures indicating turbulence in the star-forming region, which could be due to stellar winds from a nearby stellar cluster or bow shocks possibly triggered by recent merger events from satellite galaxies to the south of NGC 4449.  A BPT diagnostic diagram using [N \textsc{ii}]/H$\alpha$ flux ratios showed photoionisation to be the dominant source of ionisation, as expected for a H \textsc{ii} region. However we also find signatures of shock ionisation  in the BPT diagram from the [S \textsc{ii}]/H$\alpha$ line flux ratio, confirming the shock signatures in the velocity structure of the ionised gas. We also mapped the ionisation parameter (log $\mathcal{U}$) using the O32 diagnostic and find a radial decrease of log $\mathcal{U}$ from the peak of the continuum source.

\indent (c) What is the age of the stellar population currently ionising the gas in this region?  The age of the current ionising stellar population was mapped by comparing the equivalent widths of the recombination line (H$\alpha$) to that predicted by evolutionary synthesis models from Starburst99. The age of the stellar population in the region of study is found to vary between 20--60 Myr, which is old enough to harbour supernovae explosions in the past. The distribution of older stellar population appears to have approximately  the same shape as the apparent bow-shock in this region, suggesting that the bow-shock may in-fact be linked to the stellar population in the surrounding area. 

\indent 


\indent  The central star-forming region of NGC 4449 appears to be metal-poor compared to other regions in the galaxy. Though we suspect that it is more likely due to accretion of low-metallicity gas due to an ongoing merger-event, other hypotheses such as the outflow of metal-rich gas, or gas loss via galactic winds or supernova-driven blowout cannot be discarded. An extensive spatially-resolved study of the entire galaxy NGC 4449 would provide a complete picture presenting the interplay between star-formation, chemical abundance and gas kinematics at spatially-resolved scale in the local Universe. NGC 4449, being an excellent local analogue of high redshift galaxies, will also shed light on our understanding of the complex interplay in the high redshift Universe. 


\section*{Acknowledgements}
\indent We thank the anonymous referee for the helpful comments, specifically on the stellar properties.
 NK thanks Rob Kennicutt (Institute of Astronomy, Cambridge) and the Nehru Trust for Cambridge University for the financial support during her PhD. This research made use of the NASA/IPAC Extragalactic Database (NED) which is operated by the Jet Propulsion Laboratory, California Institute of Technology, under contract with the National Aeronautics and Space Administration; SAOImage DS9, developed by Smithsonian Astrophysical Observatory"; IRAF, distributed by the National Optical Astronomy Observatory, which is operated by the Association of Universities for Research in Astronomy (AURA) under a cooperative agreement with the National Science Foundation; Astropy, a community-developed core Python package for Astronomy \citep{Astropy2013}
; APLpy, an open-source plotting package for Python hosted at http://aplpy.github.com; and Montage, funded 
by the National Science Foundation under Grant Number ACI-1440620, and was previously funded by the National Aeronautics and 
Space Administration's Earth Science Technology Office, Computation Technologies Project, under Cooperative Agreement Number 
NCC5-626 between NASA and the California Institute of Technology.  




\bibliographystyle{mnras}
\bibliography{biblio}





\bsp	
\label{lastpage}
\end{document}